\documentstyle[preprint,aps,psfig]{revtex}
\begin{document}

\title{Factors Governing the Foldability of Proteins}
\author{D.K.Klimov and D.Thirumalai}
\address{Institute for Physical Science and Technology and 
Department of Chemistry and Biochemistry\\
University of Maryland, College Park, Maryland 20742}

\maketitle

\centerline{phone (301) 405-4803}
\centerline{fax (301) 314-9404}

\begin{abstract}

We use a three dimensional lattice model of proteins to investigate
\underline{systematically} the global properties of the polypeptide chains
that determine the folding to the native conformation starting from an
ensemble of denatured conformations. In the coarse grained description,
the polypeptide chain is modeled as a heteropolymer consisting of \(N\)
beads confined to the vertices of a simple cubic lattice. The interactions
between the beads are taken from a random Gaussian distribution of
energies, with a mean value \(B_{0}\), that corresponds to the overall
average hydrophobic interaction energy. We studied 56 sequences all with a
unique ground state (native conformation) covering two values of \(N\)
(\(15\) and \(27\)) and two values of \(B_{0}\). The smaller value of
\(B_{0}\) was chosen so that the average fraction of hydrophobic residues
corresponds to that found in natural proteins. The higher value of
\(B_{0}\) was selected with the expectation that only the fully compact
conformations would contribute to the thermodynamic behavior.  For \(N =
15\) the entire conformation space (compact as well as noncompact
structures) can be exhaustively enumerated so that the thermodynamic
properties can be exactly computed at all temperatures. The thermodynamic
properties for the 27-mer chain were calculated using slow cooling
technique together with standard Monte Carlo simulations. The kinetics of
approach to the native state for all the sequences was obtained using
Monte Carlo simulations. For all sequences we find that there are two
intrinsic characteristic temperatures, namely, \(T_{\theta}\) and
\(T_{f}\). At the temperature \(T_{\theta}\) the polypeptide chain makes a
transition to a collapsed structure while at \(T_{f}\) the chain undergoes
a transition to the native conformation. We show that foldability of
sequences can be characterized entirely in terms of these two
temperatures. It is shown that fast folding sequences have small values of
\(\sigma = (T_{\theta} - T_{f})/T_{\theta}\), whereas slow folders have
larger values of \(\sigma \) (the range of \(\sigma\) is \(0 < \sigma <
1\)). The calculated values of the folding times correlate extremely well
with \(\sigma \). An increase in \(\sigma\) from 0.1 to 0.7 can result in
an increase of 5-6 orders of magnitudes in folding times. In contrast, we
demonstrate that there is no useful correlation between folding times and
the energy gap between the native conformation and the first excited state
at any \(N\) for any value of \(B_{0}\). In particular, in the parameter
space of the model, many sequences with varying energy gaps all with
roughly the same folding time can be easily engineered. Folding sequences
in this model, therefore, can be classified based solely on the value of
\(\sigma \). Fast folders have small values of \(\sigma \) (typically less
than about \(0.1\)), moderate folders have values of \(\sigma \) in the
approximate range between \(0.1\) and \(0.6\), while for slow folders
\(\sigma \) exceeds \(0.6\). The precise boundary between these categories
depends crucially on \(N\). The range of \(\sigma \) for fast folders
decreases with the length of the chain.  At temperatures close to
\(T_{f}\) fast folders reach the native conformation via a topology
inducing 
nucleation collapse mechanism without forming any detectable
intermediates, whereas only a fraction of molecules \(\Phi (T)\) reaches
the native conformation by this process for moderate folders.  The
remaining fraction reaches the native state via three stage multipathway
process. For slow folders \(\Phi (T) \) is close to zero at all
temperatures. The simultaneous requirement of native state stability and
kinetic accessibility can be achieved at high enough temperatures for
those sequences with small values of \(\sigma \). The utility of these
results for {\em de novo} design of proteins is briefly discussed. 

\end{abstract}

\section{Introduction}

        The mechanisms, by which proteins adopt well-defined three
dimensional topological structures, have been extensively investigated
theoretically \cite{Bryn95,Dill,Hinds94,Karp92,Thirum94,Wol} 
as well as experimentally \cite{Bald,Rad}. 
The major intellectual impetus for these
studies originate in the so called Levinthal paradox \cite{Levin}. 
Since the number of conformations of even relatively short proteins is
astronomically large Levinthal suggested that it would be impossible for
proteins to reach the native conformation by a random search through all
the available conformation space.  In the last few years several groups
\cite{Bryn95,Dill,Hinds94,Karp92,Thirum94} have provided, 
largely complimentary, possible theoretical resolutions to the Levinthal
paradox.  The unifying theme that emerges from these studies, all of which
are based on certain minimal model representations of proteins
\cite{Amara,Bryn89,Chan93,Chan94,Cov,Hinds92,Honey90,Gar88,Leop,Sali94a,Sali94b,Shakh89,Shakh91,Skol90,Skol91,Zwan95,Zwan92} 
is that due to
certain intrinsic preference for native structures proteins efficiently
explore the underlying rough energy landscape. Explicit computations of
the energy landscape for certain lattice models \cite{Cam95} 
reveal that foldable sequences (those that reach the native
conformations on a relatively fast time scales, which for real proteins is
typically of the order of a second) have relatively small free energy
barriers. These and related studies \cite{Chan93,Leop,Chan95} 
have emphasized the 
importance of the connectivity
between various low energy states in determining the kinetic foldability
of proteins. Thus, it appears that in order to fully elucidate the folding
kinetics of proteins it is necessary to understand not only the low energy
spectrum, but also how the various states are connected. 

It is of interest to wonder if natural proteins have been designed so that
the requirements of kinetic foldability and stability have been
simultaneously satisfied and if so how are they encoded in the primary
sequence.  If this is the case, then it follows that because protein
folding is a self-assembly process the kinetic foldability of proteins
should be described in terms of the properties of the sequence itself.
This argument would indicate that certain intrinsic thermodynamic
characteristics associated with the sequence may well determine the
overall kinetic accessibility of the native conformation. The minimal
models are particularly suitable for addressing this question. For these
models the folding kinetic rates for every sequence can be precisely
calculated for small enough values of \(N\) - the number of beads in the
model polypeptide chain. In addition the energy spectrum for small values
of \(N\) can be explicitly enumerated for lattice models and for moderate
sized proteins it can be computed by simulation methods. Thus, these
models afford a systematic investigation of the factors governing folding
rates. 

It is perhaps useful right at the outset to say a few words about the
lattice representation of polypeptide chains. The energies employed in the
minimal models (or in other knowledge based schemes) should be thought of
as estimates of potentials of mean force after other irrelevant degrees of
freedom are integrated out. In principle, this leaves one with an
effective potential surface involving only the protein coordinates. In the
minimal models one further coarse grains this force field by eliminating
all coordinates except, perhaps, those associated with centers of the
residues.  The lattice models further confine these centers to the
vertices of a chosen lattice. These arguments show that we can at best
expect only qualitative themes to emerge from these studies. 
Nevertheless, these simulations together with other theoretical ideas have
provided testable predictions for the kinetics of refolding of proteins. 

By using the random energy model (REM), originally introduced as the
simplest mean field spin glass model \cite{Der}, as a caricature of
proteins it has been proposed that the dual requirement (kinetic
accessibility of the native conformation as well as the stability) can be
satisfied if the ratio of the folding transition temperature \(T_{f}\) to
an equilibrium glass transition temperature \(T_{g,eq}\), at which the
entropy vanishes in REM, is maximized \cite{Bryn95,Gold92}. (It has been 
noted that in order to use this
criterion in lattice models \(T_{g,eq}\) has to be replaced by a kinetic
glass transition temperature \(T_{g,kin}\) \cite{Socci94}. 
It would be desirable to clarify the relationship between \(T_{g,eq}\) 
and \(T_{g,kin}\)). 

	Based partially on lattice simulations of proteins a plausible
relationship between folding rates and the ratio of \(T_{f}\) to the
collapse transition temperature \(T_{\theta}\) was conjectured a couple of
years ago. In particular, Camacho and Thirumalai have argued that the fast
folding sequences have small values of \(\sigma =
(T_{\theta}-T_{f})/T_{\theta}\) \cite{Cam93}.  The
theoretical reason for such an expectation has been given recently
\cite{Thirum95}. The advantage of the criterion, based on the
smallness of \(\sigma \) to classify fast and slow folding sequences, is
that both \(T_{\theta}\) and \(T_{f}\) are readily calculable from
equilibrium properties. More importantly, one can deduce \(T_{\theta}\)
and \(T_{f}\) directly from experiments. 

	More recently, Sali {\em et al.} \cite{Sali94a,Sali94b} have forcefully
asserted that for the class of minimal models of the sort described here
the \underline{necessary and sufficient} condition for folding (within a
preset time scale in Monte Carlo simulations) is that the native
conformation be separated from the first excited state by a large gap
(which is presumably measured in the units of \(k_{B}T\) with \(k_{B}\)
being the Boltzmann constant and \(T\) the temperature).  However, their
studies are incomplete and rest on untested assumptions. They restricted
their conformation space to only a search among all compact structures of
a 27-bead heteropolymer.  More importantly, they did not provide for their
model the dependence of the folding times as a function of the gap to
establish the kinetic foldability of any sequence. Thus there is no direct
evidence of the dependence of the folding times for various sequences and
the gap \(\Delta = E_{1}-E_{0}\) - which in the original study has been
stated as a mathematical theorem. Notice that this gap has been defined as
the difference between the two lowest energy levels assuming that both
these correspond to compact conformations. It is, in fact, relatively
straightforward to provide counter examples to this criterion \cite{Cam95} 
casting serious doubts on the general validity of the
strict relationship between gap and folding times. Moreover, it is
extremely difficult, if not impossible, to obtain the value of the gap 
for proteins 
either experimentally or theoretically. Thus, the practical utility of
this criterion for models other than lattice systems is limited at best.

	The major purpose of this study is to critically examine the
various properties of sequences (all of which have a unique ground state)
that determine the kinetic accessibility and stability of lattice
representations of proteins. This is done by calculating the folding rates
and thermodynamic properties using \(N=15\) for a number of sequences. The
qualitative lesson from this case is verified by studying a smaller number
of sequences for the much studied case of \(N=27\). The rest of the paper
is organized as follows: In Sec. (II) the complete details of the model as
well as simulation methods are discussed. The results of this study for a
variety of cases are given in Sec. (III). The paper is concluded in Sec.
(IV) with a discussion.

\section{Description of the Model and Simulation Techniques}

\subsection{Model}

	We model proteins as chains of \(N\) successively connected beads
(residues) located at the sites of an infinite cubic lattice (Figs. 
(1,2)). To satisfy
self-avoiding condition we impose the restriction that each lattice site
can be occupied only once (or remain free). The length of the bond between
two residues is fixed and is equal to the lattice spacing \(a=1\). Any
conformation (structure, in lattice terms), which the protein can adopt,
is described by \({\bf r}_{i}\) (\(i=1,..,N\)) vectors with discrete
coordinates \(x_{i}, y_{i}, z_{i} = 0,1,2...\), which are the positions of
residues on the lattice. We assume that the only interactions,
contributing to the total energy of a protein structure, are those 
that arise due to the interactions between
residues that are far apart along a chain. We further assume that the
interactions are short ranged and can be represented by topological
contacts between residues.  A topological contact is formed when two
nonbonded beads \(i\) and \(j\) (\(\mid i-j \mid \geq 3\)) are
nearest-neighbors on the cubic lattice, i.e., \(\mid{\bf r}_{i} - {\bf
r}_{j}\mid=a\).  Thus, the total energy of a protein \(E\) is given by the
sum of the energies assigned to topological contacts found in a structure,
i.e 
\begin{equation} 
E = \sum_{i<j+3} B_{ij} \delta(r_{ij} - a),
\end{equation} 
where \(B_{ij}\) is the interaction energy
between \(i\)th and \(j\)th residues, which form a topological contact,
\(r_{ij} = \mid{\bf r}_{i} - {\bf r}_{j}\mid\) is the distance between
them, and \(\delta(0) = 1\) and 0 otherwise. In order to take into
account the heterogeneity of interactions found in real proteins 
we assume that the interaction
energies \(B_{ij}\) have a Gaussian distribution of the form 
\cite{Sali94a,Sali94b,Shakh89,Shakh91} 
\begin{equation} 
P(B_{ij}) = \frac{1}{(2\pi B^{2})^{1/2}}\exp\biggl(-
\frac{(B_{ij}-B_{0})^{2}}{2B^{2}}\biggr), 
\end{equation} 
where \(B_{0}\) is the mean value and \(B\) is the standard deviation. 
This model that has been extensively investigated theoretically 
\cite{Sali94a,Sali94b,Shakh91,Shakh90}. 
It should be stressed that the models studied here and similar lattice
models are at best caricatures of real proteins \cite{White}. The only 
objective of
these studies should be to obtain qualitative behavior which hopefully
shed light on the experiments. This, of course, requires extrapolating
from these model systems to the behavior expected in proteins in terms of
experimentally variable parameters. A tentative proposal for achieving
this has recently been given \cite{Thirum95}.

\subsection{Choice of \(B_{0}\)}

	Since the actual energy scales are not known, we set \(B\) in Eq. (2)
equal to \(1\) and thus all energies are expressed in the units of
\(B\). In contrast, we will demonstrate that the precise value of
\(B_{0}\) (or more precisely the ratio \(B_{0}/B\)) plays a crucial role. 
Negative values of \(B_{0}\) favor
random collapse of the  chain as the temperature is lowered. In addition,
the mean value \(B_{0}\) controls the nature of conformations that
constitute the low energy part of the spectrum. The extensive full
enumeration study of the conformational space for different sequences of
various lengths \(N\) indicates that as the mean value \(B_{0}\) decreases
structures with maximum number of topological contacts (compact
structures, CS) start to dominate among conformations with minimum
energies \cite{Sali94b,Klim}. Furthermore, a relationship 
can be 
obtained between the value of \(B_{0}\) and the ratio of hydrophilic and
hydrophobic residues in a sequence.
This would be relatively straightforward for random site models of 
the sort introduced recently \cite{Gar94}. 
For the random bond case, 
that is the subject 
of this and numerous previous studies, the computation of the fraction of 
hydrophobic residues is somewhat ambiguous. Nevertheless, the following 
procedure can be used to calculate their fraction. 
Since it is
natural to identify negative interaction energies as those between
hydrophobic residues and positive ones to be between hydrophilic residues,
one can specify boundary energies \(B_{H}\) and \(B_{P}\)
(\(B_{H}=-B_{P}\)) in such a way that the energies \(B_{ij}\) below \(B_{H}\)
corresponds to hydrophobic interactions and the energies above \(B_{P}\)
pertain to hydrophilic interactions. The energies \(B_{ij}\), lying
between these boundaries, are associated with mixed interactions.  If the
number of hydrophobic and hydrophilic residues in a sequences is
\(N_{H}\) and \(N_{P}\), respectively, (\(N_{H}+N_{P}=N\)), the fraction
of hydrophobic energies \(\lambda _{H}\) among \(B_{ij}\) is roughly 
\((N_{H}/N)^{2}\). 
This fraction can be also obtained by integrating the distribution (2) from
infinity to the energy \(B_{H}\). The relationship between
\(\lambda _{H}\) and \(B_{0}\) may be obtained as 
\begin{equation}
\lambda _{H} \simeq (N_{H}/N)^{2} = \int _{-\infty}^{B_{H}} 
P(B_{ij})dB_{ij}. 
\end{equation}
The precise value of \(B_{H}\) 
(and \(B_{P}\)) can easily be determined if one considers the case with 
\(B_{0}=0\), for which \(N_{H}=N_{P}\). 
Using Eqs. (2) and (3) we find that \(B_{H}=-0.675\) (for this
particular value of \(B_{0}\) one quarter of all energies \(B_{ij}\) are
below the boundary energy \(B_{H}\) and one quarter - above \(B_{P}\)). 
It is known that in natural proteins hydrophobic residues make up
approximately 54 percent of all residues in a sequence \cite{Miller}. For the
distribution in Eq. (2) this implies that the mean value \(B_{0}\) should
be approximately \(-0.1\). Most of our simulations have been performed 
with this value of \(B_{0}\). 

	We have also performed a study of the sequences with \(B_{0}=-2.0\)
that in the language of sequence composition means that hydrophobic
residues constitute about 94 percent of all residues.  The motivation for
choosing this value of \(B_{0}\) is the following. For \(B_{0}=-2.0\) the
low energy spectrum becomes more sparse \cite{Sali94b}, because as 
mentioned above the
main contribution comes from CS, whose total number is considerably less
than the number of conformations with any other number of topological 
contacts \(c\). Specifically, for
\(N=15\) the number of conformations having \(c=11\) is 3,848, \(c=10\) -
17,040, \(c=9\) - 97,216, \(c=8\) - 313,868 etc.  Studying the folding
rates of the sequences having different \(B_{0}\) enables us to assess the
role of the available conformation space and the connectivity between
various states in determining the kinetics of the folding process. The
choice of \(B_{0}=-2.0\) also allows us to compare directly our results to
previous studies found in the literature \cite{Sali94a,Sali94b}. 

\subsection{Choice of \(N\)}

        In order to fully characterize the folding scenarios it is
necessary to understand the kinetics of approach to a native conformation 
in these models 
as a function of \(N\) and temperature. It has been shown recently that
the folding of real proteins depends critically on \(N\), the
characteristic temperatures of the polypeptide chain (\(T_{\theta},
T_{f}\), and perhaps the kinetic glass transition temperature \(T_{g}\)),
viscosity, surface tension \(\gamma\) etc. \cite{Thirum95}. Thus in order to 
make the results
of the minimal models relevant to proteins it is imperative to vary \(N\)
in the simulations. 

Although one would like to understand the kinetic behavior of
foldable heteropolymers for sufficiently large \(N\) this is currently
computationally difficult. In the present study we have chosen \(N=15\)
and \(N=27\). We chose \(N=15\) because for this value one can perform
detailed kinetic study by including {\em all conformations} (compact and
noncompact). A detailed study for three dimensional (3D) lattice models
comparable to that undertaken for two dimensional (2D) systems has never
been done \cite{Dill,Chan93,Chan94,Binder}. With this value of \(N\) 
one of the
limitations of the study of Sali {\em et al.}, who restricted themselves
to compact structures only, can be overcome. From any theoretical
perspective the qualitative difference in results between \(N=15\) and
\(N=27\) should be insignificant. This is certainly the experience in
simulations of polymeric systems \cite{Binder}. Thus, we 
expect that the
qualitative aspects of the kinetics of folding should be quite similar for
\(N=15\) and \(27\). This is, in fact, the case. 

One might naively think that for
\(N=15\) the total number of conformations is not enough for folding 
times to exceed the Levinthal time, which is roughly the number of 
conformations of the polypeptide chain. 
The basis for this argument is that 
the conformation space of \(N=15\) is considerably less than for \(N=27\). 
The total number of conformations 
of the chain of \(N\) residues \(C_{N}\) or equivalently the number of all 
possible self-avoiding walks of \(N-1\) steps on a cubic lattice is 
\cite{Chan90} 
\begin{equation}
C_{N} \approx a(N-1)^{\gamma-1} Z_{\text{eff}}^{N-1}
\end{equation}
where \(Z_{\text{eff}}=4.684\), the universal exponent
\(\gamma\approx1.16\), and \(a=o(1)\). For \(N=15\) and 27 \(C_{N}\) is
approximately \(7.77\times10^{8}\) and \(4.6\times10^{17}\), respectively. 
Full enumeration (FE) of SAW which are {\em unrelated} by symmetry for 
\(N=15\)
gives \(C_{15}^{FE}=93,250,730\), which differs approximately from the 
number of {\em all} SAW by a factor of 48.  Thus, \(C_{15}\) obtained 
from Eq. (4) and
\(C_{15}^{FE}\) are consistent. The chain of 15 residues can form 3,848
CS, which belong to 3x3x2 or 4x2x2 tetragonals, whereas 27-mer chain
adopts 103,346 CS with 28 topological contacts 
\cite{Shakh90,Chan90}; all 
these CS are confined to 3x3x3 cube. 

From the above enumerations of the conformations it might be tempting to 
speculate that
virtually all sequences for \(N=15\) with a unique ground state should
fold on the Levinthal time scale of \(9.3 \times 10^{7}\) Monte Carlo
steps (MCS). However, we find that for some of the thirty two sequences
examined the maximum folding time can be larger than  \(10^{9}\) MCS
depending upon several characteristics (see below).  Thus, even if the
chain samples one conformation per MCS the bottlenecks in the energy
surface can prevent the chain from reaching the native conformation. This
implies that the number of conformations \underline{alone} cannot
determine folding times \cite{Chan95}.  In fact, in the case of 
the models of disulfide bonded
proteins it has been explicitly demonstrated that a significant reduction
of available conformation space does not guarantee a decrease in folding
times \cite{Cam95}. Thus, kinetic 
foldability is determined by several factors and
hence explicit studies of \(N=15\), where full enumeration of all
conformations is possible, should help us gain insights into the folding
of small proteins. A comparison of the results for small values of \(N\)
is also useful in assessing finite size effects. 

	The arguments given above together with explicit computations
given here reject claims \cite{KarpSali} that the 
study of short chains (\(N < 27\) in
three or two dimensions) are not of significance in illustrating the
qualitative behavior of protein folding kinetics.  Numerous studies have
shown that it is not merely the size of the conformation space, but the
connectivity between conformations, i.e. the nature of the underlying
energy landscape that allows one to distinguish between foldable and
non-foldable sequences \cite{Chan95}. The obvious advantage of 
\(N=15\) is 
that systematic thermodynamic and kinetic studies (already performed for 2D
chains of \(N\) up to 30) can be undertaken for 3D chains.

\subsection{Correlation Functions}

	For probing the thermodynamics and kinetics of protein folding we use
the overlap function (considered here as an order parameter), which is
defined as \cite{Cam93}  
\begin{equation}
\chi = 1 - \frac{1}{N^{2}-3N+2} \sum_{i\neq
j,j\pm 1} \delta(r_{ij} - r_{ij}^{N}),
\end{equation}
where \(r_{ij}^{N}\) refers to the coordinates of the native state. This 
function
measures structural similarity between the native state and the state of
interest: the smaller the value of \(\chi\) becomes the larger a given
structure resembles the native one.  Additional structural and 
kinetic information can be obtained using 
the function \(Q\), which counts the relative
number of native-like topological contacts in a structure 
\cite{Bryn89,Sali94a,Sali94b,Skol90} 
\begin{equation}
Q = \frac{c_{n}}{c_{n}^{tot}},
\end{equation}
where \(c_{n}\) is the number of native-like contacts in a given structure
and \(c_{n}^{tot}\) is the total number of contacts found in the native
structure. 

	We have calculated the relevant thermodynamic properties such as the 
total energy \(<E>\), the specific heat \(C_{v}\) with
\begin{equation}
C_{v}=\frac{<E^{2}>-<E>^{2}}{T^{2}}, 
\end{equation}
the function \(<Q>\), and the Boltzmann probability of being in the native 
state 
\begin{equation}
P(E_{0})=\frac{\exp(-\beta E_{0})}{Z},
\end{equation}
where \(Z=\sum_{i} \exp(-\beta E_{i})\) and \(\beta=1/(k_{B}T)\) (\(k_{B}\)
is set to 1 in our simulations). The brackets \(< ... >\) indicate the 
thermodynamic averages. In 
addition, the overlap function and the fluctuations in \(<\chi>\), namely
\begin{equation}
\Delta \chi   = <\chi ^{2}> - <\chi >^{2}
\end{equation}
were also calculated. The thermodynamic characteristics of the system can be 
exactly calculated for each sequence by exhaustively enumerating the 
various symmetry unrelated conformations for small enough values of 
\(N\). In particular, we calculated these quantities exactly for 
\(N=15\). For \(N=27\) we used slow cooling Monte Carlo method to calculate 
the appropriate quantities of interest. The annealing simulation 
procedure is discussed in Appendix A. 

	The parameter that distinguishes fast folding and slow
folding sequences appears to be \(\sigma =
(T_{\theta}-T_{f})/T_{\theta}\), where \(T_{\theta}\) is the collapse
transition temperature and \(T_{f}\) is the folding transition
temperature \cite{Cam93,Thirum95}.  
It is known that even in these finite 
sized 
systems \(T_{\theta}\) can be estimated using the peak in the temperature
dependence of \(C_{v}\) (cf. Eq. (7)) \cite{Skol91,Cam93,Honey92}.  
We have shown in previous
studies involving both lattice and off-lattice models that the temperature
dependence of the fluctuations in the overlap function, which serves as an
order parameter, can be used to calculate \(T_{f}\) 
\cite{Cam93,Guo}. In particular,
\(T_{f}\) corresponds to the peak in the function \(\Delta \chi \).

	For most sequences \(T_{\theta}\) and \(T_{f}\) are sufficiently
well separated that an unambiguous determination is possible by a
straightforward computation of the temperature dependence of \(C_{v}\) and
\(\Delta \chi \). However, we have generated five sequences (out of 32
for \(N=15, B_{0}=-0.1\)), for which \(C_{v}\) or \(\Delta \chi \)
appear not to have well-defined single maximum due to specific arrangement
of the energy states. For example, sequence 32 shows two maxima in the
dependence \(C_{v}(T)\) at \(T_{1}=0.28\) and \(T_{2}=0.73\) of different
amplitudes \(C_{v}(T_{1})=8.25\) and \(C_{v}(T_{2})=13.22\), respectively.
In this case, we defined \(T_{\theta}\) as a weighted average over the 
temperatures \(T_{1}\) and \(T_{2}\)
\begin{equation}
T_{\theta}=\frac{C_{v}(T_{1})T_{1}+
C_{v}(T_{2})T_{2}}{C_{v}(T_{1})+C_{v}(T_{2})}.
\end{equation}
In other instances (e.g., for sequence 10), we found that although the
dependence \(C_{v}(T)\) has a single maximum at \(T_{max}\), it also has
the interval \((T',T'')\) not including \(T_{max}\), wherein the
derivative \(\frac{dC_{v}}{dT}\) again approaches almost zero value that
gives essentially unsymmetric form to the peak of specific heat. We have
also applied Eq. (10) for calculating \(T_{\theta}\) for such sequences by
setting \(T_{1}=T_{max}\) and \(T_{2}\neq T_{max}\) corresponds to the
temperature, at which \(|\frac{dC_{v}}{dT}|\) has the smallest value
within \((T',T'')\). 

There are other ways of calculating \(T_{\theta}\) and \(T_{f}\). For
example, \(T_{\theta}\) could be directly inferred from the
temperature dependence of the radius of gyration of the polypeptide
chain. It has been shown in our earlier work on off-lattice models 
\cite{Honey92} 
that the resulting values of \(T_{\theta}\) coincide with those 
obtained from the peak in the specific heat. The folding transition
temperature is often associated with the midpoint of the temperature
dependence of the probability of being in the native
conformation. This estimate of \(T_{f}\) is in good agreement with the
peak position of the temperature dependence of \(\Delta \chi \). In
general, different order parameters can be used to calculate
\(T_{f}\). The resulting values are fairly consistent with each
other.

\subsection{Sequence Design}

	To create a database of different sequences for \(N=15\) we
generated 60 random sequences, using the mean value \(B_{0}=-0.1\) and 9
random sequences, using \(B_{0}=-2.0\).  For \(N=27\) we generated 15
random sequences with \(B_{0}=-0.1\) and 2 with \(B_{0}=-2.0\). Note that
the computational procedures for 15-mer and 27-mer sequences are similar,
except that thermodynamic quantities for \(N=27\) are calculated from slow
cooling Monte Carlo simulations.  By enumerating all possible
conformations for \(N=15\) we determined the energy spectrum for each
sequence.  The program for enumerating all the structures a protein can
adopt on a cubic lattice is based on the Martin algorithm 
\cite{Martin} 
that is supplemented by the procedure that rejects all structures related
by symmetry.  This algorithm allows us to determine the lowest (native)
energy state \(E_{0}\), its degeneracy \(g\), the coordinates of
corresponding structure(s), and the number of topological contacts \(c\)
for each sequence.  The energy levels for 10 sequences are summarized in 
Fig. (3) for \(N=15, B_{0}=-0.1\) and in Fig. (4) for \(N=27, B_{0}=-0.1\).  
The spectra for \(N=15\) were obtained by enumerating all possible 
conformations of the chain and arranging them in increasing order of 
energy.  For \(N=27\), on the other hand, the spectra for the various 
sequences are obtained by slow cooling Monte Carlo method, the details of 
which are reported in Appendix B. The results in Fig. (4) for \(N=27\) 
are instructive. For each sequence we show two columns. The left column 
gives the spectrum calculated by numerical method, whereas the right 
column is the spectrum that would be obtained if only the compact 
structures were retained. A comparison of the two columns for various 
sequences clearly reveals that for a majority of sequences the low lying 
energy levels are, in fact, noncompact. Thus, from this figure we would 
conclude that these noncompact structures would make significant 
contributions to various thermodynamic properties. This figure also shows 
that for the sequences whose native conformation is compact, 
the energy gap \(\Delta _{CS}\) calculated using compact structures 
spectrum alone exceeds the true gap. This appears to be  a general result 
and can be 
understood by noting that the lowest energy excitations for such 
sequences  are created  by 
flipping surface bonds. The resulting structure would be noncompact 
and its energy would be lower or equal to that of other compact 
structures. 
Thus, it should in general be true that when the native conformation is 
compact, \(\Delta _{CS} \geq \Delta\).  Since there are large scale
motions that are expected to be involved in protein folding the
physically relevant energy scale should be the stability gap
\cite{Honey90,GuoHoney}. There is no straightforward relationship between the
stability gap and \(\Delta _{CS}\) or \(\Delta \). 
We
rejected all sequences with \underline{nonunique} ground state from further 
analysis.

	In order to determine folding times for a range of
\(\sigma \) \((=(T_{\theta}-T_{f})/T_{\theta})\) values sequences with 
varying
spectral characteristics are required. It is known that a generic randomly
generated sequence (even with unique native state) does not fold
rapidly \cite{Bryn95,Dill,Karp92}. Thus, to expand the 
database of the sequences we used a
technique  proposed by Shakhnovich and Gutin  
\cite{Shakh93,Shakh94} to 
create a set of
optimized sequences. We should stress that in our study this was used as 
merely a technical device to generate sequences that span a rather wide 
range of \(\sigma \). Here we will briefly present the idea of this 
scheme \cite{Shakh93,Shakh94}. 
One selects an arbitrary ('target') structure corresponding to a given
initial random sequence.  Then standard Monte Carlo algorithm is applied
in the sequence space. The first step of this scheme can be described in
the following way.  Two energies \(B_{ij}\) and \(B_{kl}\) of the initial
('old') sequence picked at random are interchanged, so that the
topological contact between residues \(k\),\(l\) has the interaction
energy \(B_{kl}'=B_{ij}\) and the contact between residues \(i\),\(j\) -
the interaction energy \(B_{ij}'=B_{kl}\).  Thus, a new probe sequence,
which differs from the initial one by the energies \(B_{ij}'\) and
\(B_{kl}'\), is produced and is subject to Metropolis criterion.  To do
this, the energy of the new sequence \(E_{new}\) fitted to the target
structure is calculated and compared with \(E_{old}\) of the initial
sequence. If the new sequence provides lower energy at the target
structure, it is unconditionally accepted and \(B_{ij}=B_{ij}'\),
\(B_{kl}=B_{kl}'\). If \(E_{old}<E_{new}\), the new sequence is accepted
with the Boltzmann probability \(P=\exp(-(E_{new}-E_{old})/T)\). If the
new sequence is rejected, the initial sequence is restored. This
permutation procedure, which does not alter the composition of the
sequences, is repeated \(n\) times.  The control parameter is temperature,
and if it is sufficiently low, a series of sequences are quickly generated
after relatively small number of MCS (\(10^{4}\)), whose energies
when fitted to the target structure are remarkably low. By employing the
full enumeration procedure (or Monte Carlo simulations for \(N=27\)) one can 
verify
that these energies are actually the lowest ones in the spectrum.  In this
manner, a series of new 'optimized' sequences are created. An application
of optimization scheme allowed us to investigate essentially wider range
of values of the parameter \(\sigma\) than can be done by analyzing
sequences produced at random.

        In all, for \(N=15\) we chose 32 sequences with \(B_{0}=-0.1\) and
9 sequences with \(B_{0}=-2.0\) for detailed study. Among these with
\(B_{0}=-0.1\), 15 sequences were random and 17 - optimized. The native
structures of these sequences were CS (e.g., structure (a) in Fig. (1)) as 
well non-CS (e.g., structure (b)) and included 
from \(8\) to \(11\) topological contacts. For \(B_{0}=-2.0\) all
sequences were optimized and as suspected the native conformations 
were
all CS \cite{Sali94b}. For \(N=27\) we analyzed 15 
optimized sequences 
with \(B_{0}=-0.1\) and 2 with \(B_{0}=-2.0\).  As for \(N=15\) their 
native
conformations were CS as well as non-CS with 21-28 
topological contacts (Fig. (2)).
We believe that this choice of sequences is sufficient and we do not
expect qualitatively different behavior for this model, if a larger 
database is selected. 

	Since our objective is to compare the rates of folding for 
different sequences it is desirable to subject them to identical folding 
conditions. The equilibrium value of \(<\chi >\) measures the extent to 
which the conformation at a given temperature is exactly equivalent to a 
microscopic conformation, namely the native state. At sufficiently low 
temperature \(<\chi >\) would approach zero, but the folding time may be 
far too long. We chose to run our Monte Carlo simulations at a sequence 
dependent simulation temperature \(T_{s}\) that 
is subject to two conditions:
(a) \(T_{s}\) \underline{be less than} \(T_{f}\) for a specified sequence 
so that the native conformation has the highest occupation probability; (b) 
the value of \(<\chi(T=T_{s}) >\) be a constant for all sequences, i.e.
\begin{equation}
<\chi(T=T_{s}) >=\alpha .
\end{equation}
In our simulations we choose \(\alpha =0.21\) and this value was low 
enough so that \(T_{s}/T_{f} < 1 \) for all the sequences examined. This 
general procedure for selecting the simulation temperatures has been 
previously used in 
the literature \cite{Sali94b,Cam95}. 
For \(N=15\) \(T_{s}\) is precisely determined using Eq. (11)
because \(<\chi(T=T_{s}) >\) can be calculated exactly using the full 
enumeration procedure. The simulation temperatures for \(N=27\) are 
calculated using the protocol described in Appendix A.

\subsection{Monte Carlo Simulations and Interpretation of Folding Kinetics}

	In the present work we used standard Monte Carlo (MC) algorithm
for studying folding of different sequences to their native states.  The
local simulation dynamics includes the following moves (Fig. (5)): (i)
corner moves, which flip the position of \(j\)th residue across the
diagonal of the square formed by bonds \((j-1,j)\) and \((j,j+1)\); (ii)
crankshaft rotations, which involve changing the positions of two
successively connected beads \(j+1\) and \(j+2\) (positions of \(j\) and
\(j+1\) beads, which are nearest neighbors on a lattice, remain
unchanged); (iii) rotations of end beads, in which the end bead (\(1\) or
\(N\)) moves to any of 5 adjacent sites to the beads \(2\) or \(N-1\) (the
sites previously occupied by the beads \(1\) or \(N\) are not considered
as possible new sites).  This particular set of moves has already been
applied in Monte Carlo simulations \cite{Sali94b,Socci94}; for  
a discussion of the dependence of the kinetic results on the move set 
used in the Monte Carlo simulations see Ref. 13. In 
order to ensure maximum efficiency of
exploring conformation space it is reasonable to assign different
probabilities for the moves (i), (iii) and the move (ii).  After probing
several values we found that the probability \(p=0.2\) for the moves (i)
and (iii) (involving single residue) provides the best efficiency of
searching a native state of a test sequence by Monte Carlo algorithm.  
Qualitatively, it is clear that very small probability of
moves (ii) depletes the ability of a chain to sample different
conformations, while excessively high probability of their occurrence may
deteriorate the ability to accept moves that would lead to acquisition of 
the last few native contacts when the chain is near the native 
conformation.

	For clarity of presentation let us now describe one step of the
Monte Carlo algorithm. In the beginning, the type of move (single bead
move ((i) or (iii)) or crankshaft rotation (ii)) is selected at random
taking into account the probability \(p\) introduced above. After this a
bead in the chain is chosen at random and the possibility of performing
the selected move depending on the local configuration of the chain is
established as follows. If there is a chain turn at the bead 
\(j\) selected for
move (i) or in the case of move (ii) the beads \(j,j+3\) are lattice
nearest neighbors, a move is accomplished; otherwise, one must return to
selection of move type. Then a self-avoidance criterion is applied: if the
move results in double occupancy of lattice sites, it is rejected and a
new selection of move type has to be made. If self-avoidance criterion is
satisfied, the new conformation is adopted and Metropolis criterion is
used, i.e., the energies of old and new conformations, \(E_{old}\),
\(E_{new}\), are compared. If new conformation has lower energy, the move
is accepted and a Monte Carlo step is completed. If \(E_{new}\) is higher
than \(E_{old}\), then the Boltzmann probability
\(P=\exp(-(E_{new}-E_{old})/T)\) is calculated and compared with a random
number \(\xi \) (\( 0 < \xi < 1\)). If \(\xi \) is smaller than \(P\), the 
move is accepted and
MC step is completed.  Otherwise, the move is rejected and the previous MC
step is counted as a new one. The fraction of accepted steps on average
constitutes 5-15 percents for the entire trajectory. In principle, using
ergodic measures this percentage can be adjusted {\em a priori} to
maximize sampling rate \cite{Mount}.

	The initial conformations of all trajectories correspond to random
extended coil ('infinite temperature' conformation). After a sudden
temperature quench the chain dynamics was monitored for approximately  
\(10^{5}\) to
\(5\) x \(10^{7}\) MC steps (MCS), depending on a folding kinetics of a
given sequence.  In order to obtain the kinetics of folding for a
particular sequence the dynamics was averaged over \(M\) independent
initial conditions. For example, \(<\chi (t)>\) is calculated as
\begin{equation}
<\chi (t)> = \frac{1}{M} \sum_{i=1}^{M} \chi _{i}(t) , 
\end{equation}
where \(\chi _{i}(t)\) is the value of \(\chi \) for the \(i^{\text{th}}\) 
trajectory at time \(t\). Another important probe of folding kinetics is 
the fraction of trajectories \(P_{u}(t)\) which have not yet reached 
the native conformation at time \(t\) 
\begin{equation}
P_{u}(t)=1 - \int_{0}^{t} P_{fp}(s)ds, 
\end{equation}
where \(P_{fp}(s)\) is the probability of the first passage to the native 
structure at time \(s\) defined as
\begin{equation}
P_{fp}(s) = \frac{1}{M} \sum_{i=1}^{M} \delta (s - \tau _{1i})
\end{equation}
In Eq. (14) \(\tau _{1i}\) denotes the first passage time for the
\(i^{\text{th}}\) trajectory.  Similarly, other quantities were
calculated.  Typically, the number of trajectories \(M\) used in the
averaging varied from 100 to 800. We find that for smaller values one
cannot get reliable results at all. If \(M\) as small as \(10\) is used 
\cite{Sali94b},
one can obtain qualitatively incorrect results. The precise choice of
\(M\) (which is sequence dependent) was determined by the condition that
the resulting kinetic and thermodynamic properties should  not change
significantly with subsequent increase in \(M\). The tolerance used in
determining \(M\) was that the various quantities of interest converge to
within 5 percent. 

\subsection{Computation of Folding Rates}

	The most important goal of this paper is to obtain folding times 
for sequences at various temperatures. These times (or 
equivalently the folding rates) were calculated by analyzing the time 
dependent behavior of the dynamic quantities. It is clear that \(<\chi 
(t)>\) provides the most microscopic description of the kinetics of 
approach to the native state. The folding times reported here were 
obtained 
by a quantitative analysis of \(<\chi (t)>\). For all sequences we find 
that after a transition time \(<\chi (t)>\) can be fitted as a sum of 
exponentials, i.e.,
\begin{equation}
<\chi (t)> = a_{1} \exp(-\frac{t}{\tau_{TINC}})+a_{2} 
\exp(-\frac{t}{\tau_{f}}) 
\end{equation}
In most cases, biexponential fit like in Eq. (15) gave the best
approximation to the computed kinetic curves. However, for some sequences
it was found that a single or three exponential fit were more suitable.
The interpretation of the amplitudes \(a_{1}, a_{2} \) and the time
constants \(\tau_{TINC}, \tau_{f} \) are discussed in Sec. (III.A.3) . It 
must be
noted that the kinetic curves for very slow folding sequences (those with
large values of \(\sigma \)) do not reach the required equilibrium values.
However, even in these instances (6 out of 32 for \(N=15, B_{0}=-0.1\))
our simulations were long enough to observe the transition to the native
conformation so that an accurate estimate of the folding time can be made. 
It should be noted that the trends in folding times remain unchanged if 
the mean first passage time is substituted for \(\tau _{f}\) (or \(\tau 
_{TINC}\)) in Eq. (15).

\subsection{Monitoring Intermediates in Folding Process}

	Since the underlying energy landscape in proteins is thought to be 
rugged \cite{Bryn95,Dill,Thirum94} it is likely that there 
are low energy basins of attraction, in which the protein can get trapped 
in for arbitrary long times. Explicit construction of such landscape in 
lattice models, albeit 
in two dimensions, has revealed the presence of such states as important 
kinetic intermediates in certain folding pathways \cite{Cam95,Cam93}. In our 
simulations we 
have used the following strategy to describe the nature of intermediates 
in the folding of the various sequences. We divided each trajectory into 
two parts: the first part starts at the beginning of the trajectory 
(\(t=0\)) and ends when the native structure is formed for the first time, 
i.e. when the first passage time \(\tau_{1i}\) for the \(i^{\text{th}}\) 
trajectory is reached. We labeled the trajectory for \(0 \leq t \leq 
\tau_{1i}\) as the relaxation part. 
The remaining portion of the 
trajectory for \(\tau_{1i} \leq t \leq t_{max}\) is referred to as the 
fluctuation part, where \(t_{max}\) is the maximum time for which the
computations are done for a given sequence. 
Using the trajectories corresponding to the relaxation regime we 
calculated for each sequence the probability of occurrence of the low 
energy states \(E_{k}\), i.e.
\begin{equation}
P_{r}(E_{k})=\frac{1}{M}\sum_{i=1}^{M}\frac{1}{\tau_{1i}} 
\int_{t=0}^{\tau_{1i}} \delta(E_{i}(t)-E_{k})dt,  
\end{equation}
where \(E_{i}(t)\) is the energy at the \(i^{\text{th}}\) trajectory at 
the time \(t\). 
The state with the energy \(E_{k}\), which has the largest value of
\(P_{r}\), is defined to be a kinetic intermediate for a given sequence.
We also calculated the probability that \underline{this state} occurs in the
fluctuation parts of the trajectories. 
The fluctuation probability that the kinetic intermediates with the 
energy \(E_{k}\) (for conformations other than the native one there 
could be more than one structure with the same 
energy) are visited after the transition to the native conformation is 
defined as 
\begin{equation}
P_{fl}(E_{k})=\frac{1}{M}\sum_{i=1}^{M}\frac{1}{t_{max}-\tau_{1i}}
\int_{t=\tau_{1i}}^{t_{max}} \delta(E_{i}(t)-E_{k})dt,
\end{equation}
where \(t_{max}\) is the maximum time of simulation. 
This quantity was calculated to monitor if the chain, after reaching the 
native conformation, 
makes a transition to the same intermediate which was visited with 
overwhelming probability on the way to the native conformation.

\section{Results}

	Since this section describes in complete detail the results for a
variety of cases making it quite lengthy we provide a brief summary of its 
contents.  The general
methodology described in the previous section has been used to study in
extreme detail the kinetics of folding for \(N=15\). For this value of
\(N\) we have considered two values of \(B_{0}\) which sets the overall
strength of the hydrophobic interactions. We have chosen \(B_{0}=-0.1\)
and \(B_{0}=-2.0\). The former choice is a bit more realistic, while the
latter was chosen to contrast the role of CS versus non-CS in determining
the thermodynamics and kinetics of folding. As emphasized before the value
of \(N=15\) is about the largest value of \(N\) for which exact
enumeration studies are possible in three dimensions and for which the
kinetics can be precisely determined in terms of all allowed conformations
being explored. The results for \(N=15\) and for \(B_{0}=-0.1\) and
\(B_{0}=-2.0\) are presented in Sec. (III.A). In Sec. (III.B) we present the
results for \(N=27\), for which we have also chosen \(B_{0}=-0.1\).  The
thermodynamic properties with \(B_{0}=-2.0\) are discussed as well. A
comparison of \(N=15\) and \(N=27\) shows very similar qualitative
behavior.

\subsection{\(N = 15\): \(B_{0}=-0.1\) and \(B_{0}=-2.0\)}

\subsubsection{Thermodynamic Characteristics}

The two relevant temperatures \(T_{\theta}\) and \(T_{f}\) for each 
sequence are computed from the temperature dependence of \(C_{v}\) and 
\(\Delta \chi \), respectively. 
In addition we have computed \(<Q>\) and \(<\chi >\) as a function of 
temperature. The midpoints in the graphs of these quantities can 
sometimes be used to obtain an estimate of \(T_{f}\). In general, 
\(T_{f}\) obtained from the peak of \( \Delta \chi \) is smaller than 
that determined from the midpoint of \(<\chi >\) or \(<Q>\). 
The plots of \(<\chi >\), \( \Delta \chi \) ,\(<Q>\), and \(C_{v}\) as a
function of temperature are displayed in Fig. (6) for the sequence labeled 
14. From the graphs of \(C_{v}\) and \( \Delta \chi \)
the collapse transition temperature \(T_{\theta}\) and the folding
transition temperature \(T_{f}\) are found to be 0.65 and 0.45
respectively. The simulation temperature, \(T_{s}\), is calculated using
Eq. (11) and in this case it turns out to be 0.38. In Fig. (7a) we present
the dependence of \(T_{s}\) on the crucial parameter \( \sigma =(T_\theta
-T_f)/T_\theta \).  In addition we also display the correlation between
\(T_{s}\) and the energy gap \(\Delta \) (Fig. (7b)). From these figures it 
appears that \(T_{s}\) correlates well with \(\sigma \). 
Thus as far as the simulation
temperature is concerned it appears that \(T_{s}\) is decreasing function
of \(\sigma \). This implies that if \(\sigma \) is small then the
simulation temperature can be made higher and fast folding can therefore be
expected. This is further quantified in Sec. (III.A.4). We should emphasize 
that this correlation is only statistical in the sense that large 
(small) values of \(\sigma\) yield small (large) values of \(T_{s}\). 
However given two values of \(\sigma\) that are closely spaced it is not 
possible to predict the precise values of \(T_{s}\). 

Since the dimensionless parameter \(\sigma \) serves to distinguish between
foldable sequences and those that do not reach their native conformation
on a reasonable time scale it is interesting to see if \(\sigma \) can
correlate with the spectrum of the underlying energy function. It has been
argued that the only important parameter that is both necessary and 
sufficient to account for foldability is the gap
\(\Delta \) \cite{Sali94a,Sali94b,KarpSali}.  
The plot of \(\sigma \) as a 
function of \(\Delta \) is shown
in Fig. (8a).  In the lower panel (Fig. (8b)) we plot \(\sigma \) as a 
function of the dimensionless parameter \(\Delta /T_{s}\). This figure 
shows very clearly the lack of correlation between \(\sigma \) and 
\(\Delta \). Thus, it is seen that no clear correspondence exists even in 
these models between \(\sigma \) and \(\Delta \). This, of course, is not
surprising because \(T_{\theta}\) is determined by the entire energy
spectrum - most notably the higher energy non-compact structures. The
results in Fig. (8a) show that large values of \(\sigma \) appear to
correspond well with small values of \(\Delta \). On the other hand small
values of \(\sigma \) simultaneously corresponds to both small as well as
large values of \(\Delta \). 

\subsubsection{Contribution to Thermodynamic Properties from Non-compact 
Structures}

The number of non-compact structures even for small values of \(N\) (such 
as 15 and 27) far exceeds that of compact structures. 
It has already been mentioned that full enumeration of all self-avoiding 
structures in three dimensions becomes increasingly difficult for \(N > 
15\). Thus in the literature it has been explicitly assumed that, in 
general, the native conformation in this model is maximally compact 
and that the thermodynamic properties can be determined 
using the spectrum of compact structures alone \cite{Sali94a,Sali94b,Karp95}. 
It 
has also been argued that the only relevant aspect of the energy spectrum 
that 
determines both kinetics and thermodynamics of folding in these models is 
the gap defined as 
\begin{equation}
\Delta = E_{1} - E_{0},
\end{equation}
where \(E_{0}\) and \(E_{1}\) are the lowest energy and the 
energy of the first 
excited state in the spectrum. Since we have  
enumerated all possible conformations for \(N=15\) this can be explicitly 
checked by comparing various thermodynamic quantities computed exactly 
with those obtained by  
including only the contribution from compact structures. This has been done 
for several 
sequences and the typical results for two sequences are shown in Figs. (9) 
and (10). In Fig. (9) we present the results for \(<\chi >\) and 
\(\Delta \chi \) for \(B_{0}=-0.1\). The relatively small but 
realistic value of 
\(B_{0}\) makes the comparison between these quantities calculated using 
CS alone and the values calculated using full enumeration least 
favorable. From Fig. (9a) we find that \(T_{s}\) found from full 
enumeration is roughly one half that obtained using compact structures 
enumeration (CSE) alone. In fact, 
\(T_{s}^{CSE}\) \underline{exceeds} the collapse transition temperature 
\(T_{\theta}\) 
which implies that if simulations are performed at this temperature the 
native conformation will not be stable at all. Fig. (9b) shows that 
\(T_{f}\) (the folding transition temperature) is once again one half 
that of \(T_{f}^{CSE}\), indicating the importance of non-compact 
structures in this case. 

In Fig. (10) we show the behavior of \(<\chi > (T)\) and
\(\Delta \chi (T) \) for \(B_{0}=-2.0\) for another sequence. In this case 
the agreement between the exact results and that calculated using CS 
alone is significantly better. The difference between the two is roughly 
on the order of ten percent. These calculations clearly show that 
even in these models the importance of non-compact 
structures is dependent upon the value of \(B_{0}\). Only for large values 
of \(\mid B_{0} \mid \) the low energy spectrum is dominated by CS alone.

\subsubsection{Folding Kinetics: Kinetics of Approach to the Native 
Conformation}

We begin with a discussion of the approach to the native state starting
from an ensemble of disordered conformations. The kinetics of reaching the
native state was monitored by studying the time dependence of \(<\chi
(t)>\) averaged over several initial conditions. For all the sequences
that we have examined we find that \(<\chi (t)>\) can be fit by a sum of
exponentials after a transient time. In our previous studies using 
off-lattice models we had shown
that in general for a foldable sequence a fraction of initial population
of molecules reaches the native state directly without encountering any
discernible intermediates \cite{Guo,ThirumGuo}.  This is the case in these 
models as well, thus
further supporting the conclusions of our earlier work which was based on 
Langevin
simulations of off-lattice models. Since the data base analyzed here is
more extensive, covering a wide span of \(\sigma \), we can further classify
the meaning of the various exponential terms that arise in the time
dependence of the overlap parameter. In order to classify the various 
sequences in terms of the rapidity of folding to the native conformation 
we have used the parameter \(\sigma \) as a discrimination factor. \\
{\bf (a) Fast folding sequences (\(\sigma \lesssim 0.1\)):}\\
For these sequences the structural overlap function \(<\chi (t) >\) for 
\(t\) greater than a transient time is adequately fit by a single 
exponential, i.e., 
\begin{equation}
<\chi (t) > \simeq a_{f} \exp(-t/t_{TINC}).
\end{equation}
In these cases the folding appears to be a two state all-or-none process 
and \(\tau _{f} \approx \tau _{TINC}\), where \(\tau _{TINC}\) is the
time scale of topology inducing nucleation collapse (TINC). 
The folding and the collapse is 
almost synchronous. It has been shown in other studies that the
folding for these sequences proceeds by a TINC mechanism 
\cite{Guo,ThirumGuo,Abk1,Abk2,Otzen}.  In these studies the TINC
mechanism was established by studying the microscopic dynamics of the
trajectories. We found that once a critical number of contacts is 
formed (corresponding to a nucleus) the native conformation is reached
rapidly. 
This fast folding is clearly observed when \(\sigma\) is less that 0.1. \\
{\bf (b) Moderate folding sequences (\(0.1 \lesssim \sigma \lesssim 
0.6\)):}\\ 
These are sequences when a single
exponential fit cannot describe time course of the overlap function
\(<\chi (t) >\). We find that after an initial time, \(<\chi (t) >\) is
well fit by two exponentials. The interpretation of the fast and slow
processes have been given elsewhere in our studies of continuum models 
using Langevin simulations \cite{Guo,ThirumGuo}. 
The range of \(\sigma \) that 
characterizes 
moderate folding is \(0.1 \lesssim \sigma \lesssim 0.6\). The onset of the 
intermediate values of
\(\sigma\) is easy to obtain. This can be inferred by the smallest value
of \(\sigma\) for which biexponential fit of \(<\chi (t) >\) is
required to describe the approach to the native conformation. \\ 
{\bf (c) Slow folding sequences (\(\sigma \gtrsim 0.6\)):}\\ 
These are sequences with \( \sigma \gtrsim 0.6 \). When
\(\sigma \) gets close to unity, these sequences do not fold on
any reasonable simulation time. If \(\sigma \) is greater than almost 0.6
we once again find that multiexponential fit to \(<\chi (t) >\) is needed.
The boundary between moderate and slow folding sequences is rather
arbitrary.  In both these cases we find that the various stages of folding
can be described by a three stage multipathway mechanism (TSMM): 
The initial stage is characterized by random collapse, the second stage
corresponds to the kinetic ordering regime in which the search among
compact structures leads to native-like intermediates 
\cite{Cam93,Guo}. The final stage
corresponds to activated transition from one of the native-like structures
to the native state, which is the rate 
determining step
for folding. Thus, the transition states occur close to
the native conformation as was shown some time ago in off-lattice
simulations \cite{Honey90,Honey92}. Our earlier lattice and
off-lattice studies describe in detail the evidence for the
TSMM. Analysis of the trajectories probing the approach to the native
state (measured by \(\chi (t)\)) for the sequences studied exhibits
similar behavior.

\subsubsection{Dependence of Folding Times on \(\sigma \)}

In an earlier two dimensional lattice simulations we have suggested that
the sequences that fold fast appear to have small values of \(\sigma =
(T_{\theta } - T_{f})/T_{\theta }\) \cite{Cam93}. The reason for expecting
the relationship between \(\sigma \) and the folding time \(\tau_{f}\) has
been given recently \cite{Thirum95}. Physically if 
\(\sigma \) is small then
\(T_{\theta } \approx T_{f}\) and all possible transient structures that
are explored by the chain on its way towards the native state are of
relatively high free energy.  Consequently, all the structures that are
likely to act as \underline{traps or intermediates are effectively
destabilized} and thus folding to the native structure is rapid.  For
large \(\sigma \) \(T_{f}\) and \(T_{\theta }\) are well separated 
and hence the chain searches many compact globular states in a rough
energy landscape that leads to slow folding. When  \(\sigma \) is small
the folding process and the collapse is synchronous and this leads to
fast folding. 
Similar conclusions have been reached for 2D square lattice proteins 
\cite{Cam93,Betan}. 

In order
to test the possible relationship between the folding time and \(\sigma 
\) we have calculated \(\tau_{f}\) for the
database of sequences generated by the method described in Sec. (II).  The
results of this simulations are plotted in Fig.  (11) for \(N=15\) and
\(B_{0}=-0.1\) (solid circles), \(B_{0}=-2.0\) (open circles). The figure
shows that the folding time \(\tau_{f}\) correlates statistically
extremely well with the intrinsically thermodynamic parameter \(\sigma \). 
The sequences span a range of \(\sigma \) and consequently meaningful
conclusions can be made.  In fact, sequences with small values of \(\sigma
\) fold extremely rapidly (fast folding sequences): for \(\sigma \)
smaller than 0.1, \(\tau_{f}\) hardly exceeds \(10^{4}\) MCS. Most
sequences (i.e., having \(\sigma \) between 0.15 and 0.6) have
intermediate folding rates extending from roughly \(10^{5}\) to \(10^{7}\)
MCS (moderate folding sequences). The sequences with the largest values of
\(\sigma \) (greater than 0.6) are slow folding sequences, whose typical
folding times are above \(10^{7}\) MCS. Thus, in the range of \(\sigma \)
examined the folding rate changes by about four to five orders of
magnitude. It is important to point out that all fast folding sequences
are optimized, whereas moderate folding sequences also include random
ones. Slow folding sequences are exclusively random.  This distinction
based on folding times and its dependence on \(\sigma \) was used in
classification of the sequences in the discussion in Sec. (III.A.3). 

\subsubsection{Relationship between \(\tau_{f}\) and \(\Delta\)}

Sali {\em et al.} have recently asserted (without providing explicit
calculation of folding times) that sequences that fold rapidly and whose
native conformation is also stable are characterized by large
gap \cite{Sali94a,Sali94b}. 
The gap in their model is defined using Eq. (18). In 
order to check
this claim we plot \(\tau_{f}\) as a function of \(\Delta \) in Fig. 
(12) for \(N=15\). The corresponding plot for \(N = 27\) 
is shown in the next section.  Fig. (12) shows that this parameter
is of little relevance when used to classify folding rates of
different sequences. It appears that the sequences with large gaps
\(\Delta \) usually fold rather rapidly (about \(10^{4}-10^{5}\) MCS).
However, sequences having a small energy gap \(\Delta \) can have either very
small folding times (below \(10^{4}\) MCS) or do not reach native state
even after \(10^{8}-10^{9}\) MCS. For example, from Fig. (12) it is clear 
that if \(\tau_{f}\) is fixed at
\(10^{5}\) MCS then we can, in principle, generate a large number of
sequences with very small values of \(\Delta \) to very large
\(\Delta \) all with roughly the same \(\tau_{f}\). Thus,
\(\Delta \) alone cannot be used to discriminate between fast and
slow folding sequences. The existence of large values of \(\Delta/k_{B}T\)
being a criterion for stability follows from Boltzmann's law with proteins
being no exception.

\subsubsection{Kinetic Events in the Folding Process}

We have systematically investigated the microscopic processes that are 
involved in the folding of several sequences. This has been done by using 
the methodology for monitoring the intermediates described in Eqs. (16) and 
(17). We discuss the results of this study for the various sequences using 
the classification in terms of the parameter \(\sigma \). \\
{\bf (a) Fast folders (\(\sigma \lesssim 0.1\)):} In this case there are no 
well defined 
intermediates in the sense that the chain gets trapped in a conformation 
that is distinct from the native state for any length of time. For these 
fast folding sequences we find that the native conformation is reached by 
essentially a  TINC  mechanism, 
i.e. once a certain 
number of critical native contacts is established then the native state 
is reached rapidly \cite{Guo,ThirumGuo,Abk1}.  
By using a combination of 
Eqs. (16) and (17) we find that 
for fast folders the chain frequently visits the nearest low lying energy 
conformations even after reaching the native structure. For these 
sequences these low lying states are almost native-like (have about 90 
percent of native contacts). In terms of the underlying energy landscape 
it is clear that they belong to the same basin of attraction as the 
native conformation. \\
{\bf (b) Moderate and slow folders (\(\sigma \gtrsim 0.1\)):} These 
sequences appear to have well defined intermediates and 
their significant role makes folding in this range of \(\sigma \) quite 
distinct from the fast folders. In Fig. (13) we plot \(P_{r}\) 
and \(P_{fl}\) (see Eqs. (16,17)) for a variety of sequences. The sequences 
are arranged in 
order of increasing folding time. Recall \(P_{r}\) corresponds to the 
average probability that the 
intermediate with the energy \(E_{k}\) has the highest probability of 
occurrence  before the 
native conformation is reached for the first time and \(P_{fl}\) is the 
average  probability that this state is revisited after the native state is 
reached. This graph shows several striking features: (i) For moderate 
folders there is a finite  probability of the chain revisiting the same 
intermediate that it sampled in the approach to the native conformation. 
This seldom happens in the sequences that fold slowly. It is 
clear that slow folding sequences have well defined intermediates which 
are not visited after the chain reaches the native conformation. These 
results suggest that the rate determining step in slow folding sequences 
is the transition from one of these intermediates to the native state. 
This involves overcoming a substantial free energy barrier. The 
existence of this barrier also prevents frequent excursions from the 
native state. (ii) It is of interest to probe the nature of intermediates 
that are encountered in the folding process. In Figs. (14a) and (14b) we 
show, 
respectively, the fraction of native contacts in the most populated 
intermediates and the corresponding overlap \(\chi _{k}\) with 
native conformation for all sequences. For both fast and moderate folders 
it is clear that 
the states that are sampled have great structural similarity to the native 
conformation. In fact, in this case these conformations have roughly 80 
percent of the native contacts. However, slow folding sequences have only 
about 50 percent of native contacts in the intermediate structures. It 
also turns out that in the case of moderate folders the most populated 
intermediates prior to formation of native conformation is most often the 
first excited state whereas in slow folding sequences it is the higher 
excited states that have the largest probability of occurring. (iii) The 
rate of formation of the intermediates can be ascertained by examining 
Fig. (14c), in which the ratio of the mean time to reach the intermediate 
\(\tau _{k}\) to the folding time \(\tau _{fp}\) is plotted. 
Intermediates for fast and moderate folding sequences 
usually occur at later stages of folding than for slow folding sequences. 
In fact, for the latter cases the ration \(\tau _{k}/\tau _{fp}\) can be 
as low as 0.1. This implies that these relatively stable  misfolded 
structures are formed 
relatively early in the folding process and these off-pathway processes 
therefore slow down the folding considerably. It also follows that the 
rate determining step for slow folders occurs late in the folding process 
implying that the transition states are closer to the native structure 
\cite{Honey92}.

\subsection{\(N = 27\): \(B_{0}=-0.1\) and \(B_{0}=-2.0\)}

	For \(N = 27\) we have generated 15 sequences with \(B_{0}=-0.1\) 
using the optimized design procedure described in Sec. (II.E). In this case 
all of the sequences have been optimized so that \(\sigma \) is in the 
range of \(\sigma \lesssim 0.12\). We have studied the thermodynamic 
properties of few sequences with \(B_{0}=-2.0\). Since we have already 
established that non-compact structures make significant contribution to 
both the thermodynamic and kinetic properties for \(N=15\) it is 
necessary to include them in studying the case of \(N=27\) as well. The 
number of non-compact structures for \(N=27\) is of the order of 
\(10^{18}\) and their exact enumeration is impossible. We have, 
therefore, used slow cooling Monte Carlo method (see Appendix A) to 
calculate the thermodynamic properties in this case. The calculation 
of the kinetic processes 
have been done as before for \(N=15\). We discuss there results 
below. Since the qualitative behavior remains the same we provide a less 
detailed account for this case. 

\subsubsection{Thermodynamic Characteristics}

As before we have determined \(T_{\theta }\) and \(T_{f}\) from computing 
the temperature dependence of \(C_{v}\) and \(\Delta \chi\). The 
simulation temperature was found by requiring that the overlap function 
\(<\chi (T_{s})> = 0.21\). These temperatures were calculated using  
Monte Carlo simulations. It is interesting to compare the results for the 
overlap function obtained from MC simulations with that calculated using 
CSE only (Figs. (15,16)). The results for the temperature 
dependence of \(\Delta \chi\) for one sequence with  \(B_{0}=-0.1\) is 
given in Fig. (15b) and in Fig. (16b) \(\Delta \chi\) as a function of 
\(T\) is plotted for another sequence with \(B_{0}=-2.0\). The dotted 
line in these figures are the results obtained using the contribution of 
compact structures only. Both these sequences have large values of the 
energy gaps \(\Delta \). These figures show dramatically that the 
neglect of noncompact 
structures leads to serious errors in the determination of \(\Delta 
\chi\). Similar discrepancies are found for other thermodynamic 
quantities as well. The errors in the estimate of \(T_{f}\) in these two 
sequences are about a factor of 2 - 3. In fact, in both cases the 
estimate of \(T_{f}^{CSE}\) obtained using compact structures alone 
\underline{exceeds} 
the collapse transition temperature \(T_{\theta }\). The neglect of 
noncompact structures, even for \(B_{0}=-2.0\), results in more serious 
errors than for \(N=15\). In any event for both \(N=15\) and \(N=27\) 
restriction to compact structures alone can lead to incorrect results for
thermodynamic properties. 

	It is interesting that the discrepancy between the simulation 
temperature \(T_{s}^{MC}\) and \(T_{s}^{CSE}\) is even more dramatic for 
\(N=27\) at the value of \(B_{0}=-0.1\) (Fig. 15a). In particular, for the 
sequence 61 for which \(\Delta \chi (T)\) is displayed in Fig. (15b)  
\(T_{s}^{CSE}=3.19\), which is even larger than \(T_{\theta 
}^{MC}=1.22\), exceeds \(T_{s}^{MC}=1.17\) by almost a factor of three. 
As expected for \(B_{0}=-2.0\) the discrepancy is somewhat smaller but is 
still very significant (see Fig. (16a)). In this case \(T_{s}^{CSE}=3.60\) 
is almost twice as large as \(T_{s}^{MC}=2.06\). The value of \(T_{\theta }\)
for this sequence is \(T_{\theta }^{MC} = 2.14\).  This implies 
that, if the simulation temperatures \(T_{s}^{CSE}\) are used, the only  
conformations that are thermodynamically relevant are the random 
coil ones. These observations clearly demonstrate that the use of only 
compact structures for calculating thermodynamic quantities is in general 
totally flawed and would lead to incorrect evaluation of the folding 
rates for both \(B_{0}=-0.1\) and \(B_{0}=-2.0\) for any \(N\). It is 
worth noting that a similar conclusion has been reached 
for a two letter code model with \(N = 27\) \cite{Socci95}. 

In Fig. (17) we collect the results for the ratio of the simulation 
temperature calculated using compact structures \(T_{s}^{CSE}\) to that 
computed using all available conformations \(T_{s}\). For the case 
of \( N = 27\) we have already emphasized that the simulation temperature 
(denoted as \(T_{s}^{MC}\)) is calculated using Monte Carlo simulations 
the details of which are examined in Appendix A. In Fig. (17a) we show the 
ratio \(T_{s}^{CSE}/T_{s}^{FE}\) for \(N=15, B_{0}=-0.1\). It is clear 
that except for one sequence this ratio is greater than unity and is 
large as 2.5. The results \(T_{s}^{CSE}/T_{s}^{MC}\)  for \( N = 27, 
B_{0}=-0.1\) are displayed in Fig. (17b). Here the effects are even more 
dramatic. In all cases this ratio exceed 2.0 implying that noncompact 
structures make significant contributions in determining thermodynamic 
properties.

A further consequence of using only compact structures to determine
\(T_{s}\) (as has been done elsewhere \cite{Sali94a,Sali94b}) 
is that at these
high temperatures the native conformation has no stability. It is because
of the very large values of \(T_{s}^{CSE}\), Sali {\em et al.} find that in
many cases their native conformations are not well populated. In fact, in
most cases the probability of being in the native conformation is less
than 0.1.

\subsubsection{Folding Kinetics: \(N=27\) and \(B_{0}=-0.1\)}

Since the time scales for folding for \(N=27\)  are quite long we have 
restricted ourselves to determining the folding rates for optimized 
sequences only. Thus, we have examined 15 sequences with characteristic 
temperatures \(T_{\theta }\) and \(T_{f}\) that provide the values of 
\(\sigma \) being less than about 0.12. These sequences, according to the 
classification derived at detailed study of \(N=15\), would all be the 
fast folders. Thus, we expect that most of these sequences would 
reach 
the native conformation extremely rapidly.  
In these instances folding would appear to be a two state 
all-or-none process and the time dependence of \(<\chi (t)>\) should be 
exponential. All these expectations are borne out. In Fig. (18a) we show 
\(<\chi (t)>\) for one of the sequences from fifteen examined. It is 
obvious that \(<\chi (t)>\) is well fit by a single exponential process. 

For some sequences we do find that \(<\chi (t)>\) can be better fitted by 
a sum of two exponentials. Thus, for \(N=27\) even for these small values 
of \(\sigma \) we find that these sequences should be classified as 
moderate folders (Fig. (18b)). This is not surprising because as \(N\) 
increases the 
probability of forming misfolded structure also increases. The boundaries 
differentiating the fast and moderate folders depend on the sequence 
length. For large values of \(N\) the range of \(\sigma \) over which the 
sequences behave as fast folders decreases. Consequently, the partition 
factor \(\Phi (T)\), which is the fraction of initial population of 
molecules that reaches the native conformation {\em via } 
TINC mechanism, decreases.

\subsubsection{Dependence of Folding Times on \(\sigma \), 
\(\Delta \), and \(\Delta _{CS}\) }

The dependence of the \(\tau_{f}\) on \(\sigma \) is shown in Fig. (19). 
Even though we have examined only a small range of \(\sigma \) the general
trend that \(\tau_{f}\) is well correlated with \(\sigma \) is clear. 
Considering that one has statistical errors in determining both \(\sigma
\) and \(\tau_{f}\) the observed correlation between these quantities is,
in fact, remarkable. In this limited range of \(\sigma \) the folding time
changes by nearly a factor of 300 indicating that small changes in
\(\sigma \) (which is an intrinsic property of the sequence) can lead to
rather large changes in \(\tau_{f}\). The behavior of \(\tau_{f}\) on
\(\Delta \) is shown in Fig. (20). The trend one notices is the same as in
the case of \(N=15\). In fact, the lack of any correlation between
\(\tau_{f}\) and \(\Delta \) is even more apparent here.  As in the case
for \(N=15\) it is possible to generate sequences with arbitrary values of
the gap that would all have roughly the same folding time.  Notice that
although \(\sigma \) covers only a small range the gap extends over a 
much wider interval.  This also implicitly indicates no dependence of 
\(\sigma \) on \(\Delta \). 

If there would be any plausible relation between \(\tau _{f}\) and 
\(\Delta \) it is clear that \(\Delta \) has to be expressed in terms of 
a suitable dimensionless parameter. Unfortunately the proponents of the 
energy gap idea \cite{Sali94a,Sali94b} 
as the discriminator of folding sequences have used 
varying definitions of \(\Delta \) in different papers without providing 
the precise way this is to be made dimensionless. The energy parameter 
that would make \(\Delta \) dimensionless cannot be \(B\), the standard 
deviation in the distribution of contact energies which merely sets the 
energy scale, 
because this would mean this criterion would apply only to this model. 
The gap \(\Delta \) in Figs. (12) and (20) is 
measured in 
units of \(B\). A very natural way to make \(\Delta \) dimensionless is 
to divide it by \(k_{B}T\) which is in this case  \(k_{B}T_{s}\). In 
Fig. (21) we have presented the folding times \(\tau _{f}\) as a function 
of \(\Delta /T_{s}\). The upper panel is for \(N = 15\), while the lower 
one is for \(N = 27\). These figures demonstrate even more 
dramatically the irrelevance of \(\Delta /T_{s}\) as a parameter in 
determining folding times. In fact, this figure is almost like a scatter 
plot. Thus, it is clear that energy gap alone (measured in any suitable 
units) 
does not determine folding times in these models. 
From this it follows that the classification of sequences into fast and 
slow folders cannot be done using the value of \(\Delta \) (measured in 
any reasonable units) alone. 

In the literature it has been forcefully asserted that foldability of
sequences in this class of models is determined by \(\Delta _{CS}\),
the energy gap for the ensemble of compact structures
\cite{Sali94a,Sali94b}. Plotting \(\tau
_{f}\) as a function of \(\Delta _{CS}\) for our model appears to be
somewhat ambiguous because about half of the sequences have
non-compact native conformations. In Fig. (22) we plot \(\tau _{f}\)
as a function of \(\Delta _{CS}\). The upper panel is for \(N=15\)
and the lower panel is for \(N=27\). It is clear from this figure 
that there is no useful correlation between \(\tau _{f}\) and \(\Delta
_{CS}\). It appears that one can generate sequences with a range of
\(\Delta _{CS}\) all of which have roughly the same folding times. 
 
\subsection{Kinetic Accessibility and Stability of Native Conformation}

It is well known that many natural proteins reach their native conformation 
quite rapidly without forming any detectable intermediates. However, 
proteins are only marginally 
stable in the sense the equilibration constant \(K\) for the reaction 
\begin{equation}
U \rightleftharpoons F
\end{equation}
is only between \(10^{4} - 10^{7}\). In Eq. (20) {\em U} refers to the 
denaturated 
unfolded conformations, and {\em F} is the folded native state. Thus, 
\(\Delta G = G_{F} - G_{U} = -k_{B}T ln K\) is in the range of 
\(-12k_{B}T\) to \(-18k_{B}T\). While this is not as large as one observes 
in typical chemical reactions involving cleavage  of bonds it is still 
sufficiently large so that the 
native conformation is overwhelmingly populated relative to the ensemble 
of \underline{unfolded} states. The Helmholtz free energy of 
stabilization of the 
folded state with respect to the ensemble of denaturated conformations 
can be written as 
\begin{equation}
\beta \Delta F \simeq -\Delta E + S_{U}
\end{equation}
where \(\Delta E\) is the stabilization energy, \(S_{U}\) is the entropy 
of the ensemble of the unfolded structures. We have assumed that the 
conformational entropy associated with the native conformation is 
negligible. If the 
ensemble of denaturated structures corresponds to self-avoiding random walks 
\(S_{U}\) 
for lattice models can be estimated using Eq. (4). For cubic lattice 
\(Z_{eff}=4.684\), \(\gamma = 1.16\), and thus \(\beta \Delta F \simeq  
-\Delta E +42\) for \(N=27\). If we insist that \(\beta \Delta F \approx 
10\) this would imply that \(\Delta E (T) \approx -52 k_{B}T\). The same 
calculations would show that \(\Delta E (T) \approx -20 k_{B}T\) if the 
ensemble of structures in the denaturated states are essentially compact 
structures. This estimate is a bit more realistic. These estimates show that 
the native conformation is 
overwhelmingly populated under appropriate conditions relative to the 
compact structures. More recent 
experimental studies involving denaturant induced unfolding of few 
proteins have been used to probe the "spectrum" of low free energy 
conformations \cite{Bai}. These tools, 
while being relatively primitive, suggest that typically the equilibrium 
intermediates are also about \(6-8 k_{B}T\) higher in free energy than 
the native conformation. Thus, even in these cases, under native 
conditions, the native state is overwhelmingly (\(\geq 0.9\)) occupied.
 
Most of the simulations we have discussed so far have been done at 
temperatures below \(T_{f}\) but high enough that the \(<\chi (T_{s})>\) 
is as large as possible. This, of course, has been done for computational 
reasons and the constant value of \(<\chi (T_{s})>\) has been chosen so 
that the properties of different sequences can be compared on equal 
footing. However, at these simulation temperatures \(T_{s}\) the 
probability of occupation of the native conformation \(P_{nat}(T_{s})\) 
varies between \(0.2\) and  \(0.6\), which is significantly smaller than 
what is observed in real proteins. Notice that the calculations of  
Sali {\em et al.} \cite{Sali94a,Sali94b} have been done at such elevated 
temperatures that 
the probability of occupation of native conformation for two hundred 
sequences  examined  is usually between 
\(0.01 - 0.05\) and none exceeds \(0.4\). Thus, these authors, 
although have stated the criterion for simultaneously satisfying kinetic 
accessibility and stability (this follows from Boltzmann`s law), did not 
provide \underline{any} computational or theoretical evidence that 
\underline{any} of 
their sequences obeyed the stated criterion at any temperature. 

In light of the above arguments it is necessary to use a different 
criterion for the choice of \(T_{s}\) which would ensure stability of the 
native proteins. Accordingly, we have performed simulations for a few 
sequences with \(N=15\) and \(B_{0}=-0.1\) at the temperatures which are 
determined using the following equation
\begin{equation}
\eta (T_{s \eta}) = 1 - P_{nat}(T_{s \eta}) = c
\end{equation}
where \(P_{nat}(T_{s \eta}) \) is the probability of the chain being in 
the native conformation at \(T=T_{s \eta}\). The constant \(c\) was 
chosen to be equal to \(0.1\), which implies that probability of the 
chain being in the native conformation is \(0.9\). In order to present 
the contrast between the kinetic behavior of the sequences at 
temperatures chosen using Eq. (22) \(T_{s \eta}\) we chose three sequences 
one from fast folders (\(\sigma \lesssim 0.1\)), one from moderate folders 
(\(0.1 \lesssim \sigma \lesssim 0.6\)), and one from slow folders (\(\sigma 
\gtrsim 0.6\)). The 
ratios of the simulation temperatures for these sequences \(T_{s \eta}\) 
determined using Eq. (22) to \(T_{s}\) obtained using the overlap criterion 
are approximately \(0.5\), \(0.6\), and \(0.6\) for fast, moderate, and 
slow folders, respectively. It appears that stability of the 
native conformation (occupancy of this state \(\geq 0.8\)) for both 
\underline{fast}  
and \underline{moderate} folders can be achieved, if \underline{the 
simulation 
temperature \(T_{s}\) is taken to be one half of the folding temperature 
\(T_{f}\)}. 

The quantity \(\eta (T)\) should only be the function of \(T/T_{s \eta}\)
if the gap between the native conformation and the first excited state 
is large. More precisely, we expect this to be valid for all temperatures 
such that \(k_{B}T \ll \Delta\), where \(\Delta\) is the gap, separating 
the energy of the native conformation and the first excited state. To see 
this \(\eta (T)\) can be written as \begin{equation}
\eta (T) \simeq \frac{\exp(-\frac{\Delta}{k_{B}T} )}{1 +
\exp(-\frac{\Delta} {k_{B}T})}
\end{equation}
for \(\Delta/k_{B}T \gtrsim 1\). The temperature \(T_{s \eta}\) is 
determined from Eq. (22) and thus \(\eta (T)\) becomes
\begin{equation}
\eta (T) = f(T/T_{s \eta}) \simeq \frac{y^{1/\tau}}{1+y^{1/\tau}},
\end{equation}
where \(y=c/(1-c)\) and \(\tau = T/T_{s \eta }\). This is confirmed in Fig. 
(23), where plots of 
\(\eta \) as a function of \(T/T_{s \eta}\) are shown for three sequences. 
Two of them follow the behavior in Eq. (24) for \(T/T_{s \eta} < 1.5\), 
whereas the sequence with small gap (\(\Delta/k_{B}T_{s \eta} < 1\)) does not. 

The behavior of \(<\chi (t)>\) and the fraction of unfolded molecules 
\(P_{u}(t)\) as a 
function of time for two of the sequences is shown in Figs. (24) and
(25).  
The temperature for the third sequence was so low for Eq. (22) to be 
satisfied that the folding time for this sequence was estimated to be in 
excess of \(10^{10}\) MCS. By studying these graphs we draw the following 
conclusions: (i) The overlap function \(\chi (t)\) at the low temperature 
(Fig. (24a))  
is clearly biexponential, whereas at \(T=T_{s}\) (see Eq. (11)) the folding 
process was an all-or-none process. This is because the very small 
barrier (\(\delta E^{\ddagger} \simeq k_{B}T_{s}\)) becomes discernible at 
\(T 
\simeq T_{s \eta}\). This fact alone would yield to the prediction that 
\(\tau _{f}\) at \(T_{s \eta}\) should be about a factor of \(10\) 
larger. This is consistent with simulation results. So the emergence of 
the second component in \(P_{u}(t)\) (Fig. (24b)) is due to the activated 
transition 
from the first excited state to the native conformation. The behavior of 
\(\chi (t)\) for the moderate folding sequence is qualitatively similar to 
that at \(T=T_{s}\) except that the time constants are larger because 
\(T_{s \eta} \approx \frac{1}{2}T_{s}\) (Fig. (25)). (ii) We find that the 
ratio of 
the folding times for the three sequences at \(T=T_{s \eta}\) is roughly 
the same as found at \(T=T_{s}\). This suggest that although the overall 
folding times have increased considerably we do not expect to see 
qualitative differences in the fundamental conclusion regarding the 
statistical correlation between \(\tau _{f}\) and \(\sigma \). (iii) 
Examination of the fraction of unfolded molecules \(P_{u}(t)\) shows that 
the biexponential behavior is consistent with the kinetic 
partitioning mechanism \cite{Guo}. 
A fraction of the molecules, \(\Phi (T)\), reaches the native 
state very rapidly without forming any intermediates {\em via}  TINC
mechanism, while the remainder follows a more complex kinetic 
mechanism. The partition factor \(\Phi (T)\) is nearly unity for fast 
folding sequences at \(T=T_{f}\) leading to a two state behavior, whereas 
at low temperatures \(\Phi (T) < 1\). For fast folding sequences shown in 
Fig. (24b) \(\Phi (T=T_{s \eta})\) is approximately \(0.4\). In the case of 
moderate folders \(\Phi (T)\) is always less than one for all \(T < 
T_{f}\). This is affirmed in Fig. (25b), where we find that \(\Phi (T=T_{s 
\eta})\) is approximately \(0.19\) for the sequence with \(\sigma = 
0.19\).

\section{Conclusions and Discussion}

In recent years there have been numerous studies of lattice models of
proteins, in which the protein is modeled as a self-avoiding walk on a
three dimensional cubic lattice. A variety of interactions between the
beads on this lattice has been studied. In this work we have carried out
a systematic investigation of the kinetics and thermodynamics of a
heteropolymer chain confined to a cubic lattice under a variety of
conditions. We have, as majority of the studies have in the past, used a
random bond model to specify the interaction between the beads.
Specifically the interactions between the beads are chosen from a Gaussian
distribution of energies with a non-zero value of the mean \(B_{0}\). In
order to obtain a coherent picture of folding in this highly simplified
representation of proteins we have studied the kinetic and thermodynamic
behavior for two values of \(N\) (the number of beads in the chain),
namely \(N = 15\) and \(N = 27\). For \(N = 15\) the thermodynamic
characteristics can be exactly calculated because all possible
conformations in this case can be exhaustively enumerated, whereas for \(N
= 27\) all the quantities of interest have to be obtained by Monte Carlo
simulations. In addition the mean hydrophobic interaction \(B_{0}\) has
also been varied in this work. The kinetics of folding for a number of
sequences have been obtained by Monte Carlo simulations. The work reported
here allows us to assess the various factors that govern folding in this
model. In addition the variation in parameters can be used to address some
of the proposals made earlier in the literature. 

The exact calculations for \(N = 15\) of the thermodynamic quantities by
enumerating all the conformations clearly demonstrate the importance of
non-compact structures in determining accurately the characteristic
properties of the protein chain. For \(B_{0} = -0.1\), which is a
realistic value for proteins (see Sec. (II.B)), the values of the 
temperatures
\(T_{\theta}\) (the collapse transition temperature) and \(T_{f}\) (the
folding transition temperature) are much higher, if only the compact
structure are included. This, in turn, makes the simulation temperature
\(T_{s}\) (see Eq.  (11)) also quite high. In fact for most sequences
\(T_{s}\) determined using only the compact conformations exceeds
\(T_{\theta}\). Although the discrepancies between \(T_{s}\) determined
using CS and all the conformations for \(B_{0} = -2.0\) (a rather
unrealistic value for proteins) is lesser the resulting kinetics is
significantly affected. However, for \(N = 27\) the differences between
\(T_{s}\) found from CSE and that determined from Monte Carlo simulations 
are very
significant even for \(B_{0} = -2.0\). The two values in some instances
differ by a factor of two. Thus if the simulation temperature is chosen
using only compact structures that is much more
convenient, then one obtains a value for \(T_{s}\) that often exceeds the 
collapse transition temperature. This fact alone explains why the 
probability of occupancy of the
native state was very small (in many cases less than \(0.1\)) in the
previously reported studies in the literature \cite{Sali94a,Sali94b}. It has 
been pointed out by Chan \cite{Chan95} that the high temperatures used by 
Sali {\em et al.} \cite{Sali94a,Sali94b}  results in the
equilibrium population of the native conformation for 'folding sequences'
of only \(0.01 - 0.05\) with none exceeding \(0.4\). This was not
appreciated by Sali {\em et al.} because they used a very
crude criterion for folding. In particular a sequence was designated as a
folding sequence if at the temperature of simulation the native
conformation was reached once (a first passage time) at least four times
among ten independent trajectories of maximum duration of \(50\times
10^{6}\) MCS \cite{Sali94b}. Our studies 
indicate that ten trajectories is absolutely
inadequate for obtaining even qualitatively reliable estimate of 
\underline{any} property of interest. 

We had shown earlier in lattice and off-lattice studies 
\cite{Cam95,Cam93,Guo} that the two temperatures that are intrinsic 
to a given sequence are \(T_{\theta}\) and \(T_{f}\). This general 
thermodynamic behavior for protein-like heteropolymers has been confirmed 
recently \cite{Socci95}. 
In addition the
temperature of simulation (or experiment) should be below \(T_{f}\) so
that the folded state is the most stable. Theoretical arguments
suggest \cite{Thirum95} that the parameter \(\sigma = (T_{\theta} - 
T_{f})/T_{\theta}\) is
a useful indicator of kinetic foldability in the models of the sort
considered here, and perhaps in real proteins as well. By examining the
kinetics of approach to the native conformation we have found that roughly
(this holds good for \(N = 15\)) the kinetic foldability of sequences can
be divided into three classes depending on the value of \(\sigma \). For 
\(N = 15\) we
find that fast folders have \(\sigma \) less than about \(0.1\), moderate
folders have \(\sigma \) in the range of approximately \(0.1\) and
\(0.6\), while \(\sigma \) values greater than about \(0.6\) correspond to
slow folders. It should be emphasized that the ranges for these three 
classes of sequences depend on the length of the chain: longer chains 
have smaller range of \(\sigma \) for fast folding. Fast folding sequences 
can kinetically access their ground
state at relatively higher temperature compared to slow folding sequences.
The kinetics of approach to the native state differs significantly between
fast folding sequences and those that are moderate folding sequences. In
the former case the native conformation is reached via a  TINC 
mechanism and there are no detectable intermediates. In
this case the folding appears as an all-or-none process. The
kinetics is essentially exponential. On the other hand the moderate
folding sequences reach the native conformation (predominantly) via a
three stage multipathway mechanism as was reported in our several earlier
studies. 

The time dependence of the approach to the native conformation is
very revealing. For moderate folding sequences the simulation temperature
is lower than for sequences with small \(\sigma \). Thus if approach to
the native conformation with the same value of the overlap function is
examined we find that the moderate folding sequences follow the kinetic
partitioning mechanism (KPM) \cite{Guo,ThirumGuo}. This implies that a 
fraction of the initial
population of molecules reaches the native state via the TINC 
mechanism while the remaining one follows the three stage multipathway
process. The partition factor, \(\Phi (T)\), that determines the fraction
that follows the fast process is sequence and temperature dependent. Thus
even fast folding sequences, which would have \(\Phi (T)\) close to unity at
higher temperature, would have fractional values less than unity 
at lower temperature.
This is clearly seen in Fig. (24b). This work suggests that KPM should be a
generic feature of foldable proteins. The partition factor can be altered
by mutations, temperature, and by changing other external factors such as
pH. 

We have explicitly calculated the folding times for a number of sequences
for the various parameters (different \(N\) and \(B_{0}\)) values. This is
of great interest to examine whether there is any intrinsic property of
the sequences that can be used to predict if a particular sequence folds
rapidly or not.  It has been argued based on the random energy model for
proteins that folding sequences should have as large a value of
\(T_{f}/T_{g}\) as possible, where \(T_{g}\) is an equilibrium glass
transition temperature \cite{Bryn95,Gold92}. Kinetic studies of lattice 
models of proteins
suggest that this criterion may be  satisfied for foldable sequences 
\underline{provided} 
\(T_{g}\) is substituted by a kinetic glass transition temperature which
is defined as the temperature at which folding time scale exceeds a
certain arbitrary value \cite{Sali94b}. Scaling 
arguments have been used to suggest that
there should exist a correlation between folding times and \(\sigma \) 
\cite{Cam93,Thirum95}. 
More recently it has been emphatically stated in the form of a theorem
(without the benefit of explicit computations) that the necessary and
sufficient condition for rapid folding in the models studied here is that
there should be a large gap (presumably measured in units of \(k_{B}T\))
between the lowest energy levels. The calculations of folding times
reported here clearly show that there is no useful correlation between the
gaps and folding times for any parameter values that we have investigated.
In fact for a specified folding time one can engineer sequences with both
large and small values of the gap. Thus the precise value of the gap alone
cannot discriminate between folding sequences. On the other hand there is
a good correlation between the folding times for sequences and \(\sigma
\). It is clear that sequences with small values of \(\sigma \) have short
folding times, while those with larger values have higher folding times.
It should be emphasized here that the criterion based on \(\sigma \)
should only be used to predict trends in the folding times. 
In this sense this criterion
should be used in a qualitative manner. The major advantage of showing the
correlation between \(\sigma \) and the folding times is that \(\sigma \)
can be experimentally determined. The folding transition temperature is
nominally associated with the midpoint of the denaturation curve while
\(T_{\theta}\) is the temperature at which the protein resembles a random
coil. 

In the usual discussion of protein folding only the issue of kinetic 
foldability of sequences are raised. Since natural proteins are 
relatively stable (this does not imply that the protein does not undergo 
fluctuations in the native conformation) with respect to both the 
equilibrium intermediates and the ensemble of unfolded conformations it 
is imperative to devise the criterion for simultaneously satisfying kinetic 
accessibility of the native conformation and the associated stability. 
This paper for the first time has provided an answer to this issue. It is 
clear from our study that fast folders have small values of \(\sigma \), 
and consequently designing proteins \cite{Cam93,Fer} using this criterion 
assures kinetic 
accessibility of the native conformation at relatively high temperature. 
For example, if \(T_{\theta }\) is taken to be about \(60^{\circ }C\) then 
the fast folders would reach the native conformation rapidly even at 
temperatures as \(55^{\circ}C\) (\(\sigma \approx 0.1\)). However, at 
these temperatures the native conformation may not be very stable, i.e. 
the probability of being in the native conformation may be less than 0.5. 
On the other hand, if these fast folders are maintained at \(T \approx 
(0.5 - 0.6) T_{f} \approx (30-35)^{\circ} C \) for \(\sigma 
\approx 0.1\) and \(T_{\theta } \approx 60^{\circ}C\) then both kinetic 
accessibility as well as stability can be simultaneously satisfied. 

Our simulations also suggest that the dual criterion can also be 
satisfied by using moderate folders. In these cases we find that the 
native state is reached relatively rapidly at low enough temperatures 
(\(T \approx (25 - 30)^{\circ} C\) assuming \(T_{\theta} \approx 
60^{\circ}C\)) at which the excursions to other conformations are not 
very likely. In the extreme case of very slow folders we find (see Fig. 
(26)) that the average fluctuation probability of leaving the native 
conformation after initially reaching it is small, i.e. the 
stability condition is easily satisfied. In these cases however the 
kinetic accessibility is, in general, not satisfied. From these 
observations it follows that in order to satisfy the dual criterion it is 
desirable to engineer fast folders (small values of \(\sigma \)) and 
perform folding at temperatures around \((0.5-0.6)T_{f}\). 
Alternatively, one can use moderate folders at temperatures around 
\(0.8T_{f}\). Since moderate folders are more easily generated (by a 
random process) it is tempting to suggest that many natural proteins 
specifically large single domain proteins may be moderate folders. 

Finally, we address the applicability of the results obtained here to real
proteins. Since many features that are known to be important in real
proteins (such as side chains, the possibility of forming secondary
structures etc.) are not contained in these models the direct applicability
to real proteins is not clear. Nevertheless simulations based on other more
realistic minimal models together with theoretical arguments can be used to
suggest that the scenarios that have emerged from this and other related
studies should be observed experimentally. In particular it appears that the
kinetic partitioning mechanism which for most generic sequences is a
convolution of the  topology inducing nucleation collapse mechanism
and the three
stage multipathway kinetics should be a very general feature of protein
folding in vitro. The theoretical ideas developed based on these minimal
models also suggest that the partition factor \(\Phi (T)\) is a property of 
the intrinsic
sequence as well as external factors such as temperature, pH etc. In fact
recent experiments on chymotrypsin inhibitor 2 (CI2) suggest that this
KPM has indeed been observed 
\cite{Otzen}. In this
particular case Otzen {\em et al.} have observed that CI2 reaches the 
native
state immediately following collapse. This, in the picture suggested here
and elsewhere \cite{Guo}, 
would imply that in the case of CI2 under the 
conditions of
their experiment (\(T = 25^{\circ}C\) and \(\text{pH} = 6.2\)) the native 
conformation is
accessed via a  TINC mechanism. Moreover, the 
partition factor 
\(\Phi (T = 25^{\circ}C)\) is close to unity making CI2 under these 
conditions a
fast folder. If we assume that \(T=25^{\circ}C \approx  T_{f}/2\) then it
follows that for CI2 the value of \(\sigma \) is roughly 0.15. On the other
hand these authors have also noted that for barnase the rate limiting step
comes closer to the native state involving the rearrangement of the
hydrophobic core. They suggest in the form of a figure (see Fig. (3) of 
Ref. 52) that barnase follows a three stage kinetics 
with the rate determining
step being the final stage. In terms of the physical picture suggested here
this can be interpreted to mean that for barnase \(\Phi (T)\) is small making
it either a moderate folder or even a slow folder. If this were the case our
theoretical picture would suggest \(\sigma \) for barnase is bigger which 
is a
consequence of the fact that it is a larger protein. These observations are
consistent with the experimental conclusions of Otzen {\em et al.} 
which are
perhaps the first experiments that seem to provide some confirmation of the
theoretical ideas that have emerged from the minimal model studies. It is
clear that a more detailed comparison of the entire kinetics for various
proteins under differing experimental conditions is required to fully
validate the general conclusions based on the kinetic partitioning mechanism
together with the classification of sequences based on the values of the
parameter \(\sigma \). 

\acknowledgments
This work was supported by a grant from the Airforce Office of Scientific 
Research (through grant number F496209410106) and the National Science 
Foundation. One of us (DT) is grateful to 
P.G. Wolynes and J.N. Onuchic for a number of interesting discussions. 

\appendix
\section{}

It is clear from our studies for \(N = 15\), for which all conformations
(compact and noncompact) can be exactly enumerated that the neglect
of noncompact structures leads to qualitatively incorrect results for
thermodynamic and kinetic properties at all temperatures. However for \(N =
27\) the enumeration of all possible conformations is out of reach. Thus one
has to resort to numerical methods to obtain as accurate results as
possible. We have used a combination of slow cooling technique in
conjunction with Metropolis Monte Carlo method to calculate several
thermodynamic characteristics of the system. We have used similar
algorithms in our studies on continuum models of \(\beta \)-barrel
and four helix bundles \cite{Guo,Helix}. 
In this appendix we describe the 
algorithms that
have been used to obtain the collapse transition temperature \(T_{\theta}\)
and the folding transition temperature \(T_{f}\). 

For a given sequence we first estimate the approximate temperature range
which includes \(T_{\theta}\) and \(T_{f}\) (and also \(T_{s}\), at 
which \(<\chi (T_{s})>=0.21\)). To obtain
this temperature interval, \(T_{l} < T_{s} < T_{f} < T_{\theta} < 
T_{h}\), we quench the system starting
from a random unfolded conformation (from an arbitrary self-avoiding
conformation) to a high enough temperature \(T_{h}\). In the first phase of
the algorithm a few trajectories are generated so that approximate
estimates of \(T_{\theta}\) and \(T_{f}\) can be made. This 
allows us to pin
the interval (\(T_{l},T_{h}\)). This temperature interval spans the peaks of
\(C_{v}\) and \(\Delta \chi \) from which the characteristic temperatures are
obtained. Starting from a given initial conformation at \(T_{h}\) the
following linear cooling schedule was applied to lower the temperature
\begin{equation}
T_{i}=T_h-i \Delta T
\end{equation}
where \(i\) labels the \(i^{\text{th}}\) step in the slow cooling process
and \(\Delta T\) is the cooling rate. The initial
conformation for the \((i+1)^{\text{th}}\) iteration is the final 
conformation
at the end of the \(i^{\text{th}}\) step. At each step of the slow 
cooling process we run Monte Carlo simulations for the time \(\tau _{max} 
\). For the computation of
equilibrium quantities we reject the conformations for a certain length of
time \(\tau _{eq}\) immediately after quench to a new temperature. This
allows for the chain to equilibrate at that temperature. The precise value
of \(\tau _{eq}\) clearly is temperature dependent. In our simulations we
chose a large enough value of \(\tau _{eq}\) so even at the lowest
temperatures it is sufficient for equilibration. Thus at each temperature
the time averaged property of any quantity of interest, \(A\), can be
calculated as
\begin{equation}
A_{j}(T_{i})=\frac{1}{\tau _{max}-\tau _{eq}}\int_{\tau _{eq}}^{\tau _{max
}}A_{j}(s,T_{i})ds,
\end{equation}
where \(j\) labels the initial trajectory generated at \(T_{h}\). The true
thermodynamic value \(<~A(T)>\) is obtained by averaging over an ensemble
of initial conditions, i.e. 
\begin{equation}
<A(T_{i})>=\frac{1}{M}\sum_{j=1}^{M}A_{j}(T_{i})
\end{equation}
In using Eqs. (A2) and (A3) we have assumed that the weights associated
with the initial trajectories are uniform. This will only be valid provided
there is not significant amount of multivalley structure in the potential
energy surface. Since for \(N = 27\) we have used only optimized sequences
this assumption appears reasonable. Furthermore we have also checked that
the same values of the thermodynamic properties are obtained in the limit
of long times in the kinetic simulations as well. This again assures us
that the computations based on the above procedure yield accurate values
of all quantities of interest for the sequences examined. 
 
In our simulations we chose \(\tau _{eq} = 50,000\) MCS, \(\tau _{max} =
10^{6}\) MCS, and the cooling rate \(\Delta T = 0.01\). This value of
\(\Delta T\) was small enough so that the changes in the thermodynamic
quantities in between two successive temperatures were less than the
fluctuations at these temperatures. Specifically this choice ensures us
that \(<\chi (T_{i})>-<\chi (T_{i}-\Delta T)> \ll (\Delta \chi
(T_{i}))^{1/2}\). After the determination of \(<\chi (T_{i})>\), 
\(<E(T_{i})>\), 
\(C_{v}(T_{i})\), \(\Delta \chi (T_{i})\) for all \(i (=0,1,2...)\) using the
above protocol they were fitted to appropriate polynomials or in the case
of \(<\chi (T)>\) to sum of two hyperbolic tangents. These functions were
used to obtain \(T_{\theta}\) and \(T_{f}\) for fifteen sequences with
\(B_{0}=-0.1\) and for two sequences with \(B_{0}=-2.0\). The later
calculations were done merely to check the role of non-compact structures
in determining thermodynamic properties for \(N=27\) at a larger value of 
the overall hydrophobic interaction.

\section{}

In this appendix we describe the methods used to obtain the low energy
spectrum of the sequences with \(N = 27\). The sequences for this case
were all optimized using the procedure described in Sec. (II.E). For a
given primary sequence we first determined a set of low energy structures
using the algorithm described in Appendix A. The lowest of these energy
structure was then used as the target structure and an optimized sequence
that is fitted to this target structure was generated. Typically the
target structures contained between 22 to 28 (the maximum being 28)
topological contacts.  This procedure was repeated with another primary
sequence until all the sequences are generated. 

The lowest energy obtained in the design procedure was assumed to be the 
ground state of
the designed sequence. We did not, during the course of our simulations,
find any counter examples to this. Thus the native energy is determined
in the process of sequence design itself. In order to obtain the other
energy levels we used a slow cooling Monte Carlo simulation starting from
a high temperature \(T_{h}\). The details of this protocol are given in
Appendix A. During the course of this slow cooling simulations the energy
\(E(t, T_{i})\) is monitored for all values of \(t\) and \(T\). These 
energies were
then arranged in increasing order thus yielding the spectrum for one given
sequence. In order to ensure that the energies are truly equilibrium
spectrum this procedure was repeated for several distinct initial
conditions (namely, for \(M=100\)). The simulation time for each sequence
ranged from \(5 \times 10^{7}\) to \(6 \times 10^{7}\) MCS. 

The procedure described here works best for the determination of the gap
\( \Delta \), when the temperature \(T_{l}\) is low enough so that one
essentially has a two level system. Since the sequences are all optimized
most of the values of \( \Delta \) are reasonably large that this never is
the problem. There were few sequences with relatively small values of \(
\Delta \). In these cases we also performed slow warming simulations
starting from \(T_{l}\). Once again the energy as a function of time was
recorded and we obtained exactly the same results for \( \Delta \). Thus
it is extremely unlikely that there are any significant errors at least in
the determination of \( \Delta \).

\begin{figure}

Fig. 1.  Native structures for two sequences with \(N = 15, B_{0}=-0.1\). 
These structures are the lowest energy structures obtained by full 
enumeration of all the possible conformations. (a) This corresponds to a 
sequence for which the lowest energy structure displayed is fully 
compact. It contains 11 topological contacts which are ternary 
interactions between between two beads that are separated by at least two 
other beads along the backbone. (b) An example of the lowest energy 
non-compact structure containing only 10 topological contacts. In both 
cases the native state is non-degenerate. Such contacts between 
non-bonded residues are shown as dashed lines.

Fig. 2. Representation of the native structures for two sequences with 
\(N = 27, B_{0}=-0.1\). The native structures were determined by a 
combination of slow cooling procedure and standard 
Monte Carlo algorithm. The details are given in Appendix A. (a) For this 
sequence the native structure is fully compact and contains 28 topological 
contacts. (b) An example of the native structure that is non-compact 
containing 22 topological contacts. Such contacts between non-bonded 
residues are shown as dashed lines.

Fig. 3. Lower part of the energy spectrum for ten sequences. The value of 
\(N= 15\) and \(B_{0}=-0.1\). The spectra were obtained by explicitly 
enumerating all the possible conformations of the chain for each 
sequence. The value of the energy gap separating the two lowest energy 
levels and the number of the sequence in the database are given below 
each spectrum.

Fig. 4. Lower part of energy spectra for 10 optimized sequences (\(N =27,
B_{0}=-0.1\)) labeled 61 through 70. The left columns of each sequence
are the energy levels obtained from slow cooling Monte Carlo
simulations. The columns on the right represent the energy spectra of
compact
structures. Below each pair of spectrum columns the energy gap \(\Delta \)
calculated from Monte Carlo simulations (upper quantity)  and \(\Delta _{CS}\) calculated
from the spectra of compact structures (lower quantity) are indicated. It is seen that for
several sequences even the lowest energy in the spectrum of compact
structures belongs to continuous part of the "true" spectrum. These
spectra clearly show that in general non-compact structures are crucial 
for the thermodynamics (and kinetics). It is also obvious that \(\Delta  
_{CS} \geq \Delta \) for the sequences with the native conformation being 
compact structure. This should be true 
for such sequences in this model for all \(N\) and all negative values of
\(B_{0}\).

Fig. 5. The set of moves used in the Monte Carlo simulations: (i) Corner 
flip; (ii) Crankshaft rotation; (iii) Rotation of end residue.

Fig. 6. Temperature dependence of thermodynamic quantities for sequence 14
(\(N = 15, B_{0}=-0.1\)) calculated using full enumeration of all
conformations: (a) overlap function \(<\chi >(T)\); (b) fluctuations in
overlap function \(\Delta \chi (T)\); (c) function \(<Q>(T)\);  (d)
specific heat \(C_{v}\). The peaks in the graphs of \(\Delta \chi \) and
\(C_{v}\) correspond to the folding transition and collapse temperatures,
\(T_{f}\), \(T_{\theta }\), respectively. 

Fig. 7. (a) Dependence of the simulation temperature \(T_{s}\) determined
using Eq. (11) on the parameter \(\sigma = (T_{\theta} -
T_{f})/T_{\theta}\). \(T_{\theta}\) and \(T_{f}\) are calculated from the
temperature dependence of \(C_{v}\) and \(\Delta\chi (T)\), respectively. 
(b) Variation of \(T_{s}\) with the energy gap \(\Delta \) between the
ground state and the first excited state for 32 sequences. The energy gap
is determined from full enumeration of all possible conformations and is
hence exact.  The data are for (\(N =15, B_{0}=-0.1\)).

Fig. 8. (a) The dependence of \(\sigma (= (T_{\theta} - 
T_{f})/T_{\theta}\)) on the energy gap \(\Delta \) for various sequences 
with (\(N =15, B_{0}=-0.1\)). This figure shows that there is no trend 
between \(\sigma \) and \(\Delta \) because the determination of \(\sigma 
\) requires the knowledge of the entire energy spectrum. (b) Plot of 
\(\sigma \) versus \(\Delta \) expressed in units of \(T_{s}\) - the 
simulation temperature.  This was done because for any useful purposes 
\(\Delta \) has to be measured in dimensionless form. The relevant energy 
is  \(k_{B}T_{s}\). The parameter \(B\) (see Eq. (2)) sets the energy 
scale in 
terms of which everything is measured. This figure shows even more 
dramatically the lack of correlation between \(\sigma \) and \(\Delta \).

Fig. 9. The upper panel (a) shows the overlap function \(<\chi >\) (Eq. 
(5)) as a function of temperature for the sequence labeled 10 with \(N =15,
B_{0}=-0.1\). The solid line is obtained by calculating \(<\chi >\) using 
the enumeration of all conformations and the dotted line is the plot of 
\(<\chi >\) calculated using only the set of compact structures. The 
dashed line is given by the constant value of \(<\chi >=0.21\). The 
intersection of the solid line with the dashed line determines the 
simulation temperature, \(T_{s}\). The constant value of \(<\chi >=0.21\) 
is chosen so that \(T_{s} < T_{f}\) for all sequences. The lower panel 
displays the temperature dependence of \(\Delta \chi \). As in Fig. (9a) 
the solid line is the exact result obtained by full enumeration while the 
dotted line is the corresponding result for \(\Delta \chi \) using 
compact structures only. Note that the folding temperature \(T_{f}\)  
obtained from the peak of the solid line is considerably lower than  the 
peak in the dotted line. For all the sequences examined \(T_{f}\) is also 
less than the midpoint of \(<\chi >\) or \(<Q>\) (see Fig. (6c)). 

Fig. 10. Same as Fig. (9) except the curves are for a sequence labeled 91 
which has \(N =15, B_{0}=-2.0\). In this case Fig. (10b) shows that the 
difference in \(T_{f}\) (i.e., between the peaks in  the solid curve (full 
enumeration) and 
the dotted curve (compact structures only)) is less than for the case with 
\(B_{0}=-0.1\). 

Fig. 11. Dependence of the folding time \(\tau _{f}\) on \(\sigma = 
(T_{\theta} - T_{f})/T_{\theta}\) for a number of sequences all with \(N 
= 15\). The folding time for each sequence is obtained by fitting \(<\chi 
(t)>\) to an appropriate number of exponential functions as described in 
Sec. (II.G). The full circles correspond to \(B_{0}=-0.1\) while the 
empty circles are for \(B_{0}=-2.0\). There are 32 sequences with 
\(B_{0}=-0.1\) and 9 sequences with \(B_{0}=-2.0\). The trend from both 
these parameter values is clear. The smaller values of \(\sigma \) have 
smaller folding times. The folding times decreases by nearly five orders 
of magnitude upon reducing \(\sigma \) from \(0.8\) to around \(0.1\). 
For the same range of \(\sigma \)  the folding time is considerably 
greater for \(B_{0}=-2.0\) than for \(B_{0}=-0.1\). 

Fig. 12. Dependence of the folding time \(\tau _{f}\) on the gap 
\(\Delta \) which is the difference in energy between the native 
conformation and the first excited state. The database of sequences is 
\underline{exactly} the same as that used in Fig. (11). As in Fig. (11) the 
full circles are for  \(B_{0}=-0.1\) and the empty circles corresponds to 
\(B_{0}=-2.0\). This figure shows that there is no useful correlation 
between \(\tau _{f}\) and \(\Delta \). For a given folding time \(\tau 
_{f}\) one can engineer sequences spanning a wide range of gap (both 
quite small and large). 

Fig. 13. The kinetic probabilities \(P_{r}(E_{k})\) (cf. Eq. (16)) 
(shown as empty bars) corresponding to the largest probabilities that a 
state with the energy \(E_{k}\) occurs before the native conformation is 
reached. The sequences displayed are moderate folders (\(0.1 \lesssim \sigma 
\lesssim 0.6\)) and slow folders (\(\sigma \gtrsim 0.6\)). All the sequences 
correspond to  \(N =15, B_{0}=-0.1\). The sequences are arranged in  
increasing order of folding time. For most of the sequences with 
relatively large folding times there is a finite probability of 
visiting an intermediate. The fluctuation probabilities \(P_{fl}(E_{k})\) 
(Eq. (17)), which probe the probability that the same kinetic 
intermediate is visited after the chain reaches the native conformation, 
are shown under hatched bars. These probabilities show that most slower 
folding sequences rarely visit the kinetic intermediate after initially 
reaching the native conformation at least on the maximum time scale of 
the simulations. 

Fig. 14. Plots of various properties of the kinetic intermediates for 
the 32 sequences with \(N =15, B_{0}=-0.1\). The sequences are arranged 
in order of increasing folding time. (a) Fraction of native contacts 
in the intermediate. The sequences which fold fast have large fraction of 
native contacts in the intermediate (\(\gtrsim 0.7\)) while slow folding 
sequences have typically less than 50\% of native contacts. These are 
misfolded structures which act as traps thus slowing down the folding 
process. (b) The value of the overlap function \(\chi _{k}\) between the 
kinetic intermediate and the native conformation for the sequences 
arranged in order of increasing folding time. Since the structural 
overlap function is a more microscopic order parameter than \(Q\) this 
plot is 
more informative than Fig. (14a). The trends basically are the same. 
Intermediates for fast folding sequences have the smallest value of 
\(\chi _{k}\) (are most native-like) and slow folding sequences have 
large value of \(\chi _{k}\). (c) Plot of the ratio of the average first 
passage time to reach the kinetic intermediate \(\tau _{k}\) to the mean 
first passage time \(\tau _{fp}\) to reach the native conformation. The 
mean first passage time  \(\tau _{fp}\) is obtained from the fraction of 
unfolded molecules \(P_{u}(t)\) as \(\tau _{fp} = \int_{0}^{\infty} 
P_{u}(t)dt\). The figure clearly shows that for relatively fast folding 
sequences the kinetic intermediate which is very native-like is reached 
on time scales comparable to \(\tau _{fp}\). For slow folding sequences 
the kinetic intermediate, which does not share many features in common 
with the native conformation, is reached very rapidly. For these 
sequences the rate determining step is the transition from these 
structures to the native state. 

Fig. 15. (a) The temperature dependence of the structural overlap function 
for a sequence labeled 61 with \(N =27,B_{0}=-0.1\). The results obtained 
by using the slow cooling Monte Carlo simulations (see Appendix A for 
details) are shown in squares. The solid line through these squares 
represents the fit of the simulation data by hyperbolic tangents. The 
dotted line represents exact calculation of the temperature dependence of 
\(<\chi >\) using only the ensemble of compact structures. The simulation 
temperature is determined from the intersection of the simulation results 
with the dashed line with \(<\chi >=0.21\). (b) The fluctuations in the 
overlap function \(\Delta \chi \) as a function of temperature. The 
squares represent the results from the slow cooling Monte Carlo 
simulations whereas the dotted line shows the corresponding results 
obtained using the contributions from the compact structures only. The 
solid line is a polynomial fit through the simulation data. 

Fig. 16. Same as Fig. (15) except it is for a sequence labeled 81 which 
has \(N =27, B_{0}=-2.0\). In this case the compact structures are 
expected to dominate. Although the results here for the both the 
temperature dependence of \(<\chi >\) and \(\Delta \chi \) are closer to 
the simulation data than for sequence 61 (Fig. (15)) there are 
significant differences in the estimation of \(T_{s}\) and \(T_{f}\) between 
the simulation results and those obtained from compact structures 
enumeration. This trend is seen for all sequences. 

Fig. 17. (a) Ratio of simulation temperatures obtained using compact
structures only \(T_{s}^{CSE}\) in conjunction of Eq. (11) to that
obtained for the entire ensemble of conformations \(T_{s}=T_{s}^{FE}\)
for \(N =15, B_{0}=-0.1\) for a variety of sequences. (b) Same as Fig.
(17a) except that the simulation temperature \(T_{s}=T_{s}^{MC}\) is
obtained using slow cooling Monte Carlo simulations as described in
Appendix A. These figures show that \(T_{s}^{CSE}\) is always greater  
than \(T_{s}\) and often is much greater than \(T_{s}\). We have excluded
those sequences for which the native conformation is non-compact. The   
sequences are arranged in order of increasing folding time.

Fig. 18. (a) The time dependence of the overlap function \(<\chi (t)>\) 
for the sequences labeled 69 with \(N =27, B_{0}=-0.1\). The simulation 
temperature is \(T_{s}=1.07\) and the value of \(\sigma \approx 0\). The 
overlap function has been calculated using Eq. (12) by averaging over 200 
initial independent conditions. The solid line is a single exponential 
fit to the simulation data. In this case the folding appears to be an 
all-or-none process. (b) Same as (a) except the plot shows \(<\chi (t)>\)   
for the sequences labeled 65. The simulation
temperature is \(T_{s}=0.77\) and the value of \(\sigma = 0.075\). The 
solid line is a biexponential fit to the simulation data. In this case 
folding appears to be a three stage multipathway process.

Fig. 19. The variation of the folding time \(\tau _{f}\) with \(\sigma 
= (T_{\theta} - T_{f})/T_{\theta}\) for \(N =27, B_{0}=-0.1\). There are 
fifteen sequences for which \(\tau _{f}\) has been calculated using the 
procedure described in Sec. (II.G). The basic trend here is the same as 
for \(N=15\) (see Fig. (11)). There is a dramatic increase in folding 
time as \(\sigma \) changes from \(0.02\) to \(0.1\). Since all the 
sequences used were optimized the range of \(\sigma \) examined here is 
considerably less than for \(N=15\). 

Fig. 20. Dependence of the folding time \(\tau _{f}\) on the gap 
\(\Delta \) for the same set of sequences as in Fig. (19). The gap 
\(\Delta \), which is the energy difference between the native 
conformation and the first excited state, was calculated using the 
procedure given in Appendix B. It is clear that as for \(N=15\) (see Fig. 
(12)) there is no useful correlation between \(\tau _{f}\) and \(\Delta 
\). 

Fig. 21. Plots of the folding time \(\tau _{f}\) as a function of the  
energy gap \(\Delta \) divided by the simulation temperature \(T_{s}\).
(a) The solid circles correspond to  \(B_{0}=-0.1\) whereas the open
circles are for  \(B_{0}=-2.0\). The value of \(N\) is \(15\). (b) This   
plot is for \(N =27, B_{0}=-0.1\). These figures reaffirm even more
emphatically the lack of any relationship between  \(\tau _{f}\) and the
energy gap expressed in suitable dimensionless units.

Fig. 22. Dependence of the folding time \(\tau _{f}\) on the energy gap
\(\Delta _{CS}\). (a) The solid circles correspond to \(B_{0}=-0.1\)
and open circles are for \(B_{0}=-2.0\). The value of \(N=15\). The
database of sequences is the same as in Fig. (11). (b) This is for
\(N=27\) and \(B_{0}=-0.1\). The
database of sequences is the same as in Fig. (19). 
These figures show that the folding times
do not correlate with the energy gap even when restricted to the
ensemble of compact structures.

Fig. 23. Temperature dependence of \(\eta \) (cf. Eq. (22)) which gives 
the probability that the chain is not in the native conformation. Notice 
that the temperature is measured in units of \(T/T_{s\eta }\). For 
sequences with a large enough gaps (i.e. \(\Delta /k_{B}T_{s}\gg 1 \)) 
\(\eta (T)\) is only a function of \(T/T_{s\eta }\) (see Eq. (24)). This 
is verified for the sequences labeled 10 and 62 for \(T/T_{s\eta } \lesssim 
1.5\). The temperature \(T_{s\eta }\) is determined from the 
intersection of \(\eta (T)\) with the straight dashed line giving the 
constant value of \(\eta =0.1\). Thus, at these temperatures the 
probability of being in the native conformation is \(0.9\). 

Fig. 24. (a) Time dependence of the overlap function \(<\chi (t)>\) for 
the sequence labeled 62 with \(N =15, B_{0}=-0.1\). The simulation 
temperature \(T_{s\eta }=0.47\) was determined using Eq. (22). The 
equilibrium value of \(<\chi >\) from exact enumeration is \(0.028\) at 
\(T=T_{s\eta }\). At this temperature, at which the probability of the 
chain being in the native state is \(0.9\), \(<\chi (t)>\) shows a 
biexponential behavior. This fast folding sequence (\(\sigma \approx 0.0\)) 
exhibits an exponential time dependence at \(T=T_{s}=0.76\). (b) Plot of 
the fraction of unfolded molecules \(P_{u}(t)\) (see Eq. (13)) as a 
function of the number of Monte Carlo steps. The solid line is a 
biexponential fit to the simulation data. At this low temperature a 
fraction of molecules get trapped in the kinetic intermediate and the 
transition from this intermediate to the native conformation represents 
the slow step in the folding process. 

Fig. 25. (a) Same as Fig. (24a) except it is for the sequence labeled 
10, whose \(\sigma \) is \(0.19\). This is a moderate folding sequence in 
our computational scheme. The dashed line corresponds to the equilibrium 
value of \(<\chi >\) at \(T=T_{s\eta }=0.25\). On the time scale of the 
simulation (\(3\times 10^{7}\) MCS ) the equilibrium
value is not reached. (b) Time dependence of the fraction of unfolded 
molecules \(P_{u}(t)\). It is clear that \(P_{u}(t)\) has not decayed 
significantly on the time scale of the simulations. The simulation 
temperature \(T_{s\eta }\) is so low that only 24\% of the initial 
population of molecules has reached the native state on the time scale of 
\(3\times 10^{7}\) MCS.  

Fig. 26. Probability that the native conformation is populated 
\(P(E_{0})\) for 32 sequences with \(N =15, B_{0}=-0.1\) at the 
simulation temperatures \(T_{s}\) (cf. Eq. (11)). The sequences are 
arranged in order of increasing folding time. The probability ranges from 
about \(0.2\) to \(0.6\). It is clear that some of the slow folding 
sequences make less frequent transitions to other states from the native 
state whereas the fast folding sequences often visit native-like 
conformations.

\end{figure}

\newpage
\begin{center}
\begin{minipage}{15cm} 
\[
\psfig{figure=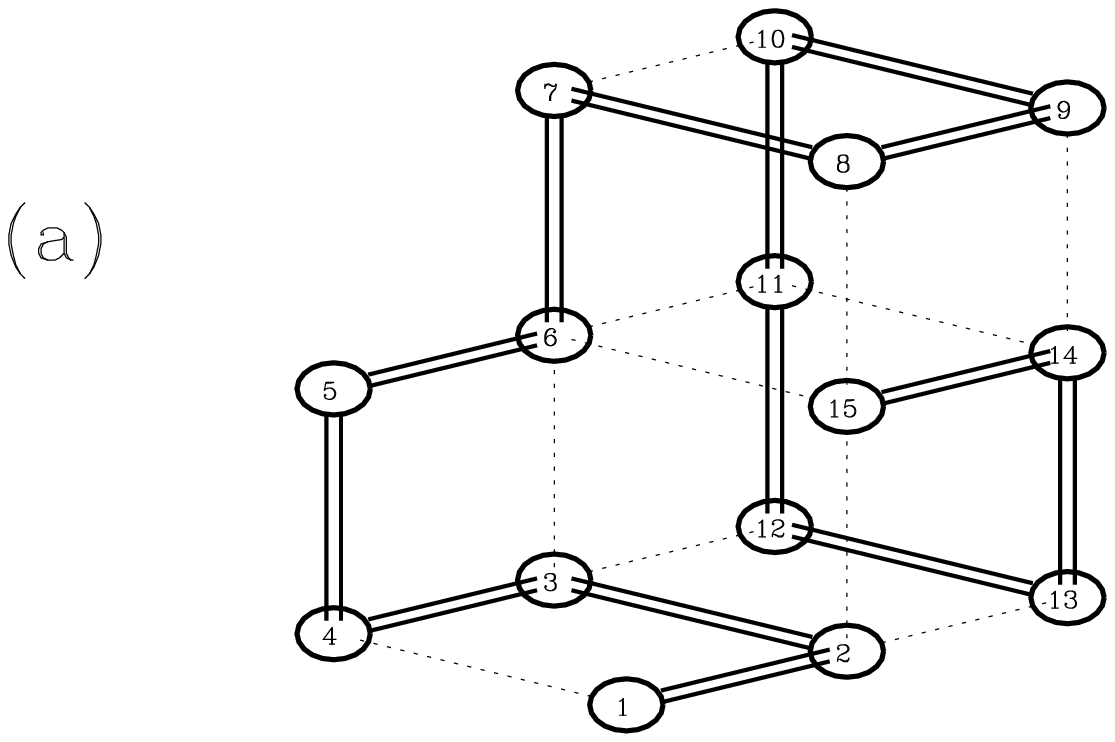,height=10cm,width=14cm}
\]
\end{minipage}   
\end{center}

\begin{center}
\begin{minipage}{15cm}
\[
\psfig{figure=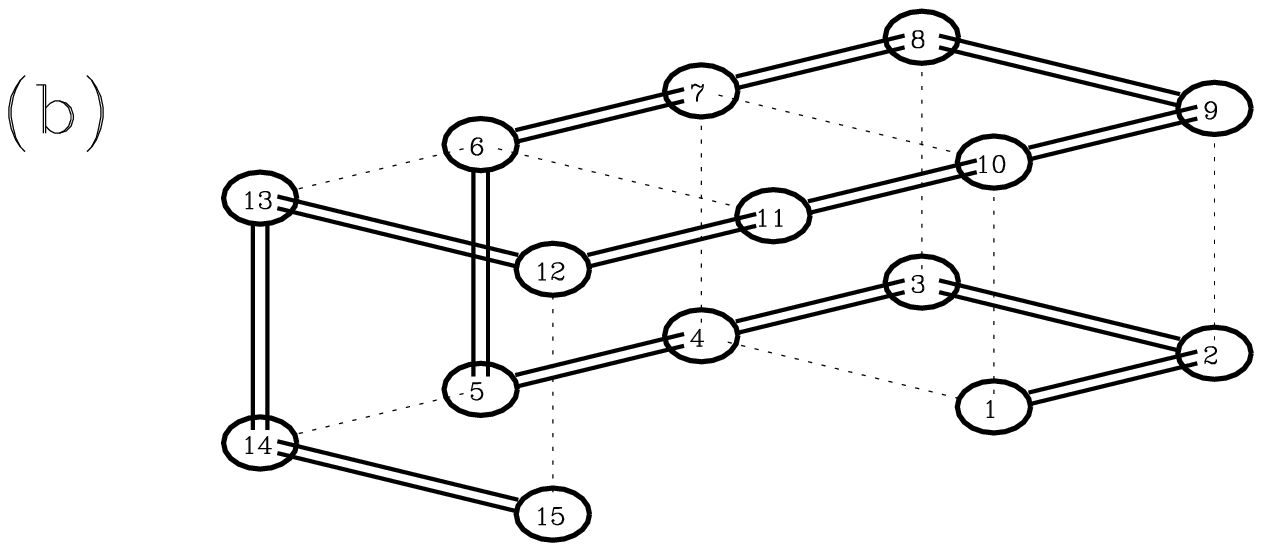,height=10cm,width=14cm}
\]
\end{minipage}

{\bf \large Fig. 1(a,b)}\\
\end{center}

\newpage

\begin{center}
\begin{minipage}{15cm}
\[
\psfig{figure=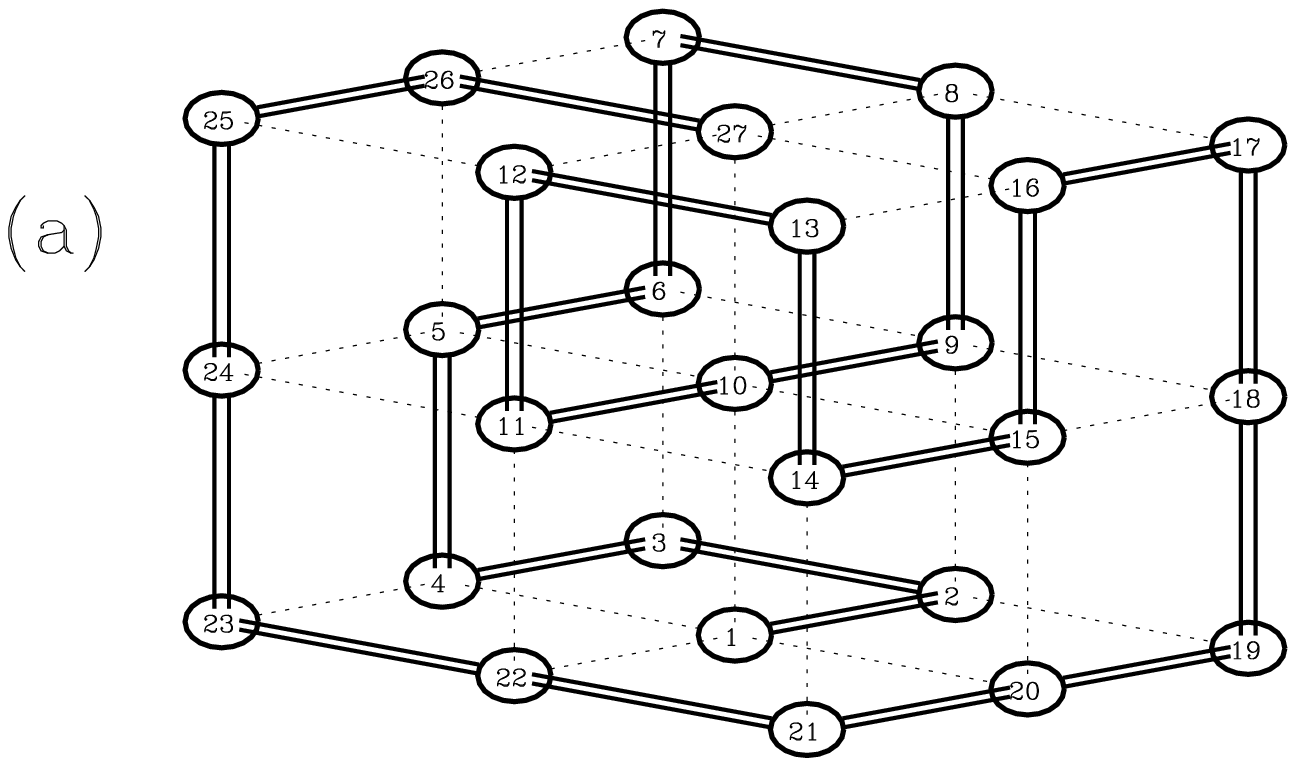,height=10cm,width=14cm}
\]
\end{minipage}
\end{center}

\begin{center}
\begin{minipage}{15cm}
\[
\psfig{figure=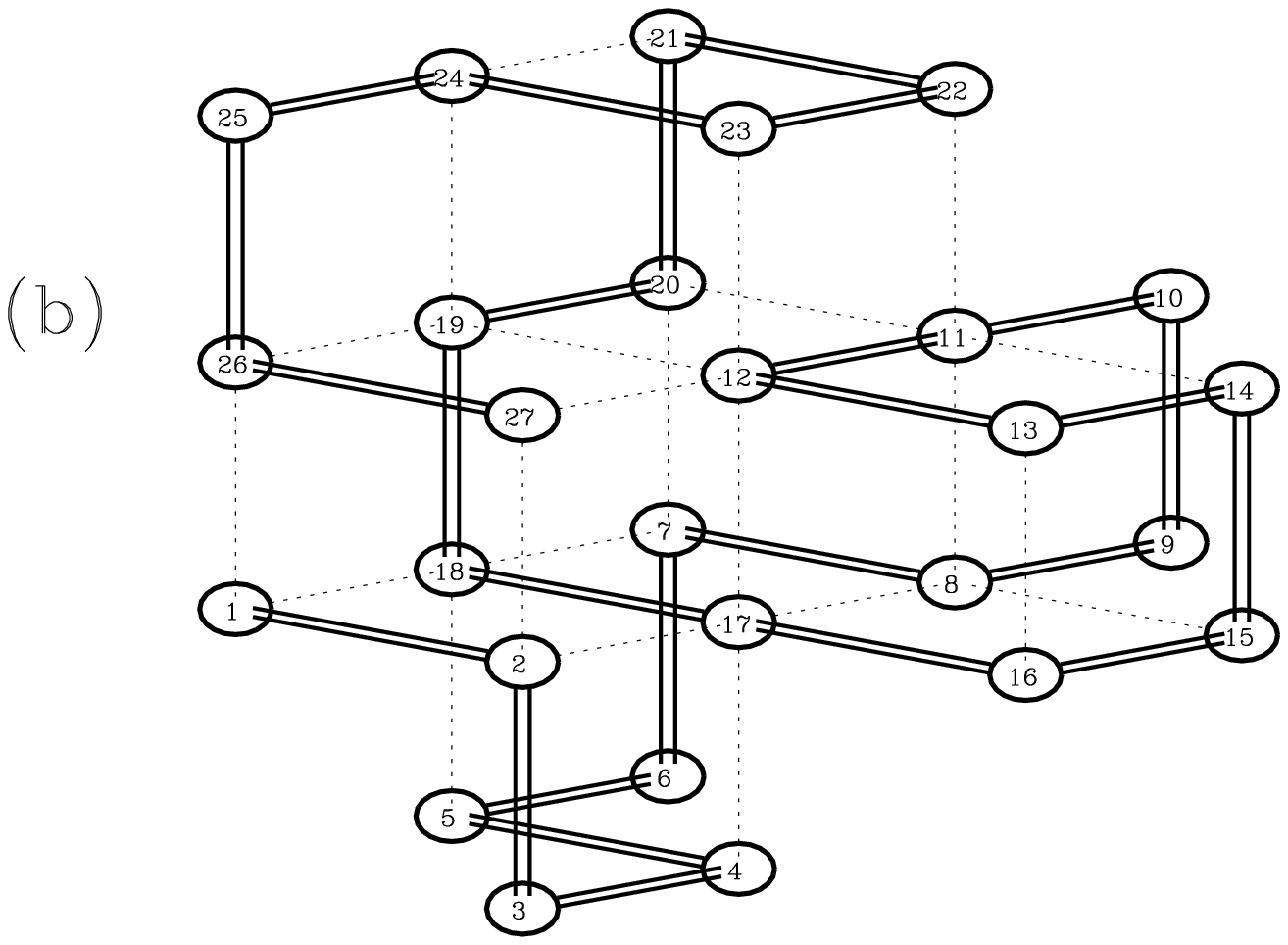,height=10cm,width=14cm}
\]
\end{minipage}

{\bf \large Fig. 2(a,b)}\\
\end{center}

\newpage

\hspace{10cm}
  
\begin{center}
\begin{minipage}{18.5cm}
\[
\psfig{figure=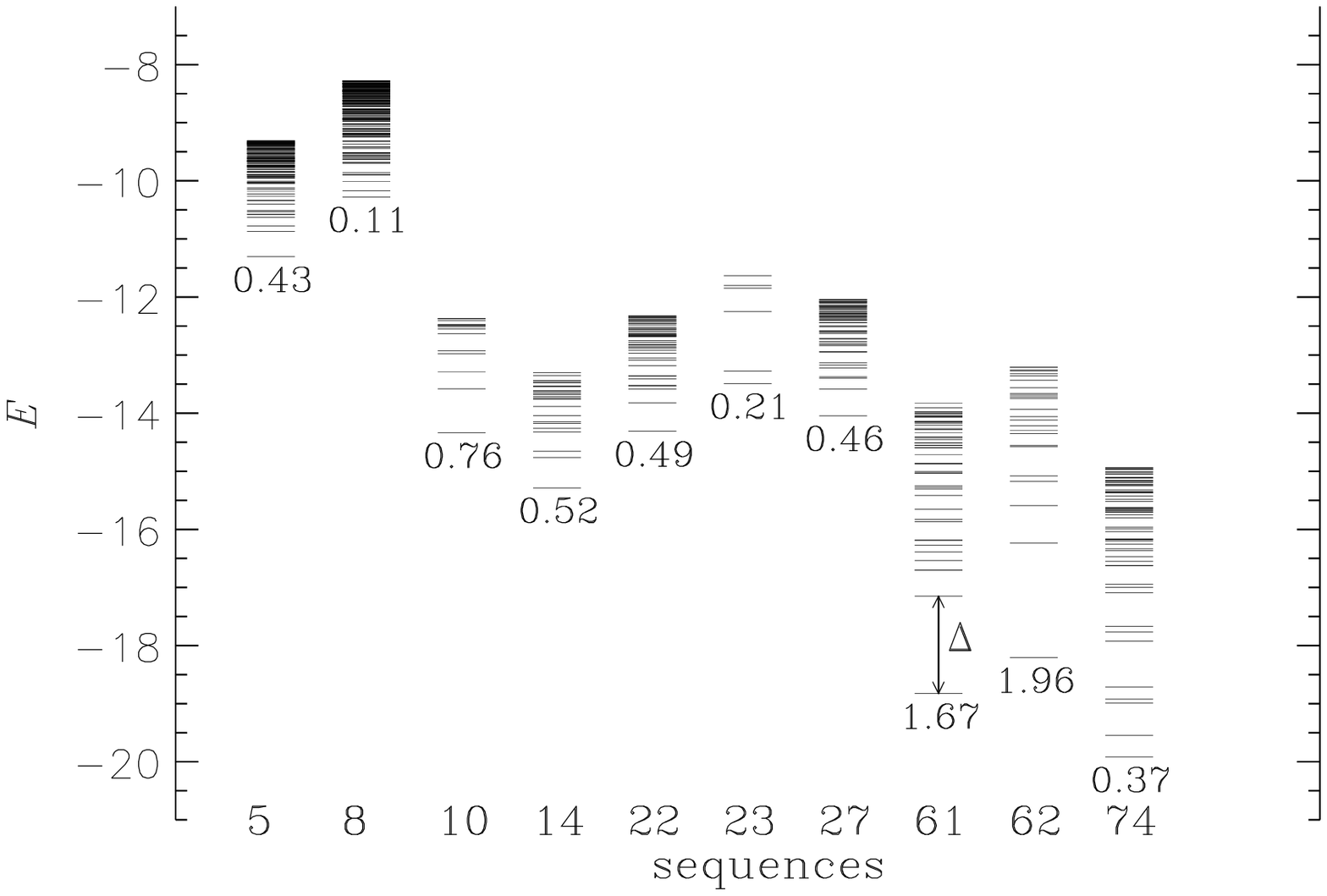,height=13.5cm,width=18.5cm}
\]
\end{minipage}

{\bf \large Fig. 3}\\
\end{center}

\newpage
  
\hspace{10cm} 

\begin{center}
\begin{minipage}{18.5cm}
\[
\psfig{figure=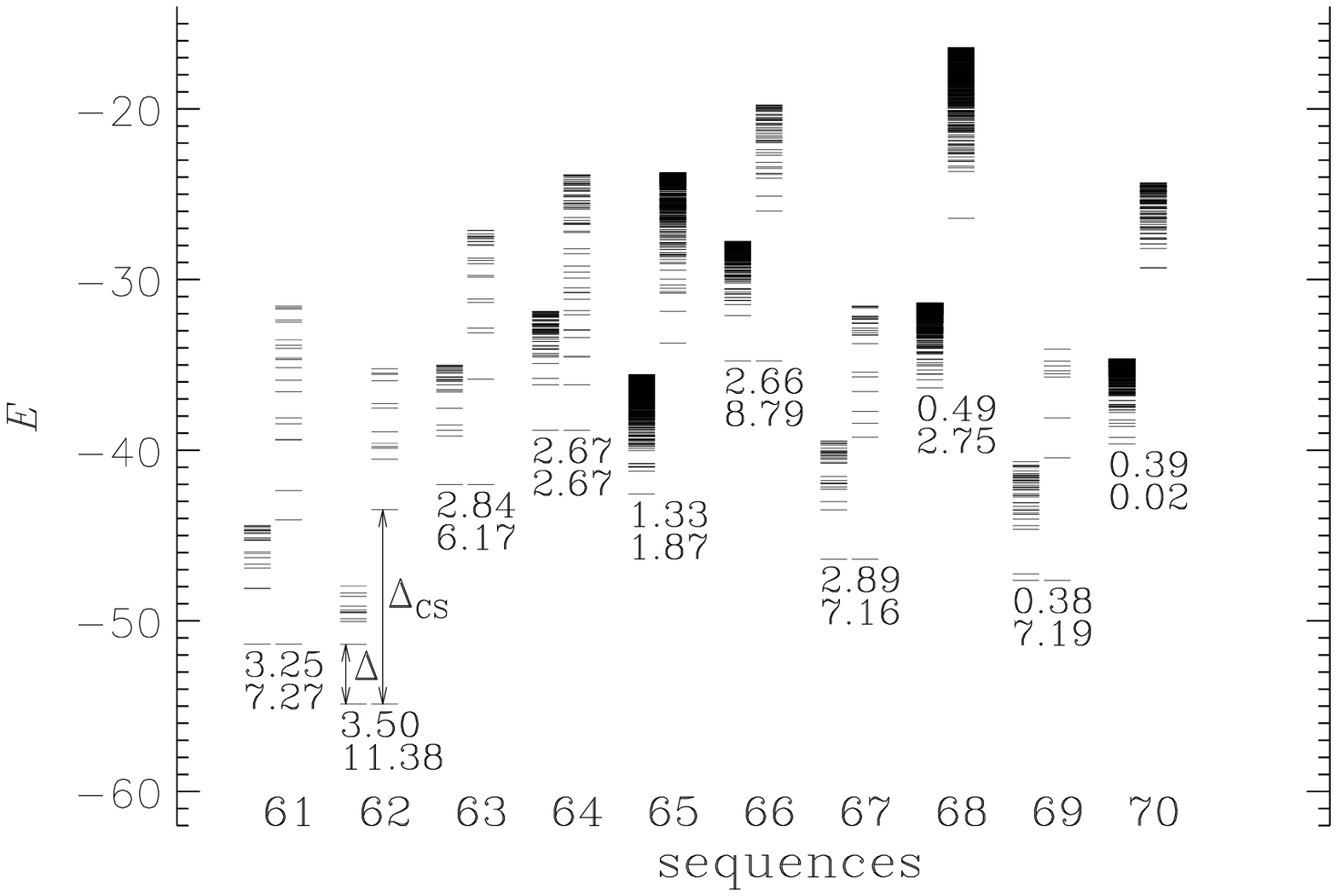,height=13.5cm,width=18.5cm}
\]
\end{minipage}

{\bf \large Fig. 4}\\
\end{center}

\newpage

\hspace{10cm}

\begin{center}
\begin{minipage}{18.5cm}
\[
\psfig{figure=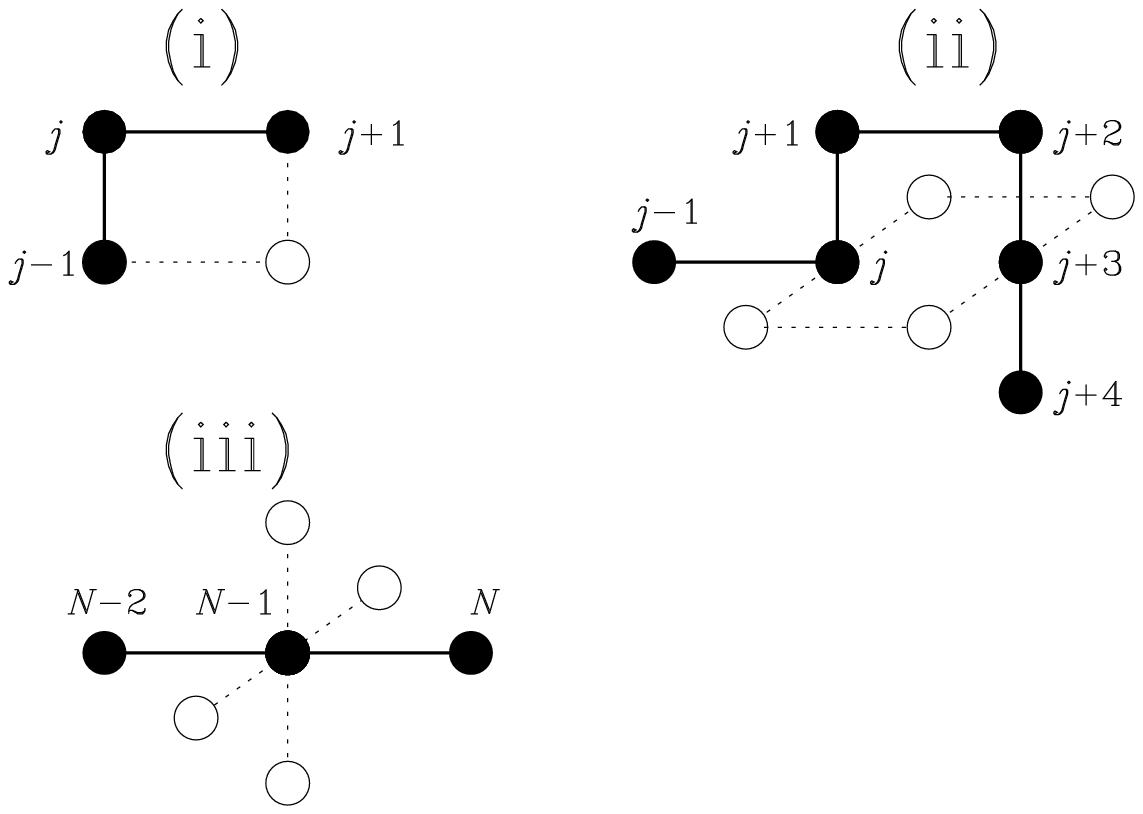,height=13.5cm,width=18.5cm}
\]
\end{minipage}

{\bf \large Fig. 5}\\
\end{center}

\newpage

\begin{center}
\begin{minipage}{15cm}
\[
\psfig{figure=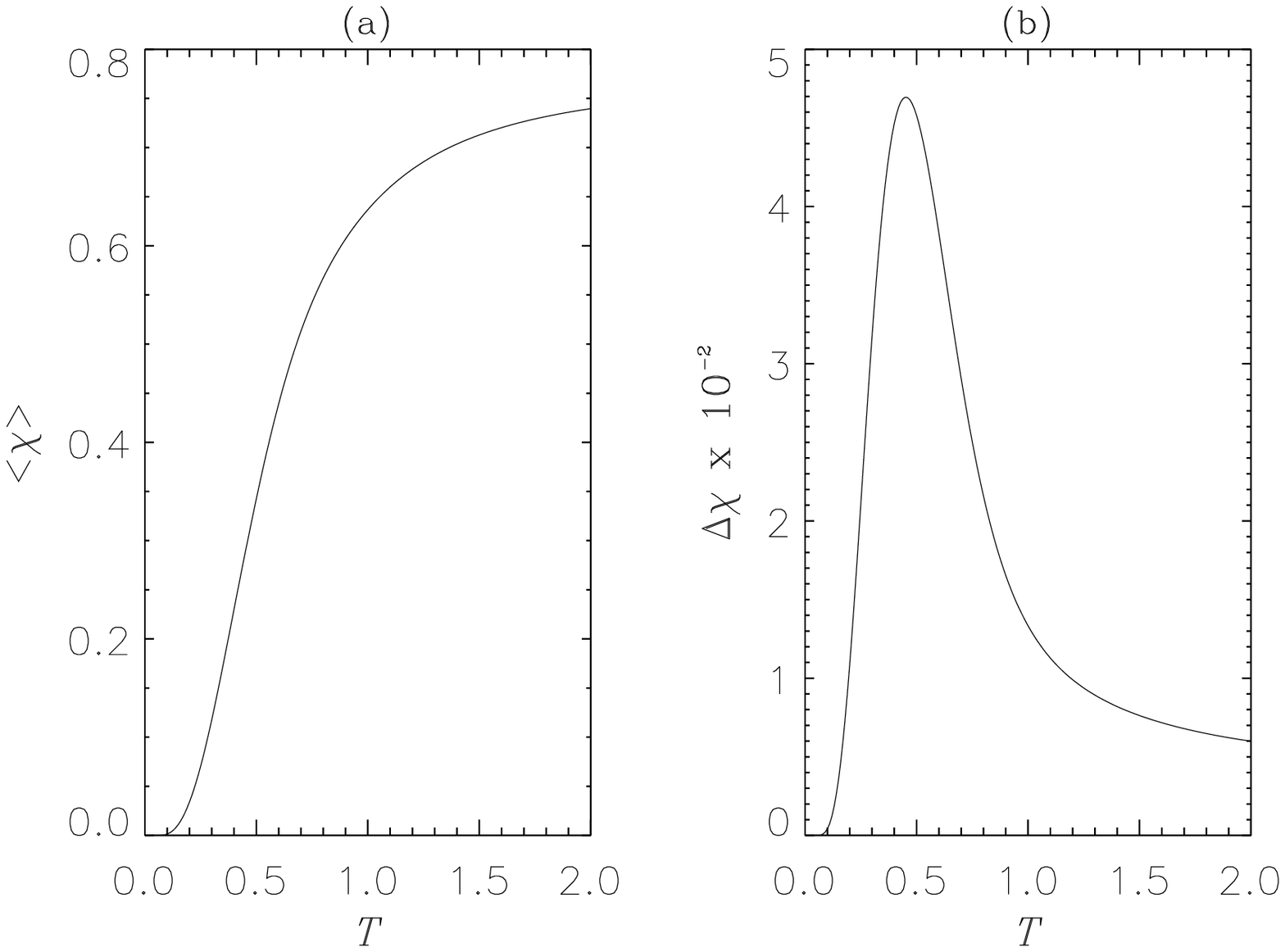,height=10cm,width=14cm}
\]
\end{minipage}
\end{center}

\begin{center}
\begin{minipage}{15cm}
\[
\psfig{figure=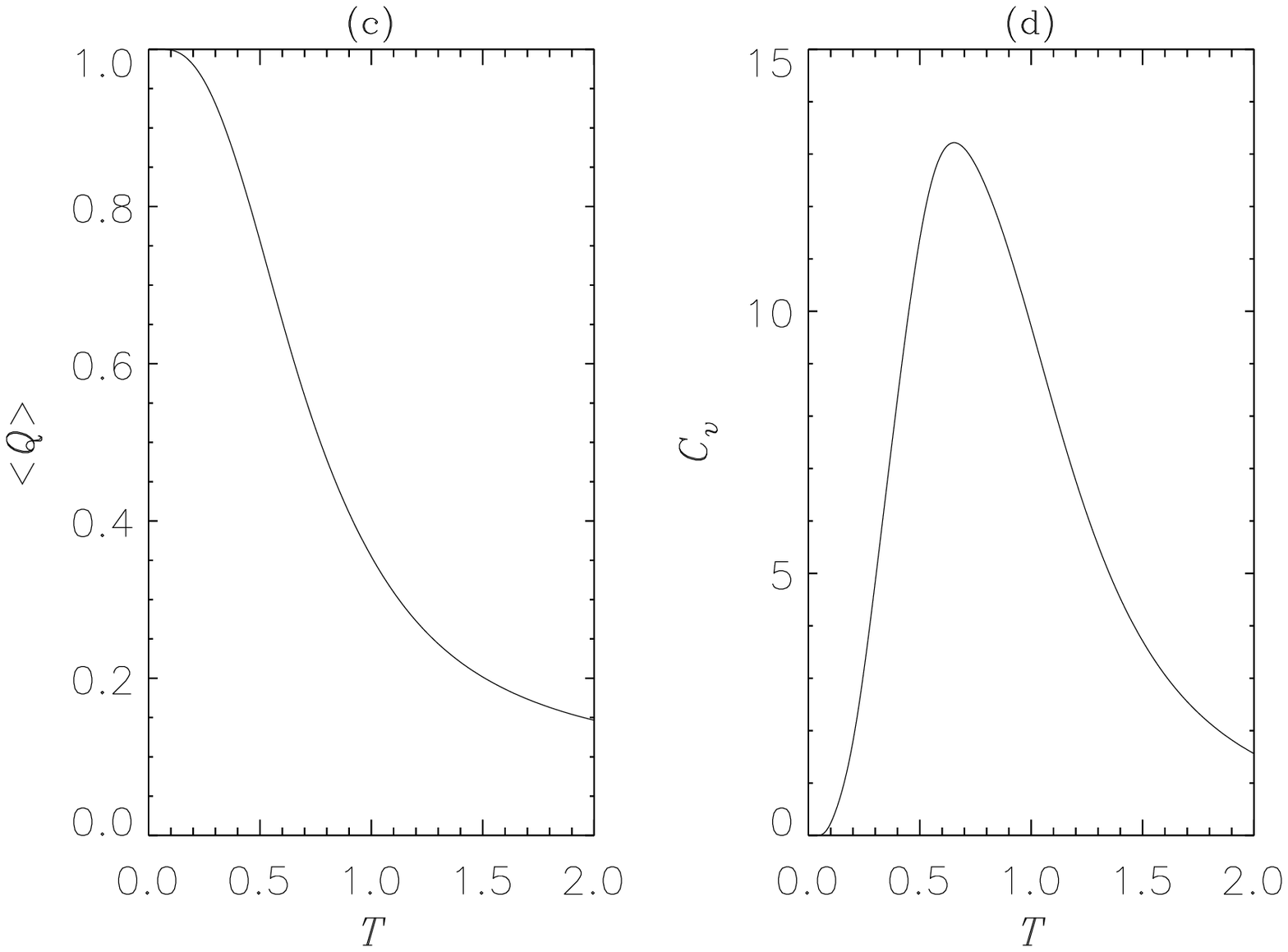,height=10cm,width=14cm}
\]
\end{minipage}

{\bf \large Fig. 6}\\
\end{center}

\newpage

\begin{center}
\begin{minipage}{15cm}
\[
\psfig{figure=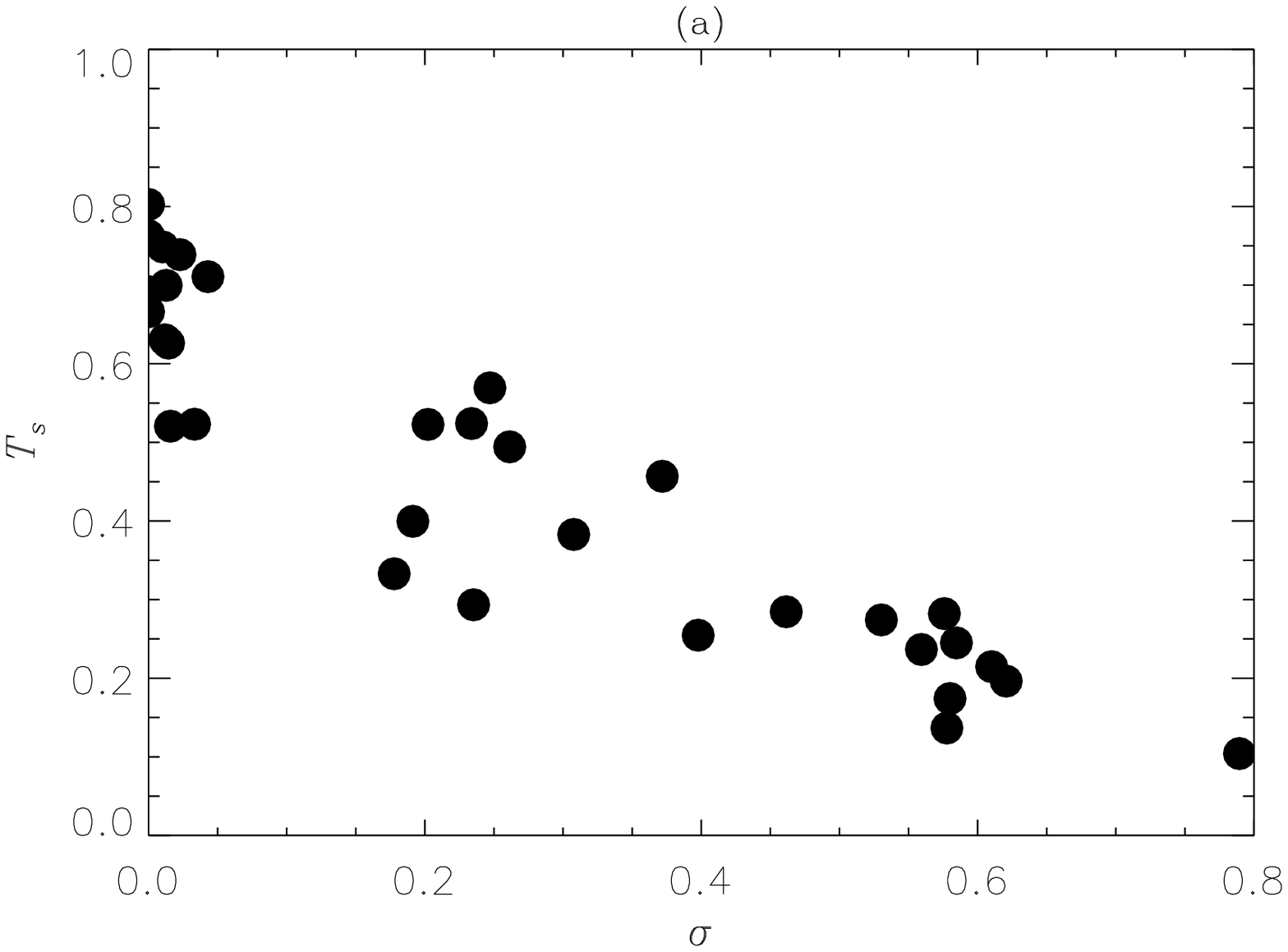,height=10cm,width=14cm}
\]
\end{minipage}
\end{center}

\begin{center}
\begin{minipage}{15cm}
\[
\psfig{figure=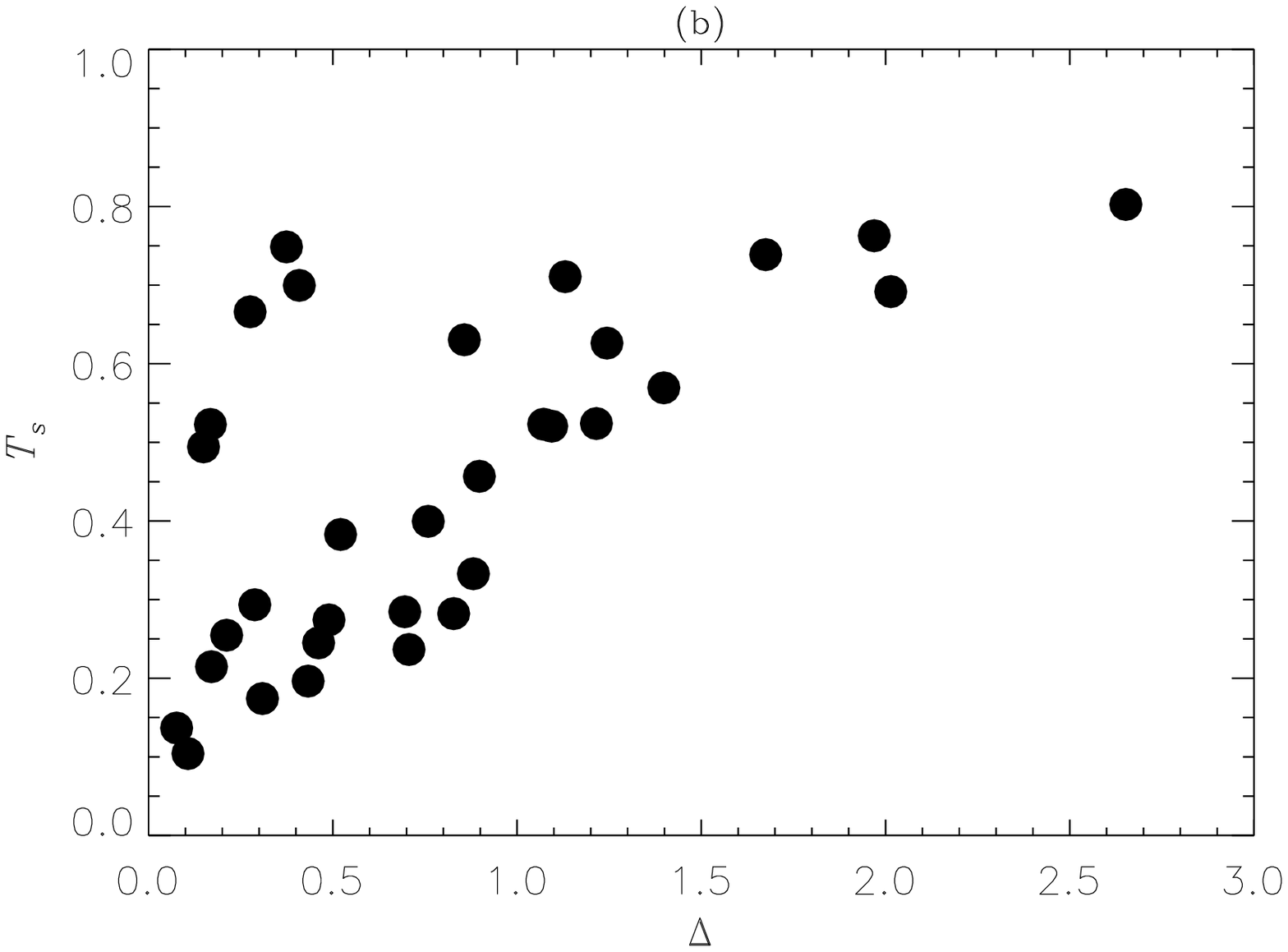,height=10cm,width=14cm}
\]
\end{minipage}

{\bf \large Fig. 7}\\
\end{center}  

\newpage

\begin{center}
\begin{minipage}{15cm}
\[
\psfig{figure=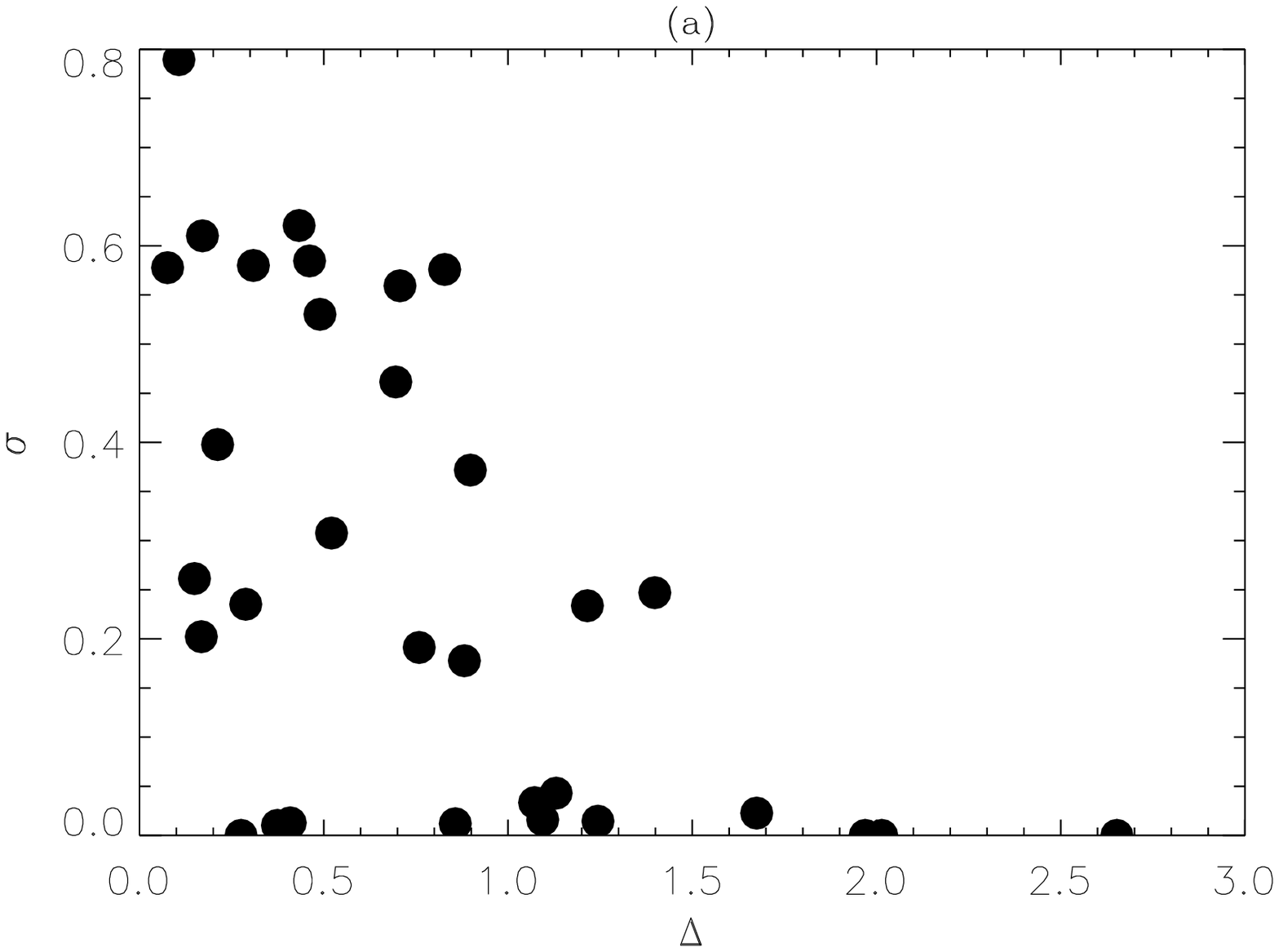,height=10cm,width=14cm}
\]
\end{minipage}
\end{center}

\begin{center}
\begin{minipage}{15cm}
\[
\psfig{figure=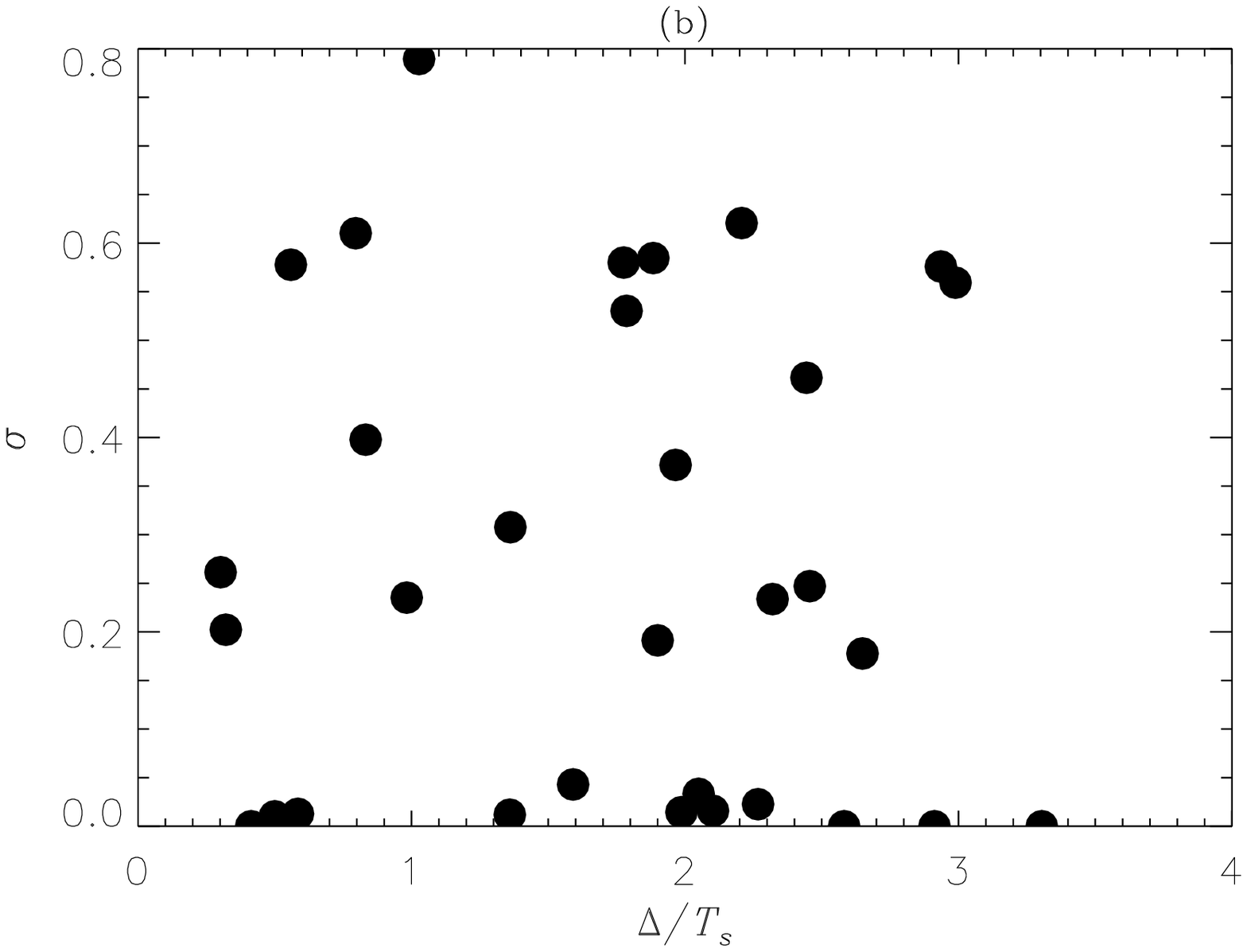,height=10cm,width=14cm}
\]
\end{minipage}

{\bf \large Fig. 8}\\
\end{center}

\newpage

\begin{center}
\begin{minipage}{15cm}
\[
\psfig{figure=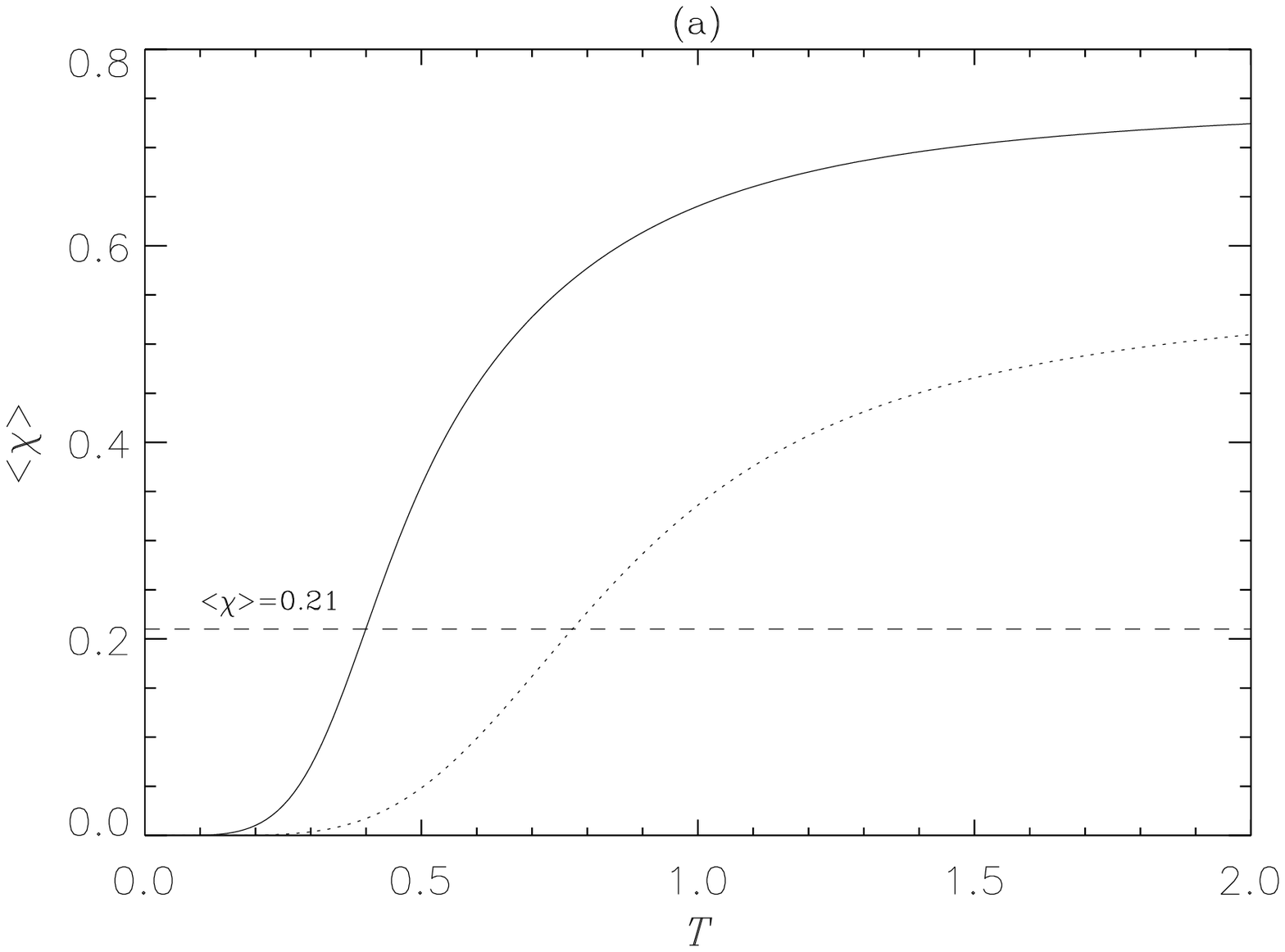,height=10cm,width=14cm}
\]
\end{minipage}
\end{center}

\begin{center}
\begin{minipage}{15cm}
\[
\psfig{figure=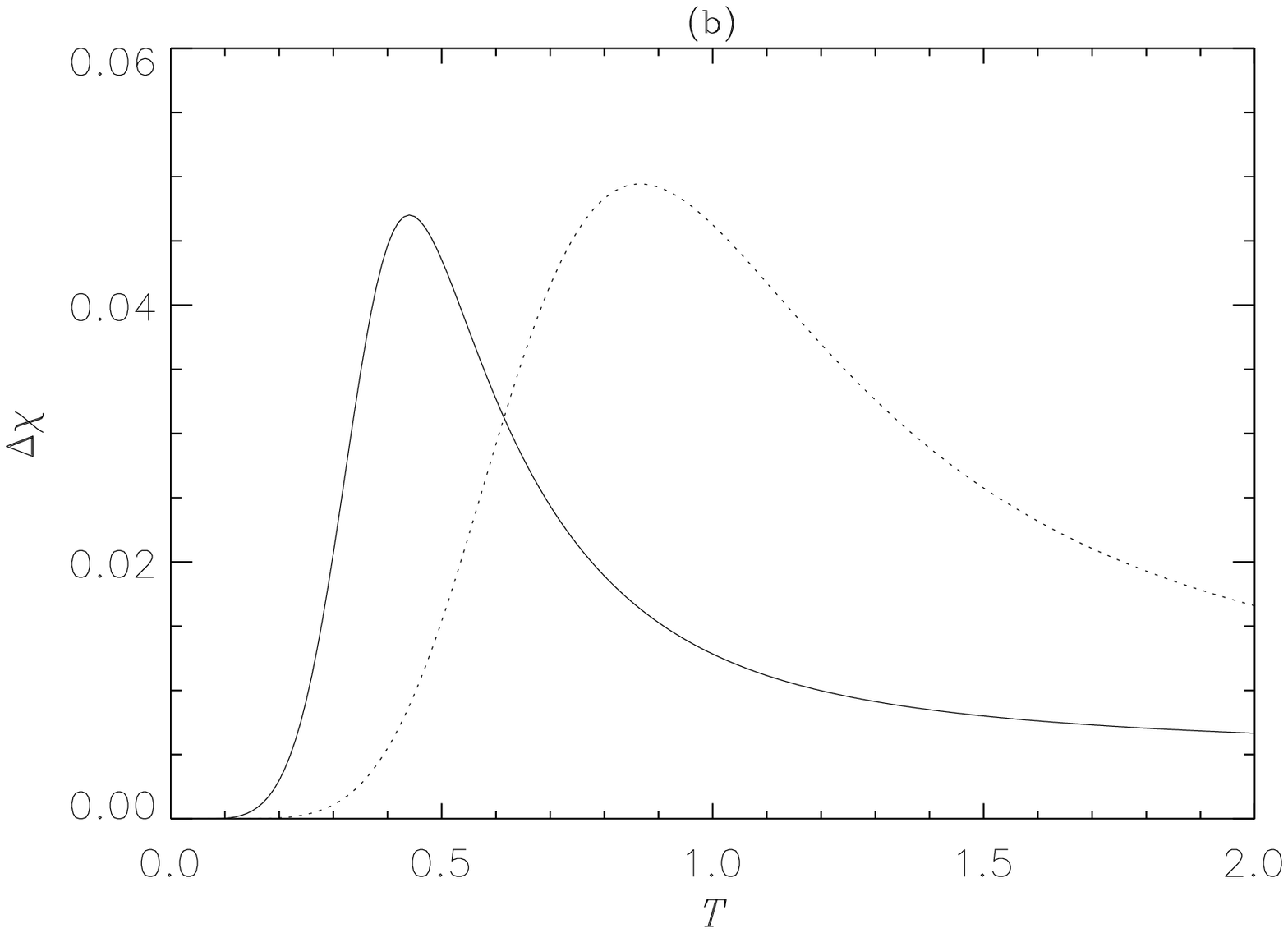,height=10cm,width=14cm}
\]
\end{minipage}

{\bf \large Fig. 9}\\
\end{center}

\newpage

\begin{center}
\begin{minipage}{15cm}
\[
\psfig{figure=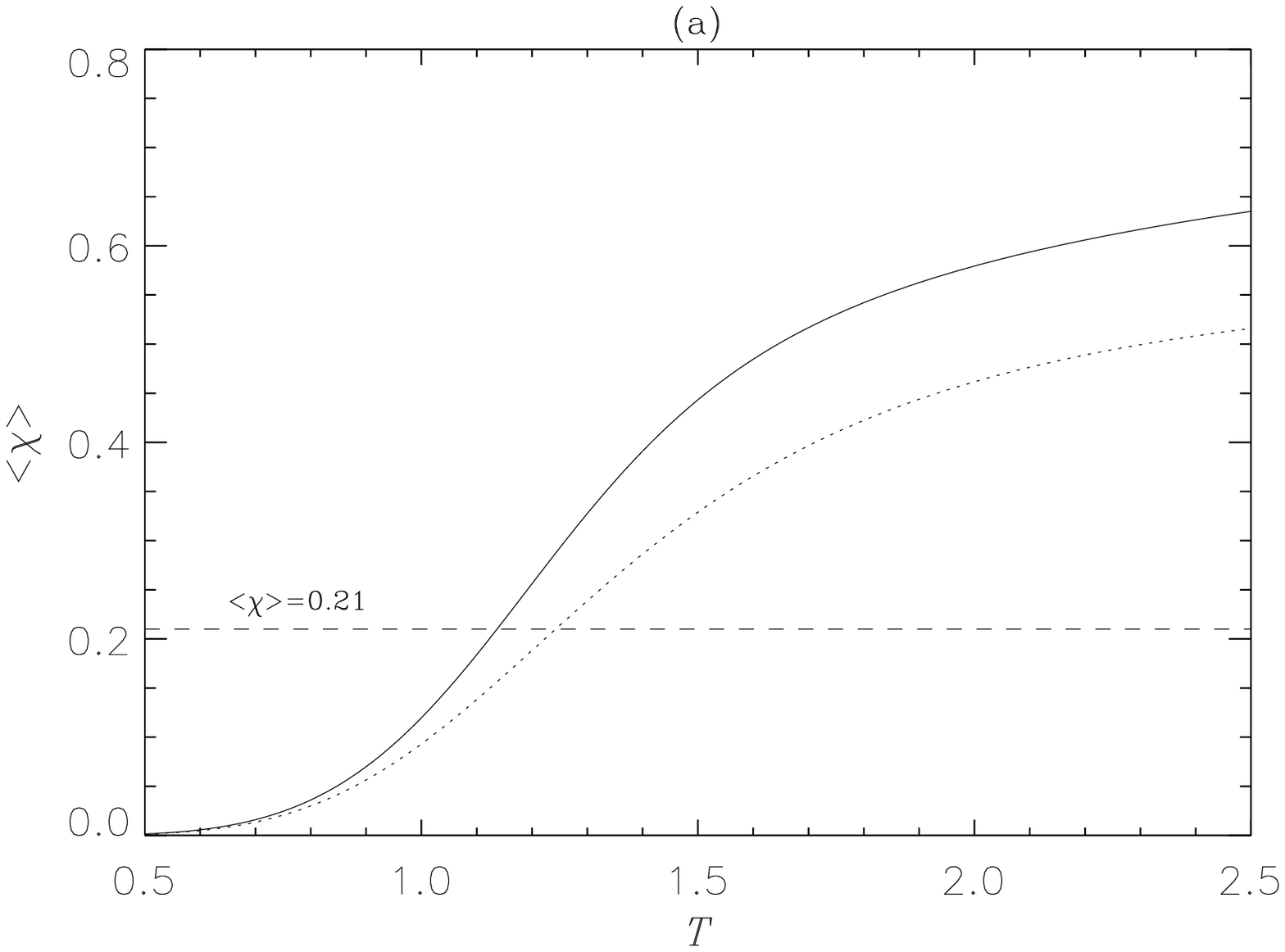,height=10cm,width=14cm}
\]
\end{minipage}
\end{center}  

\begin{center}
\begin{minipage}{15cm}
\[
\psfig{figure=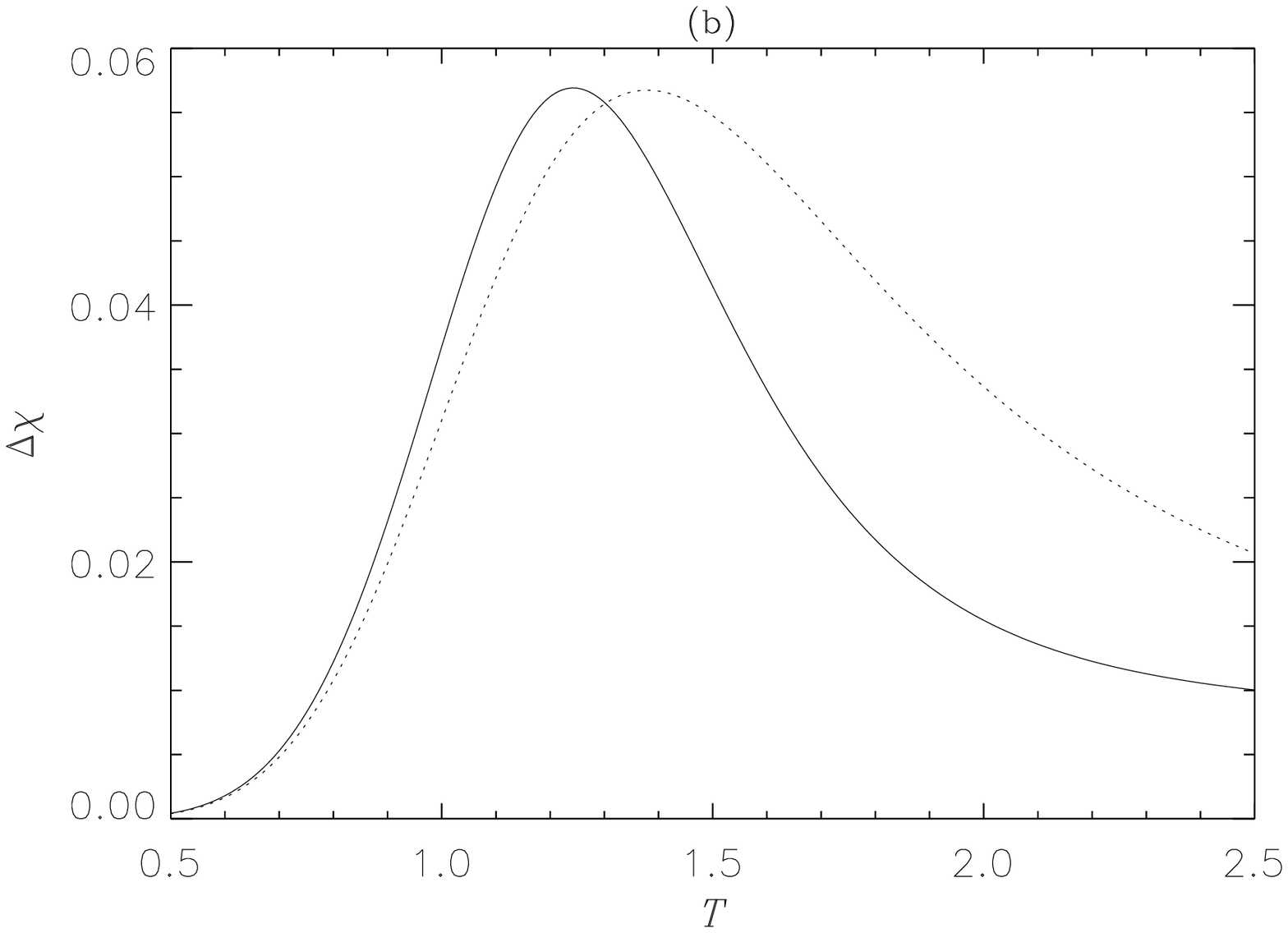,height=10cm,width=14cm}
\]
\end{minipage}

{\bf \large Fig. 10}\\
\end{center}

\newpage

\hspace{10cm}

\begin{center}
\begin{minipage}{18.5cm}
\[
\psfig{figure=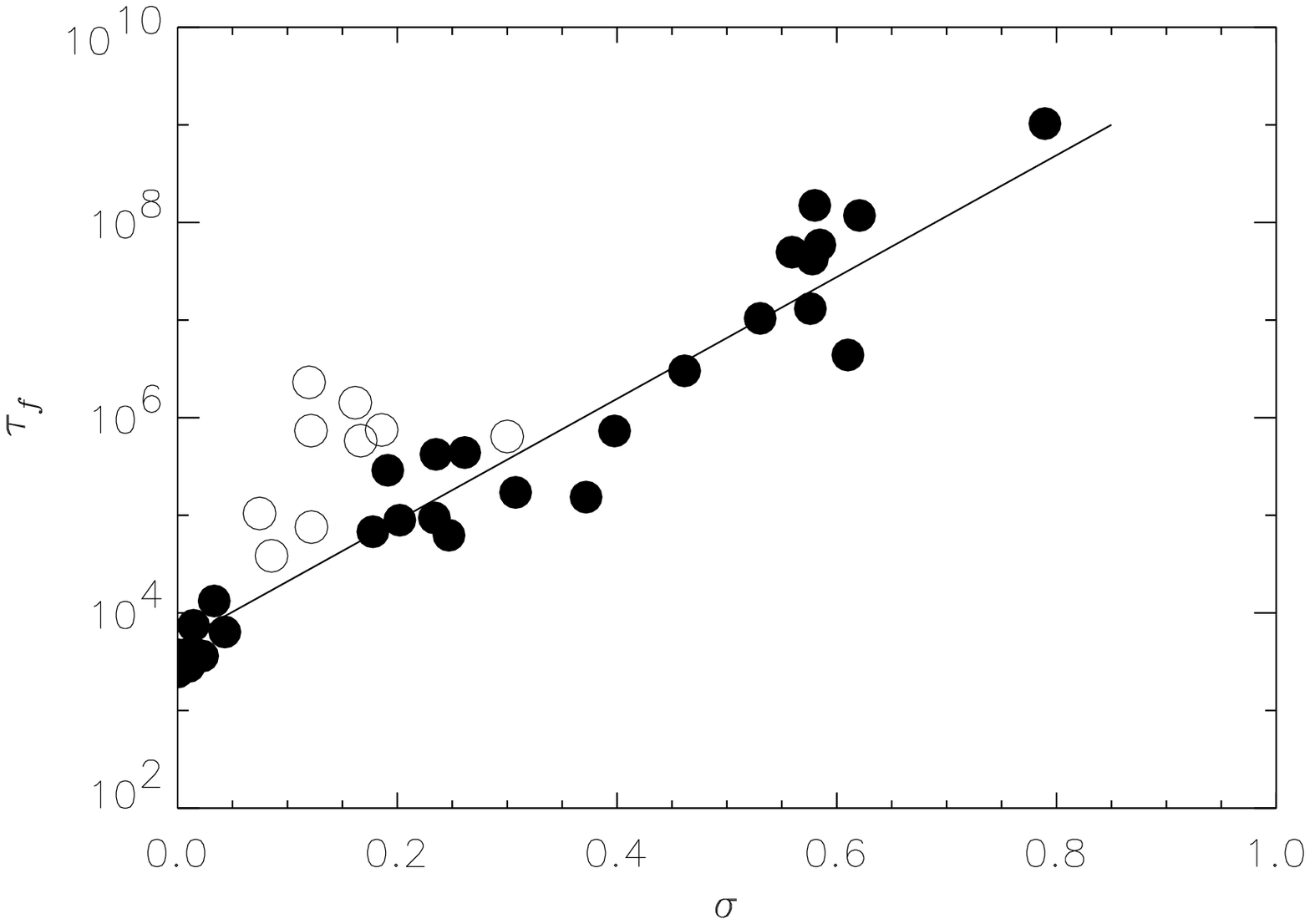,height=13.5cm,width=18.5cm}
\]
\end{minipage}

{\bf \large Fig. 11}\\ 
\end{center}

\newpage

\hspace{10cm}

\begin{center}
\begin{minipage}{18.5cm}
\[
\psfig{figure=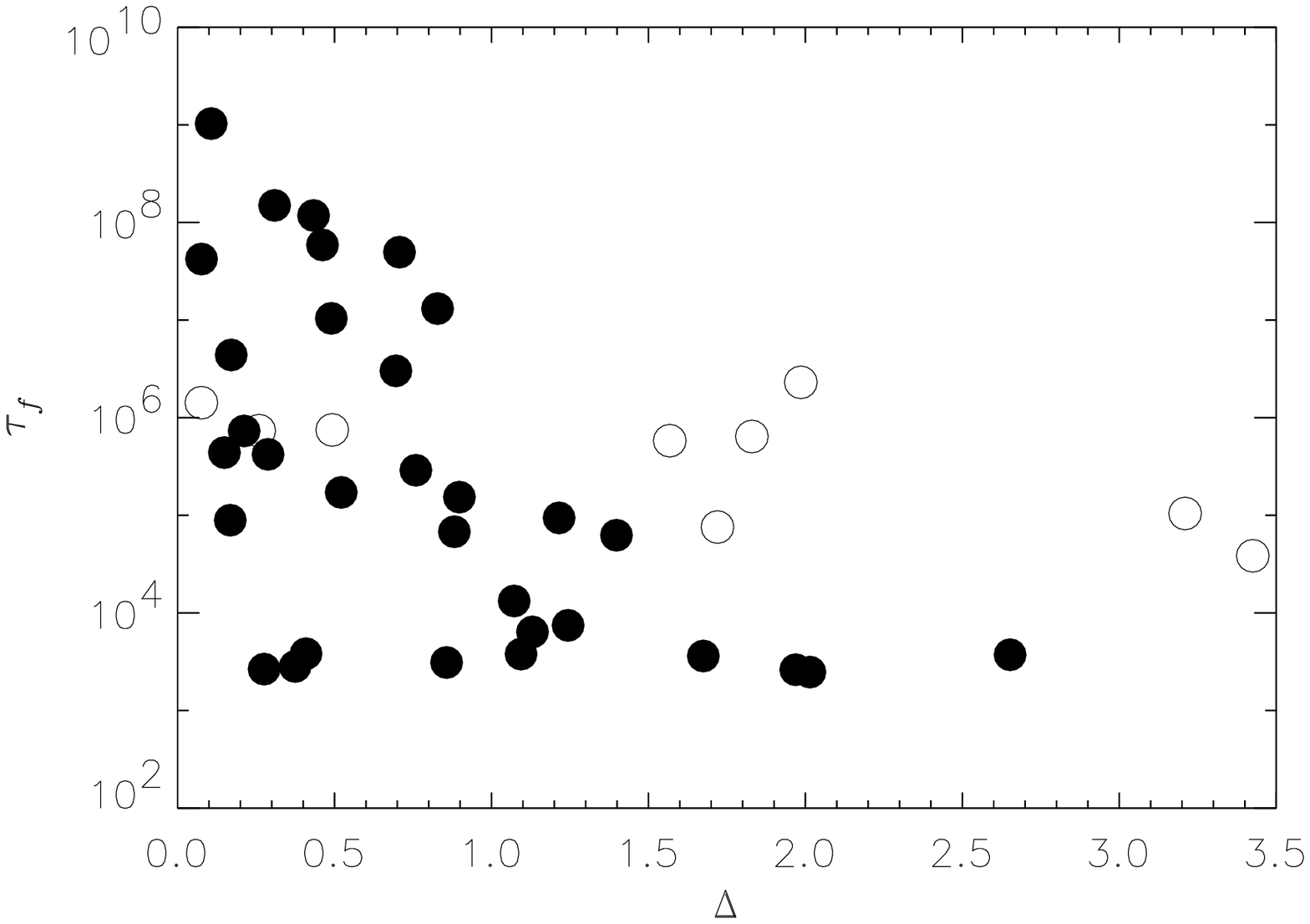,height=13.5cm,width=18.5cm}
\]
\end{minipage}

{\bf \large Fig. 12}\\ 
\end{center}

\newpage

\hspace{10cm}

\begin{center}
\begin{minipage}{18.5cm}
\[
\psfig{figure=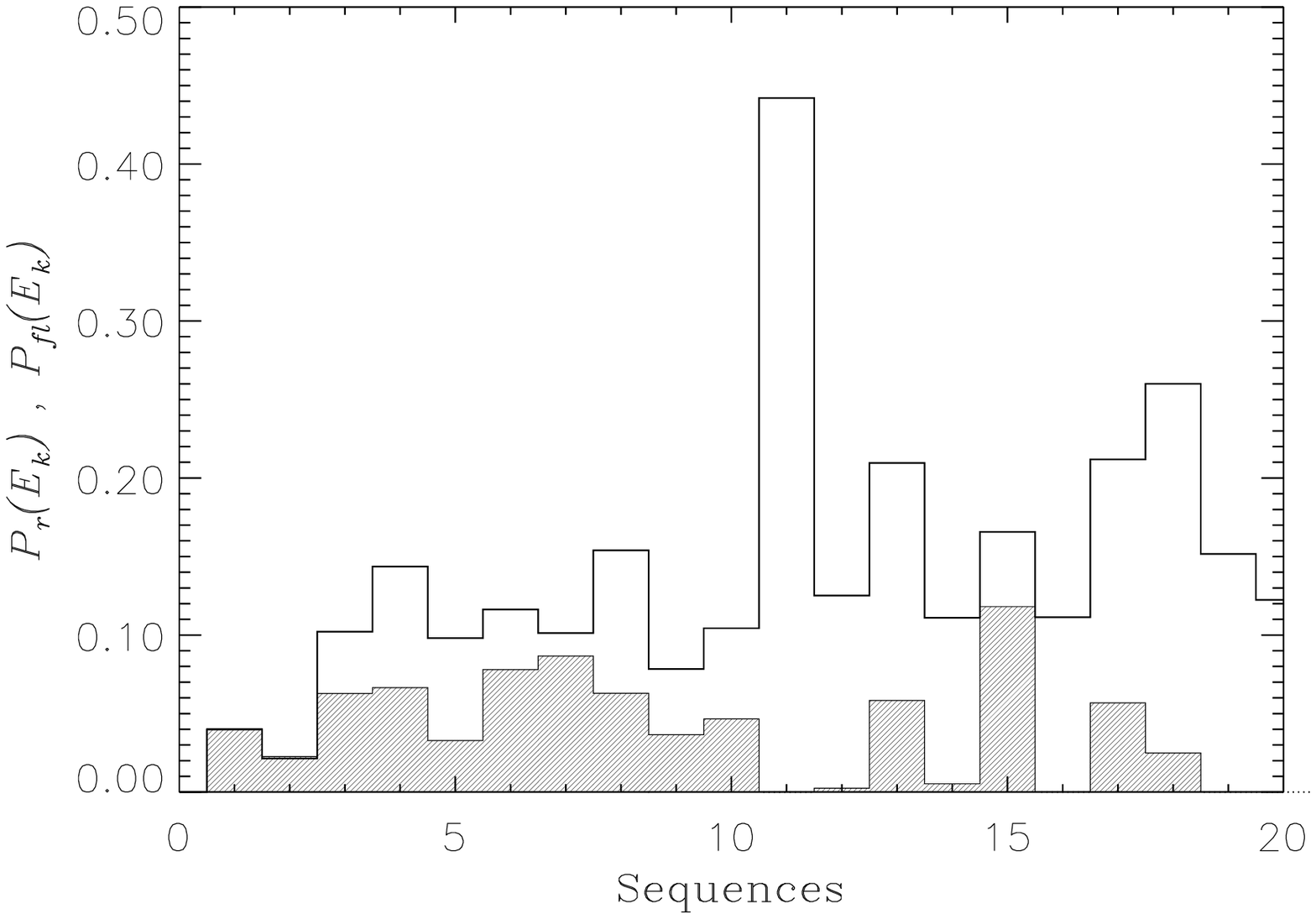,height=13.5cm,width=18.5cm}
\]
\end{minipage}

{\bf \large Fig. 13}\\
\end{center}

\newpage

\begin{center}
\begin{minipage}{7cm}
\[
\psfig{figure=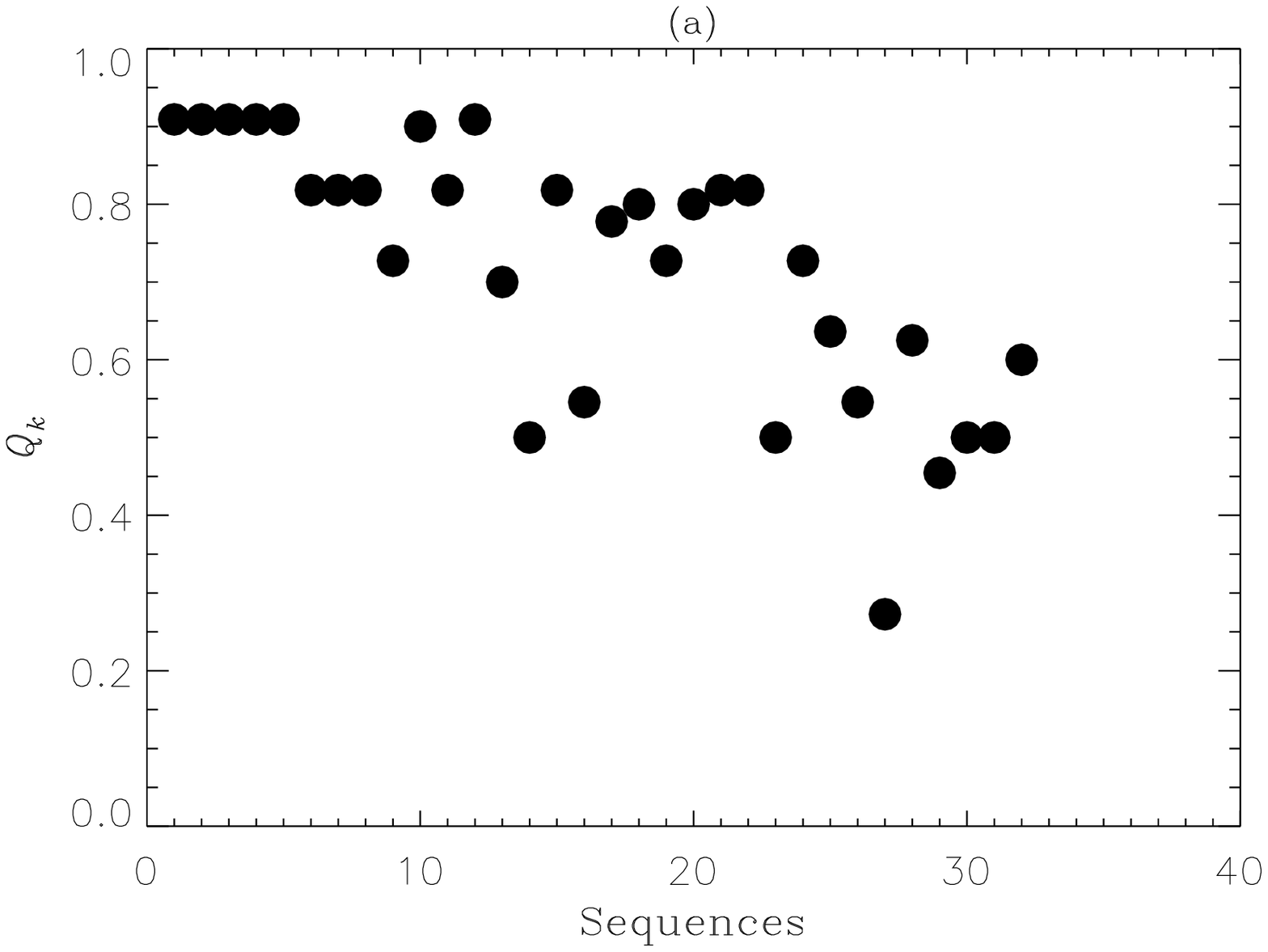,height=6cm,width=11cm}
\]
\end{minipage}
\end{center}

\begin{center}
\begin{minipage}{7cm}
\[
\psfig{figure=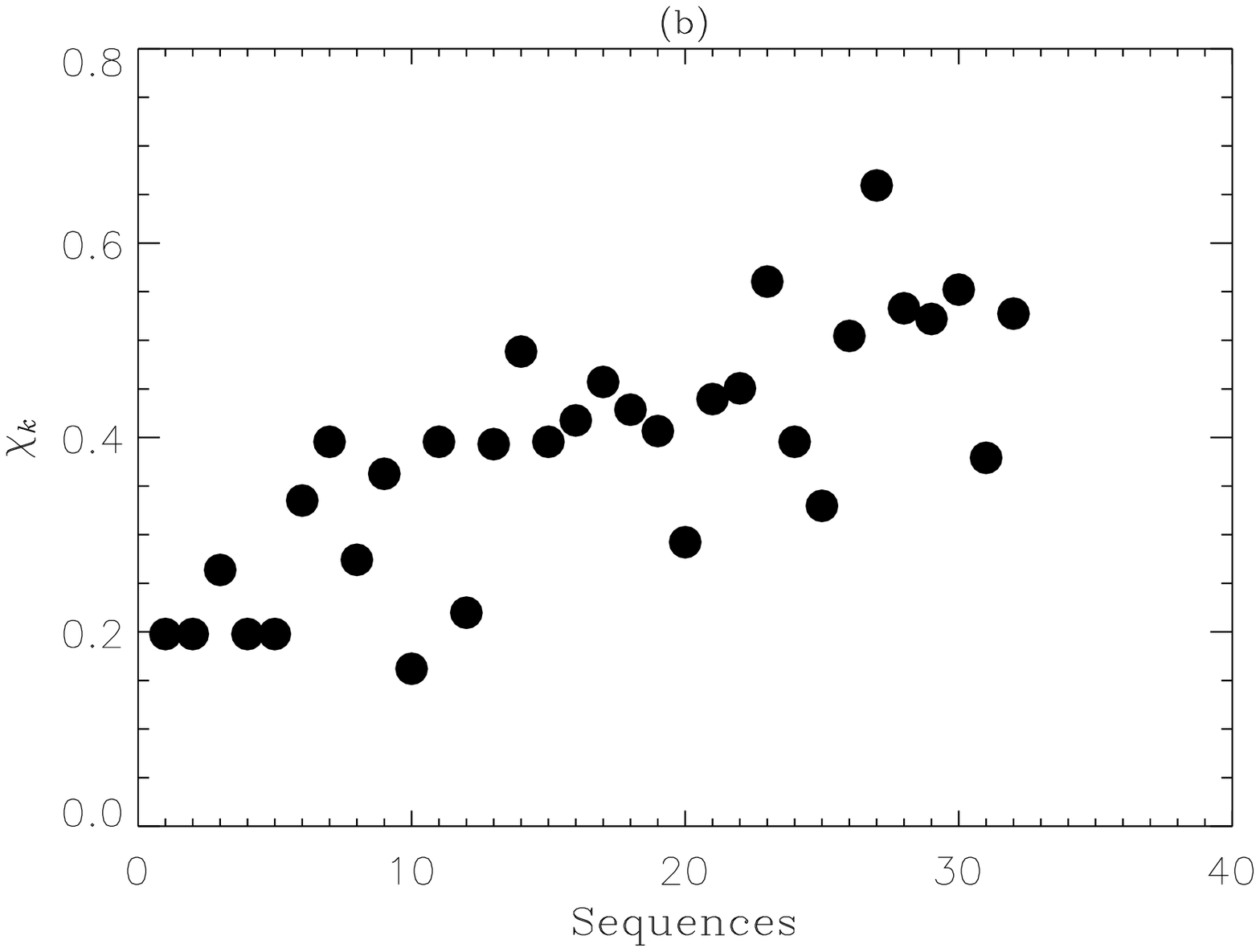,height=6cm,width=11cm}
\]
\end{minipage}

\end{center}           

\begin{center}
\begin{minipage}{7cm}
\[
\psfig{figure=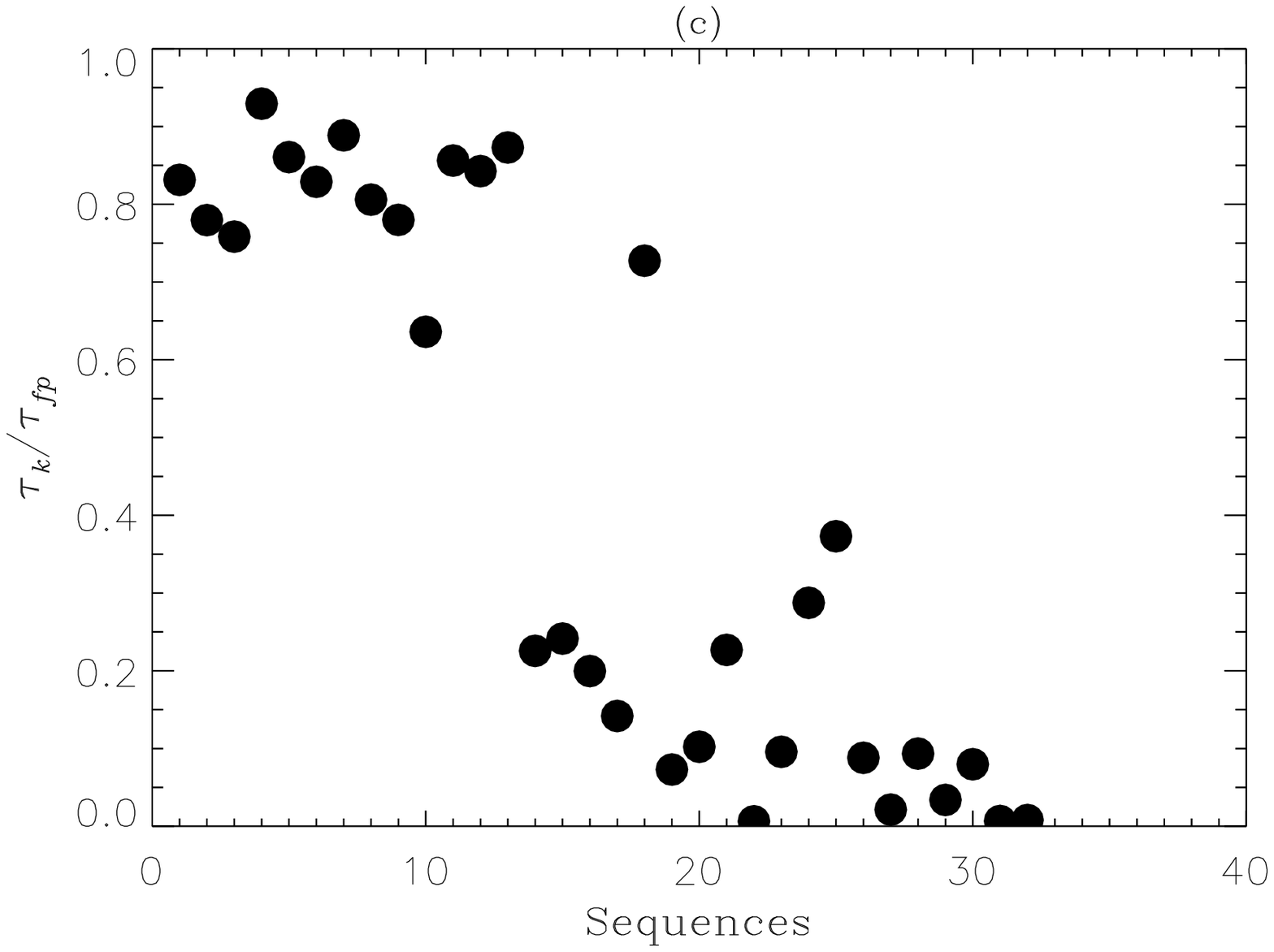,height=6cm,width=11cm}
\]
\end{minipage}

{\bf \large \hspace{4.5cm} Fig. 14}\\

\end{center}

\newpage

\begin{center}
\begin{minipage}{15cm}
\[
\psfig{figure=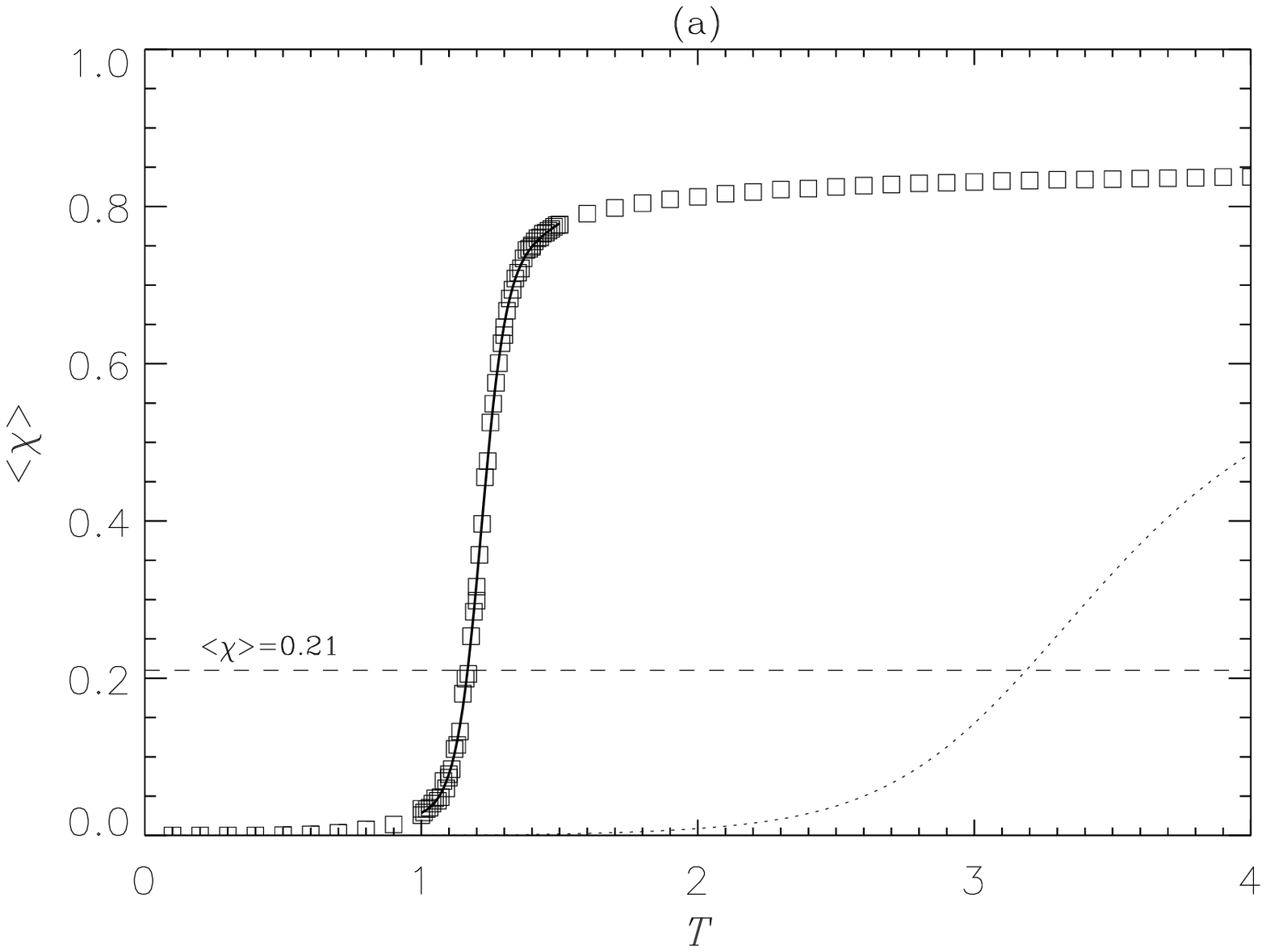,height=10cm,width=14cm}
\]
\end{minipage}
\end{center}

\begin{center}
\begin{minipage}{15cm}
\[
\psfig{figure=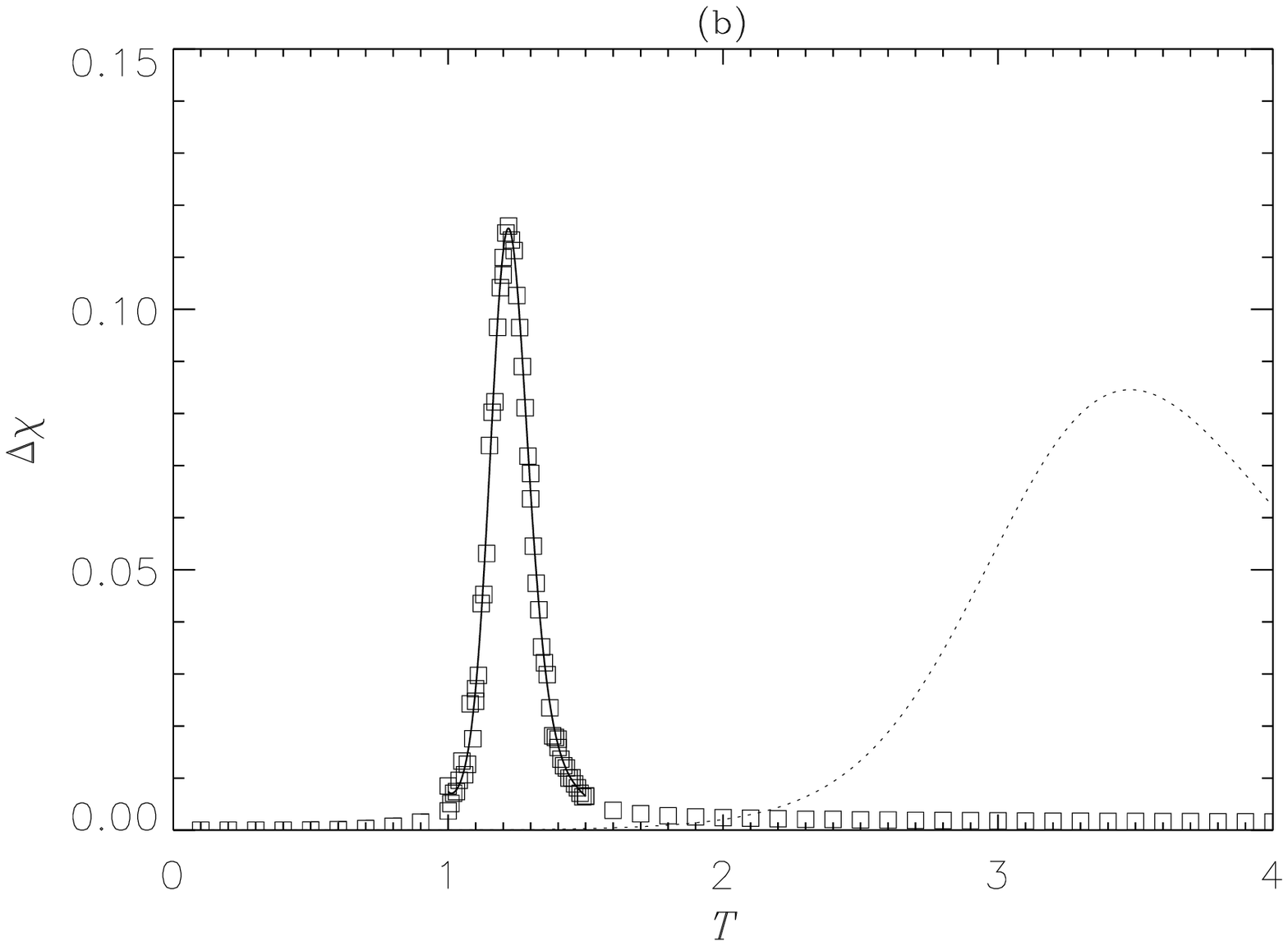,height=10cm,width=14cm}
\]
\end{minipage}

{\bf \large Fig. 15}\\
\end{center}

\newpage

\begin{center}
\begin{minipage}{15cm}
\[
\psfig{figure=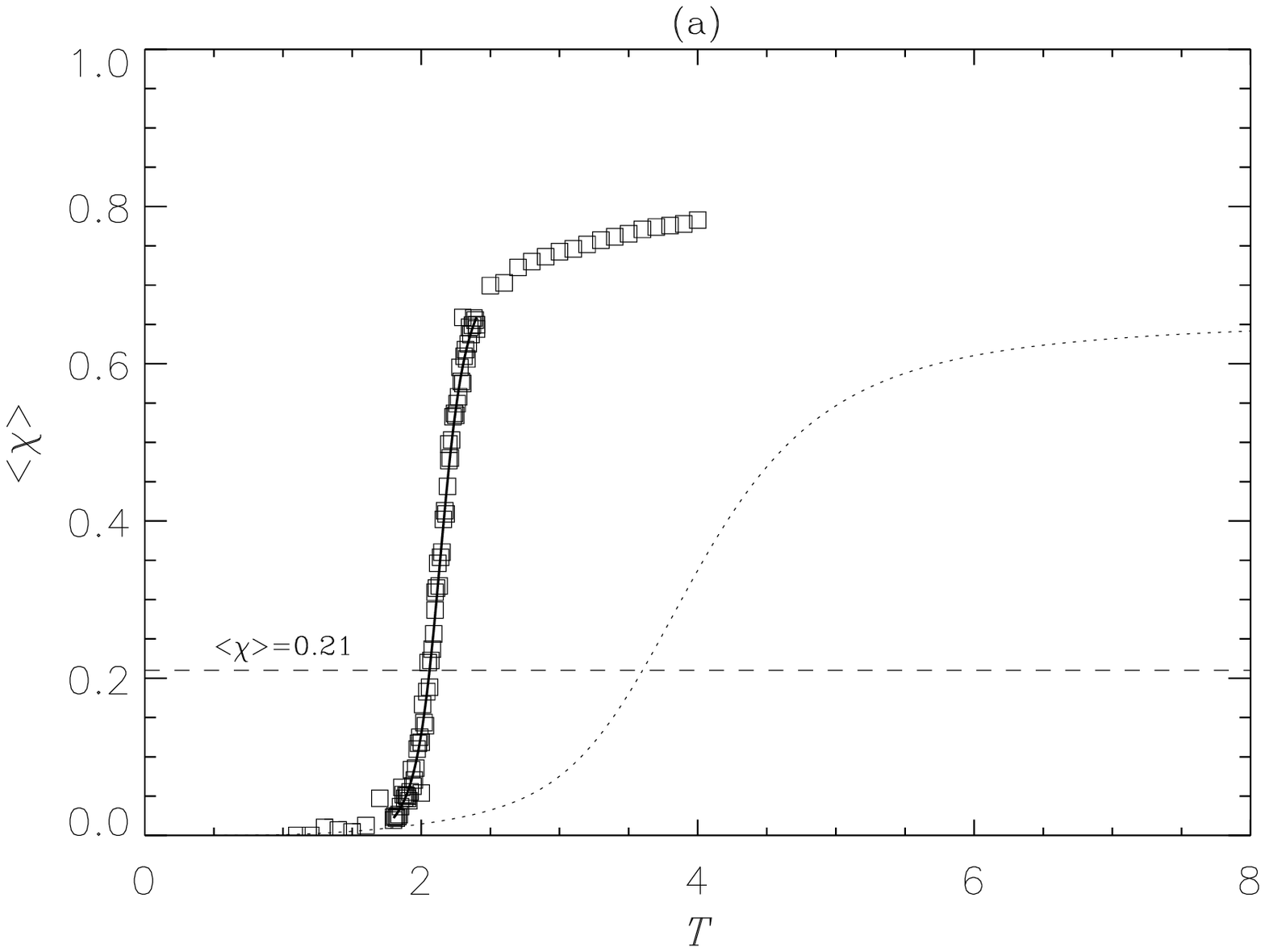,height=10cm,width=14cm}
\]
\end{minipage}
\end{center}

\begin{center}
\begin{minipage}{15cm}
\[
\psfig{figure=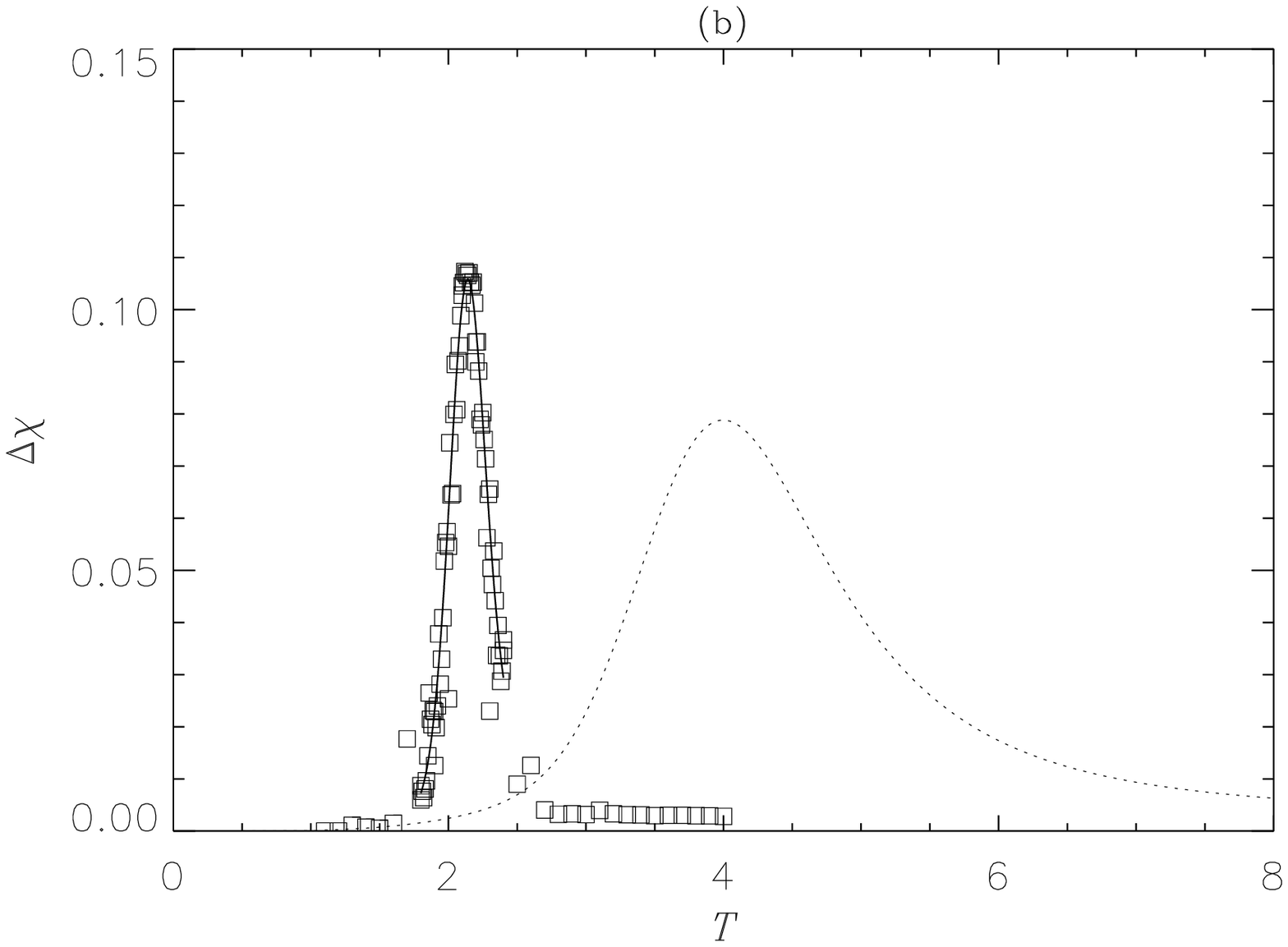,height=10cm,width=14cm}
\]
\end{minipage}

{\bf \large Fig. 16}\\
\end{center}

\newpage
\begin{center}
\begin{minipage}{15cm}
\[
\psfig{figure=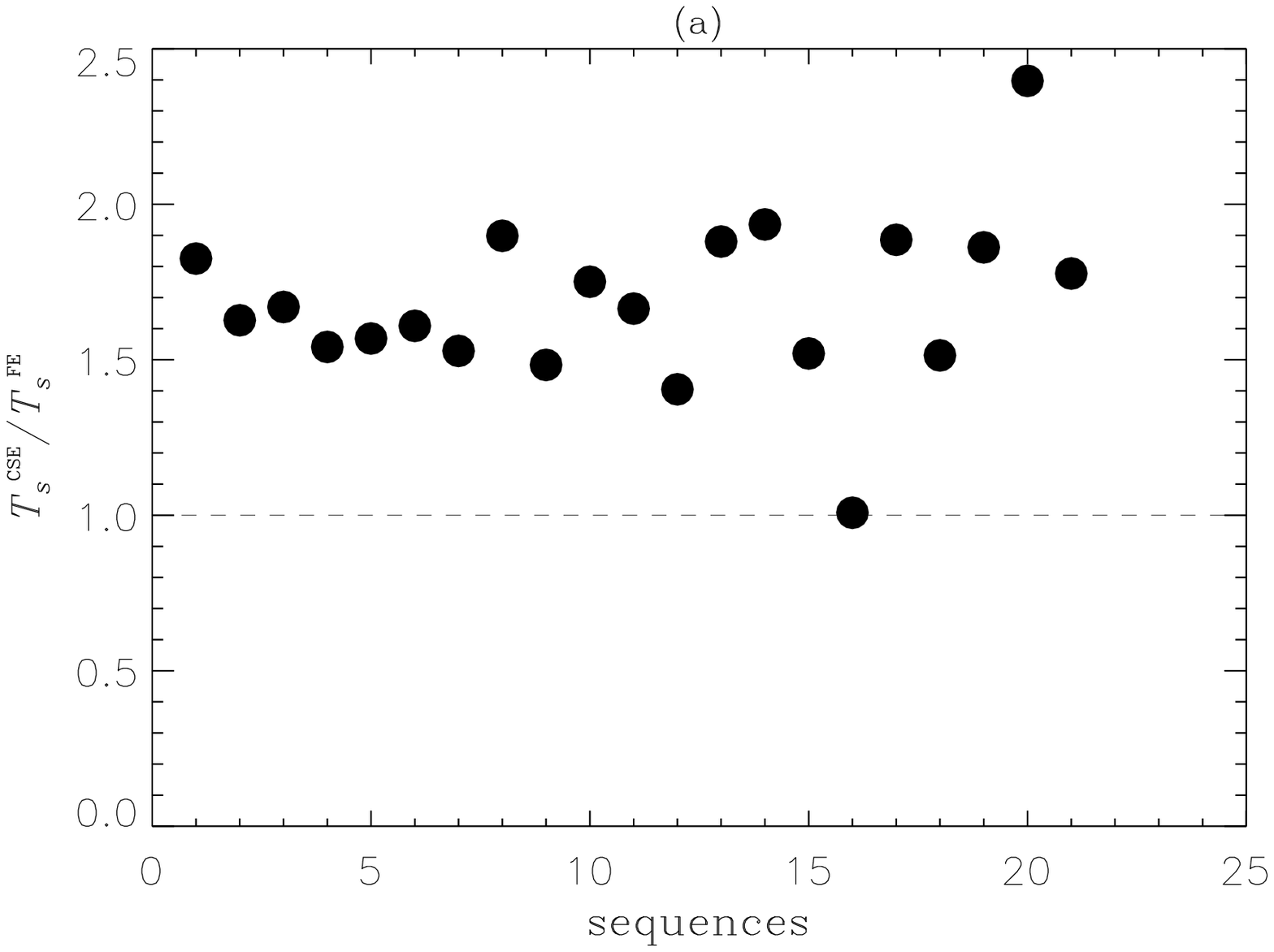,height=10cm,width=14cm}
\]
\end{minipage}
\end{center}

\begin{center}
\begin{minipage}{15cm}
\[
\psfig{figure=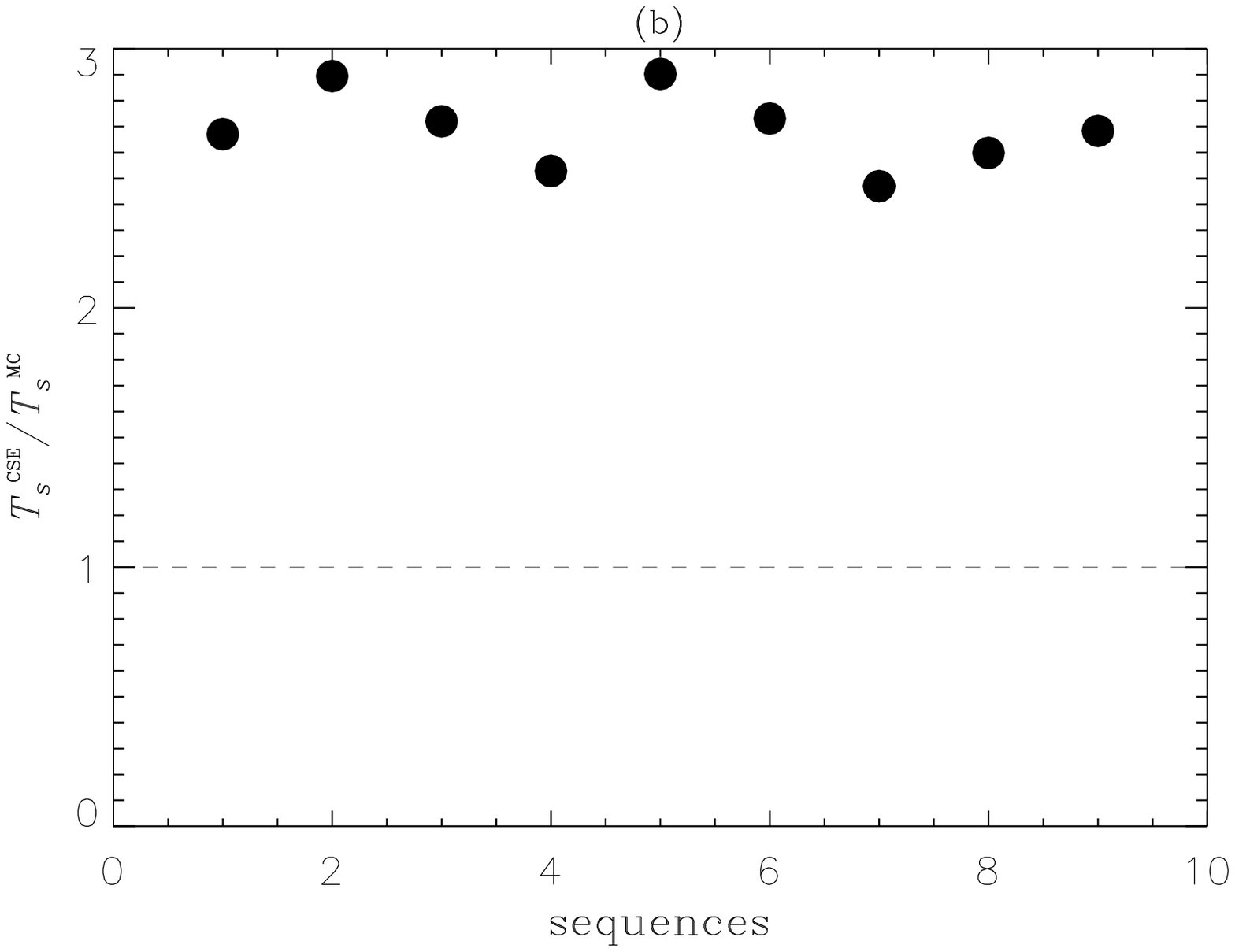,height=10cm,width=14cm}
\]
\end{minipage}

{\bf \large Fig. 17}\\
\end{center}

\newpage

\begin{center}
\begin{minipage}{15cm}
\[
\psfig{figure=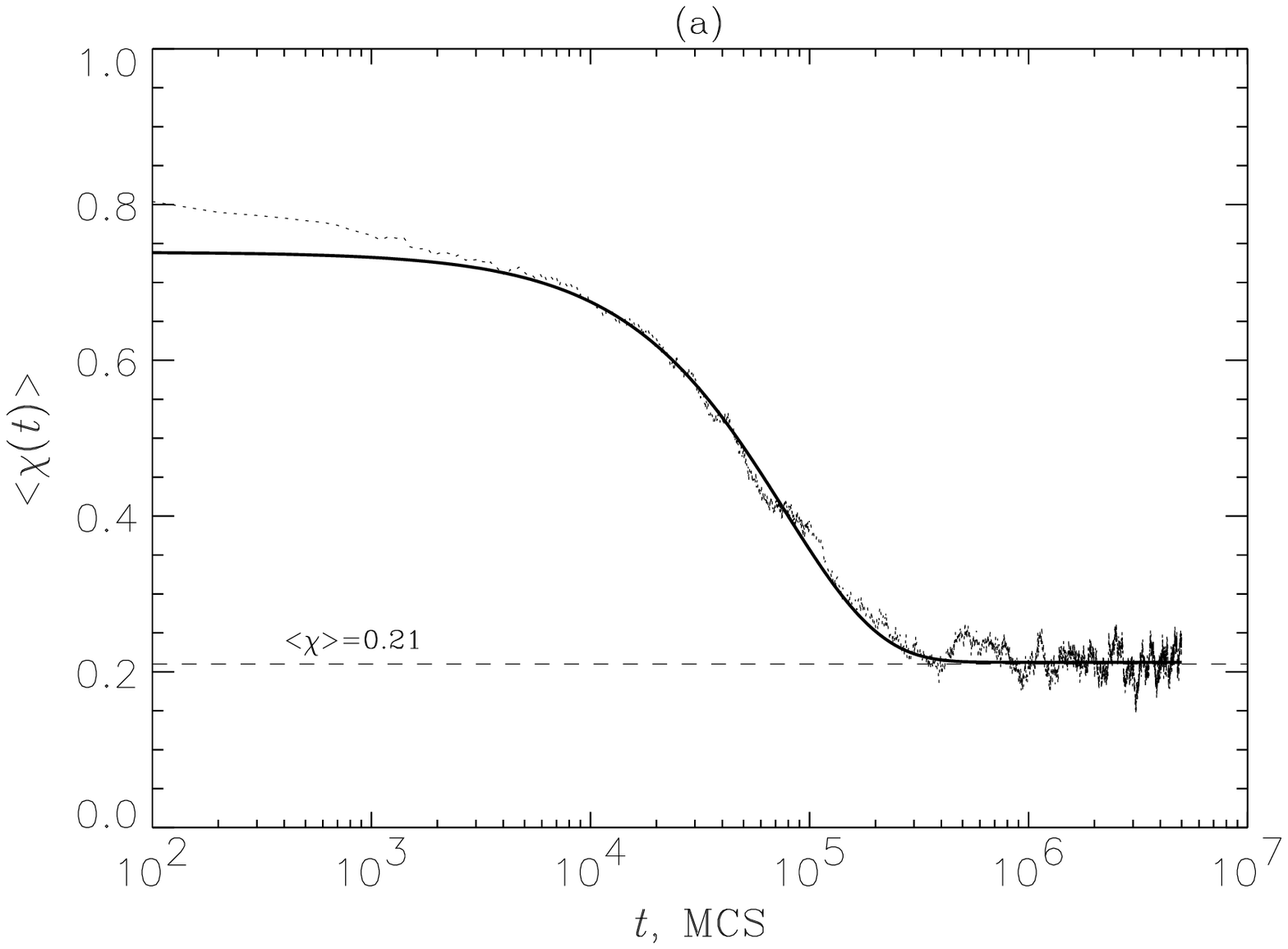,height=10cm,width=14cm}
\]
\end{minipage}
\end{center}

\begin{center}
\begin{minipage}{15cm}
\[
\psfig{figure=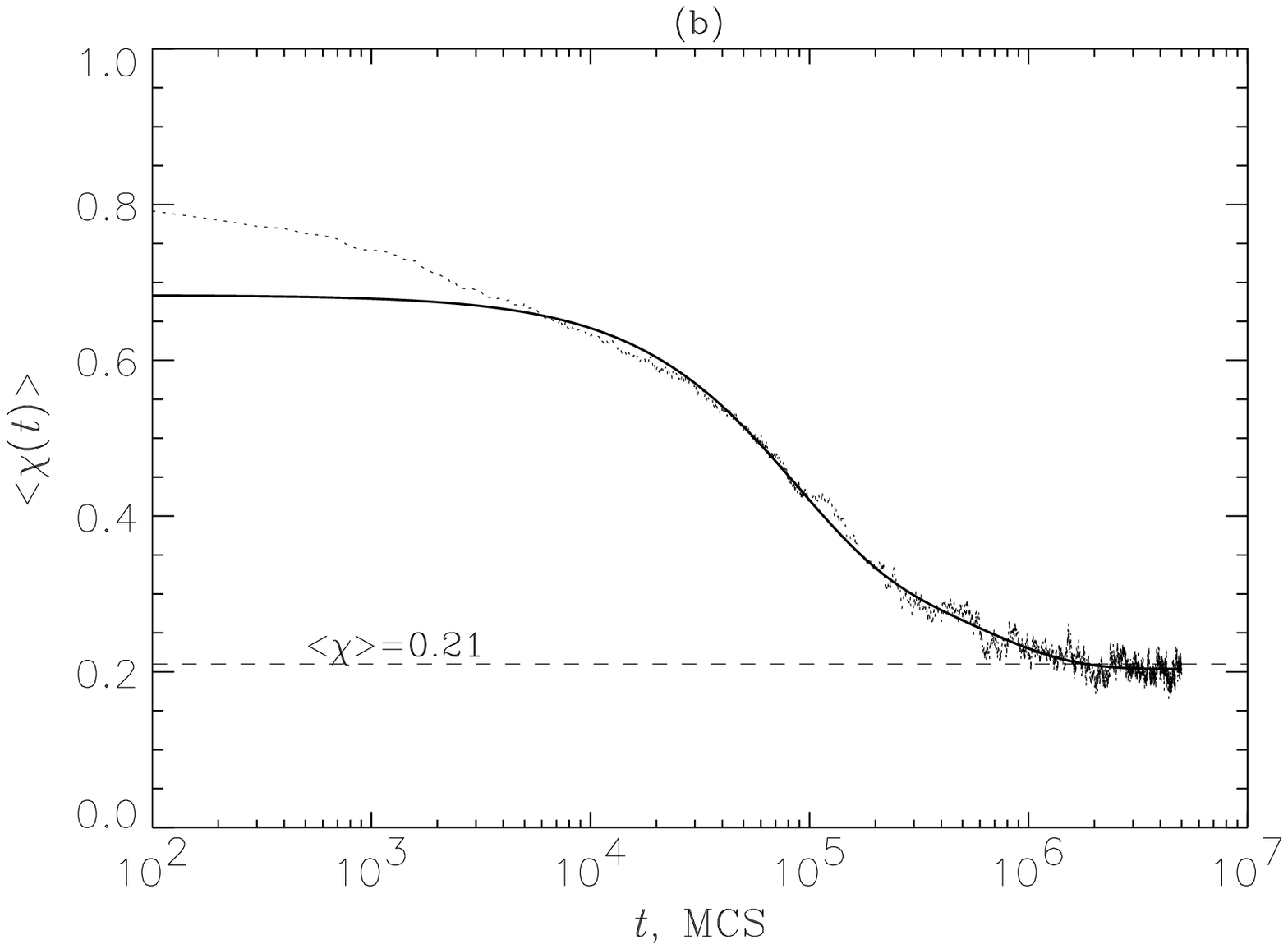,height=10cm,width=14cm}
\]
\end{minipage}

{\bf \large Fig. 18}\\
\end{center}
 
\newpage

\hspace{10cm}

\begin{center}
\begin{minipage}{18.5cm}
\[
\psfig{figure=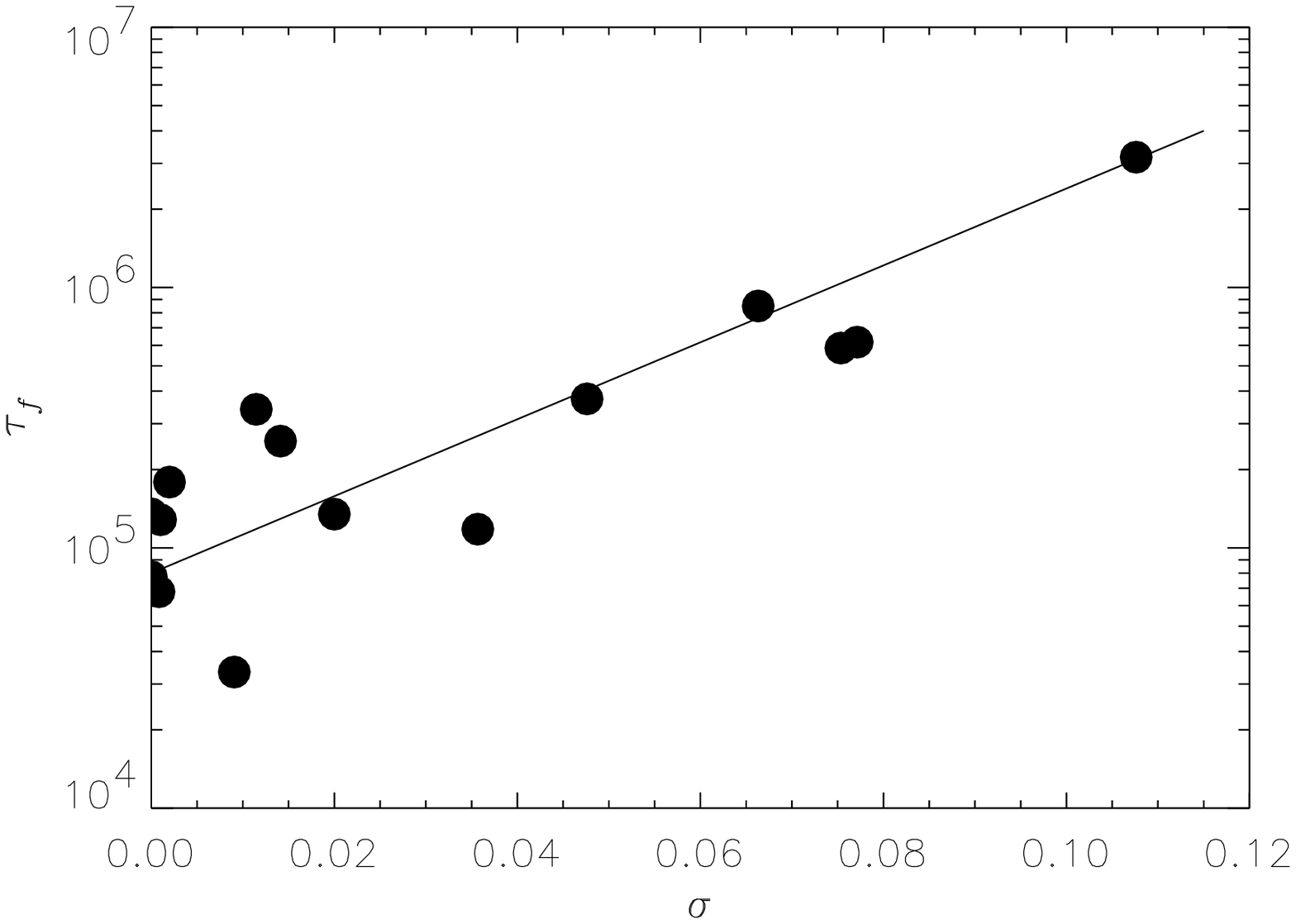,height=13.5cm,width=18.5cm}
\]
\end{minipage}

{\bf \large Fig. 19}\\
\end{center}
 
\newpage

\hspace{10cm}

\begin{center}
\begin{minipage}{18.5cm}
\[
\psfig{figure=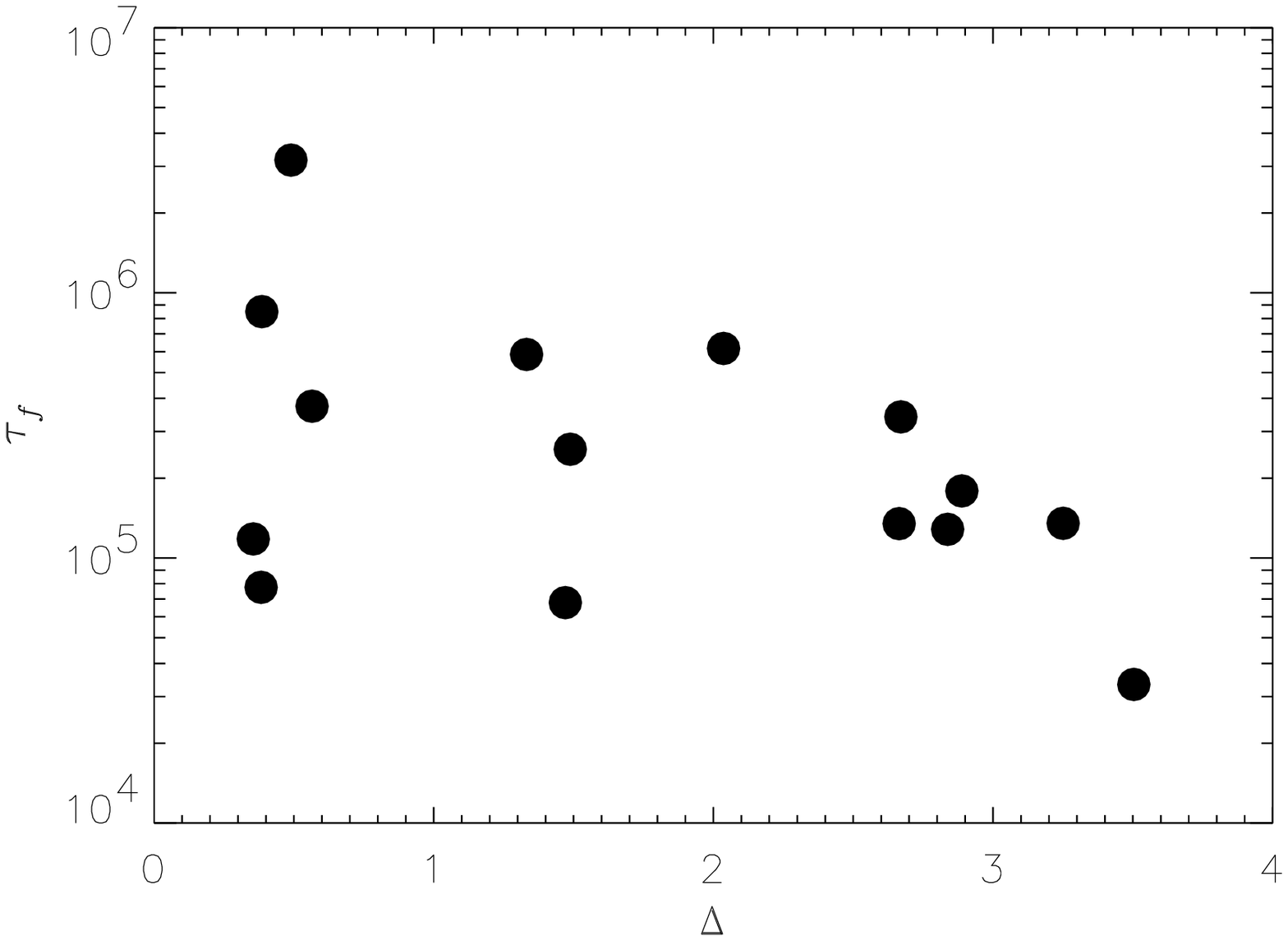,height=13.5cm,width=18.5cm}
\]
\end{minipage}

{\bf \large Fig. 20}\\
\end{center}

\newpage

\begin{center}
\begin{minipage}{15cm}
\[
\psfig{figure=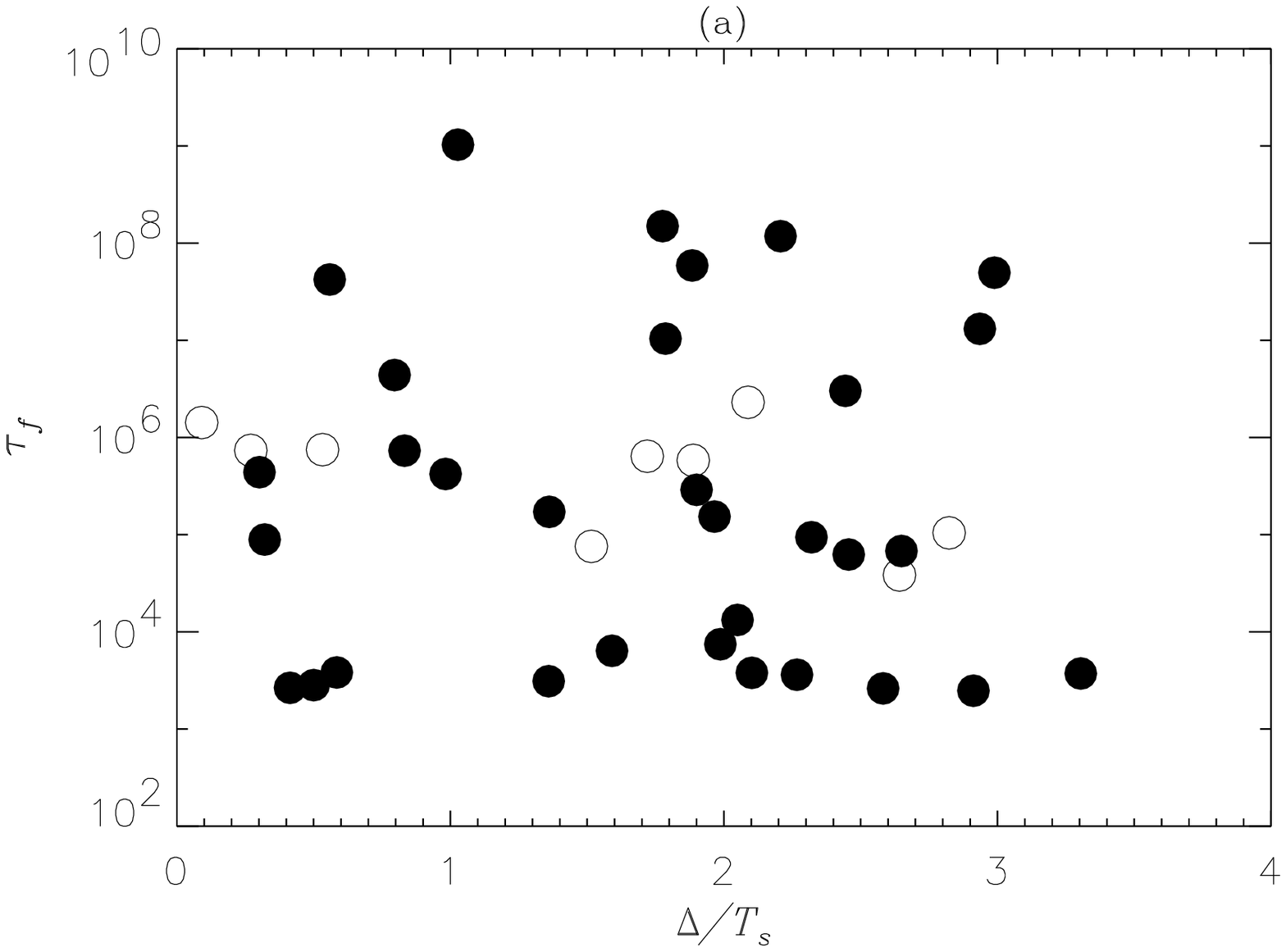,height=10cm,width=14cm}
\]
\end{minipage}
\end{center}  

\begin{center}
\begin{minipage}{15cm}
\[
\psfig{figure=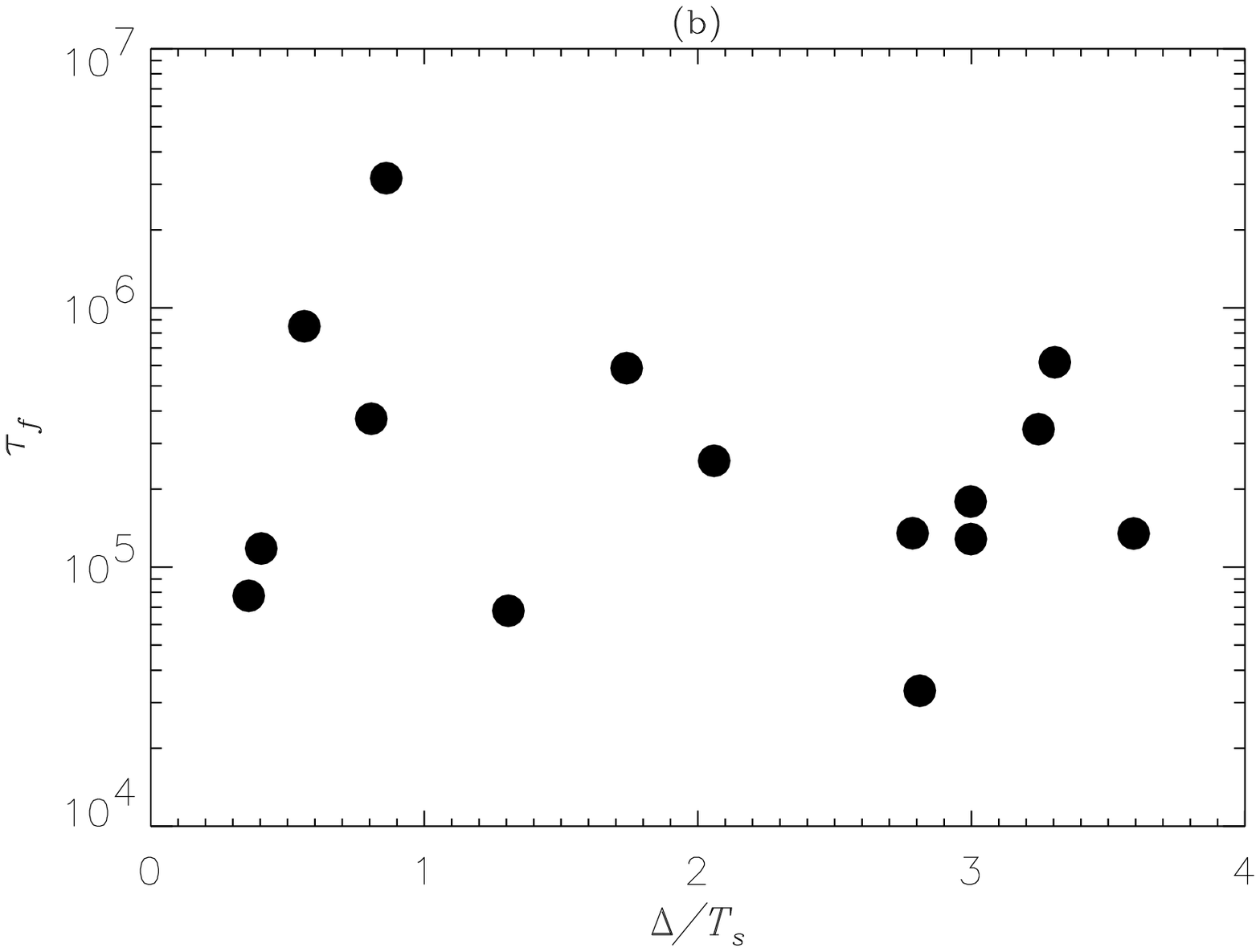,height=10cm,width=14cm}
\]
\end{minipage}

{\bf \large Fig. 21}\\
\end{center}

\newpage

\begin{center}
\begin{minipage}{15cm}
\[
\psfig{figure=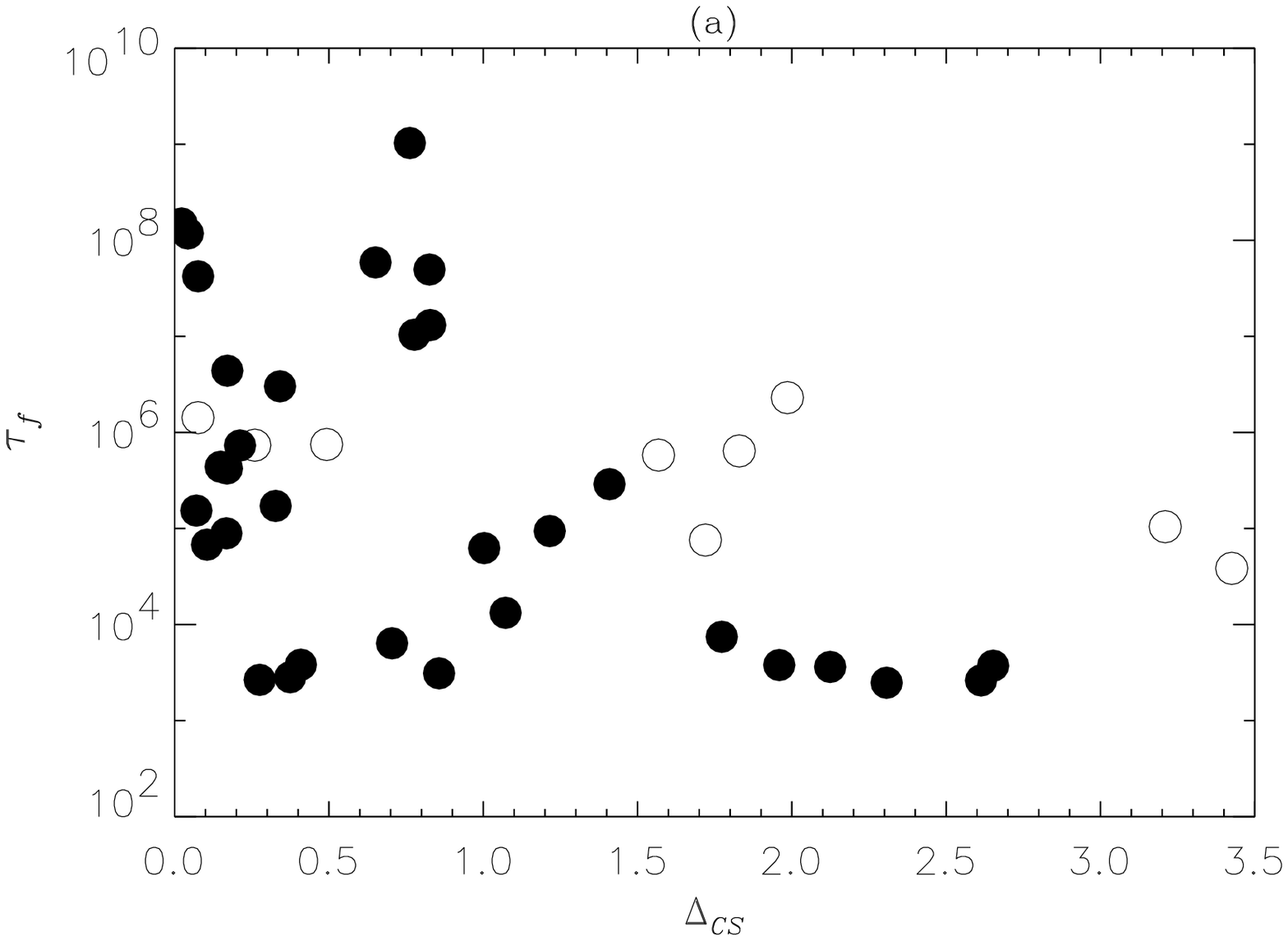,height=10cm,width=14cm}
\]
\end{minipage}
\end{center}  

\begin{center}
\begin{minipage}{15cm}
\[
\psfig{figure=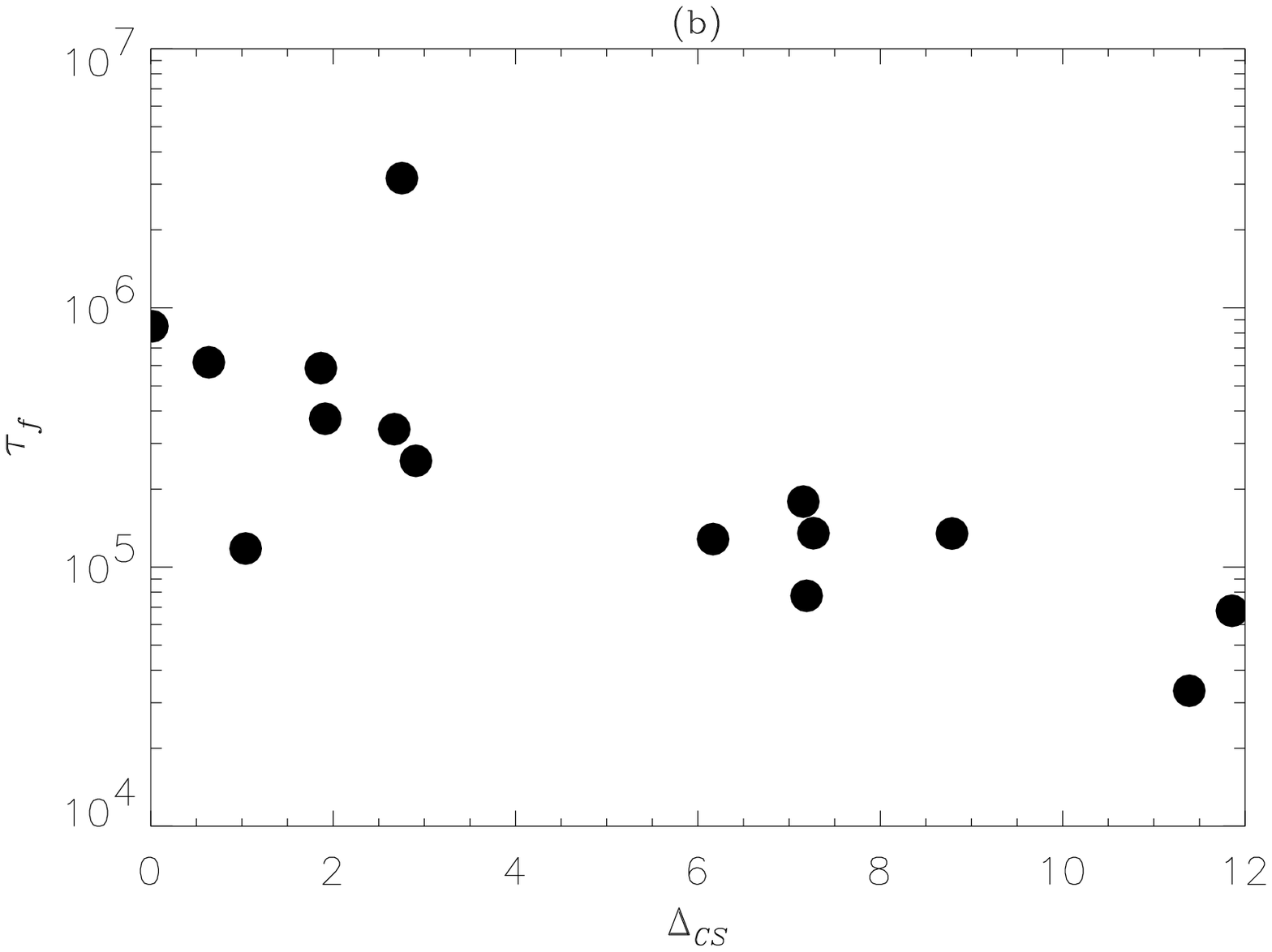,height=10cm,width=14cm}
\]
\end{minipage}

{\bf \large Fig. 22}\\
\end{center}

\newpage

\hspace{10cm}

\begin{center}
\begin{minipage}{18.5cm}
\[
\psfig{figure=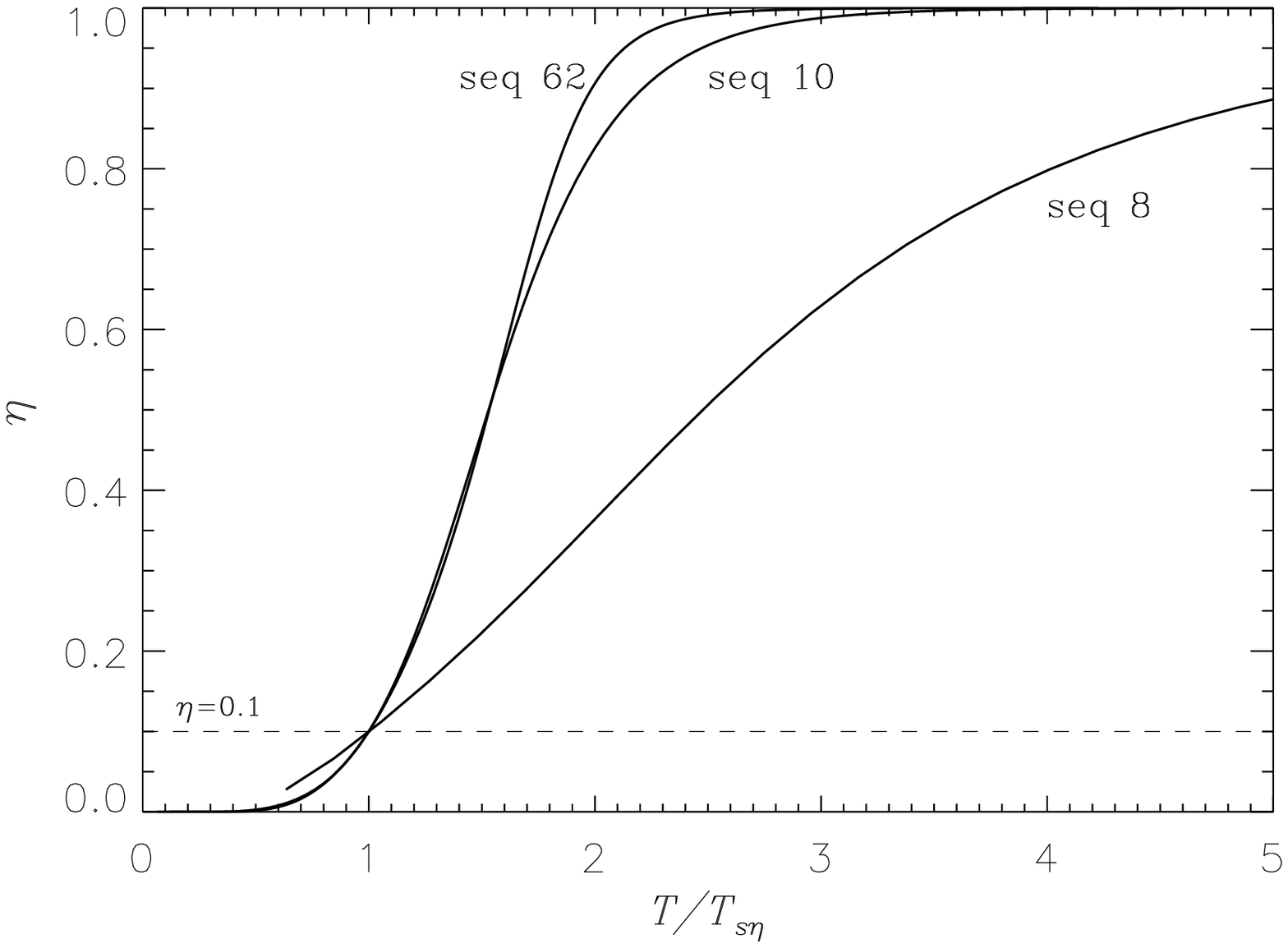,height=13.5cm,width=18.5cm}
\]
\end{minipage}

{\bf \large Fig. 23}\\
\end{center}

\newpage
 
\begin{center}
\begin{minipage}{15cm}
\[
\psfig{figure=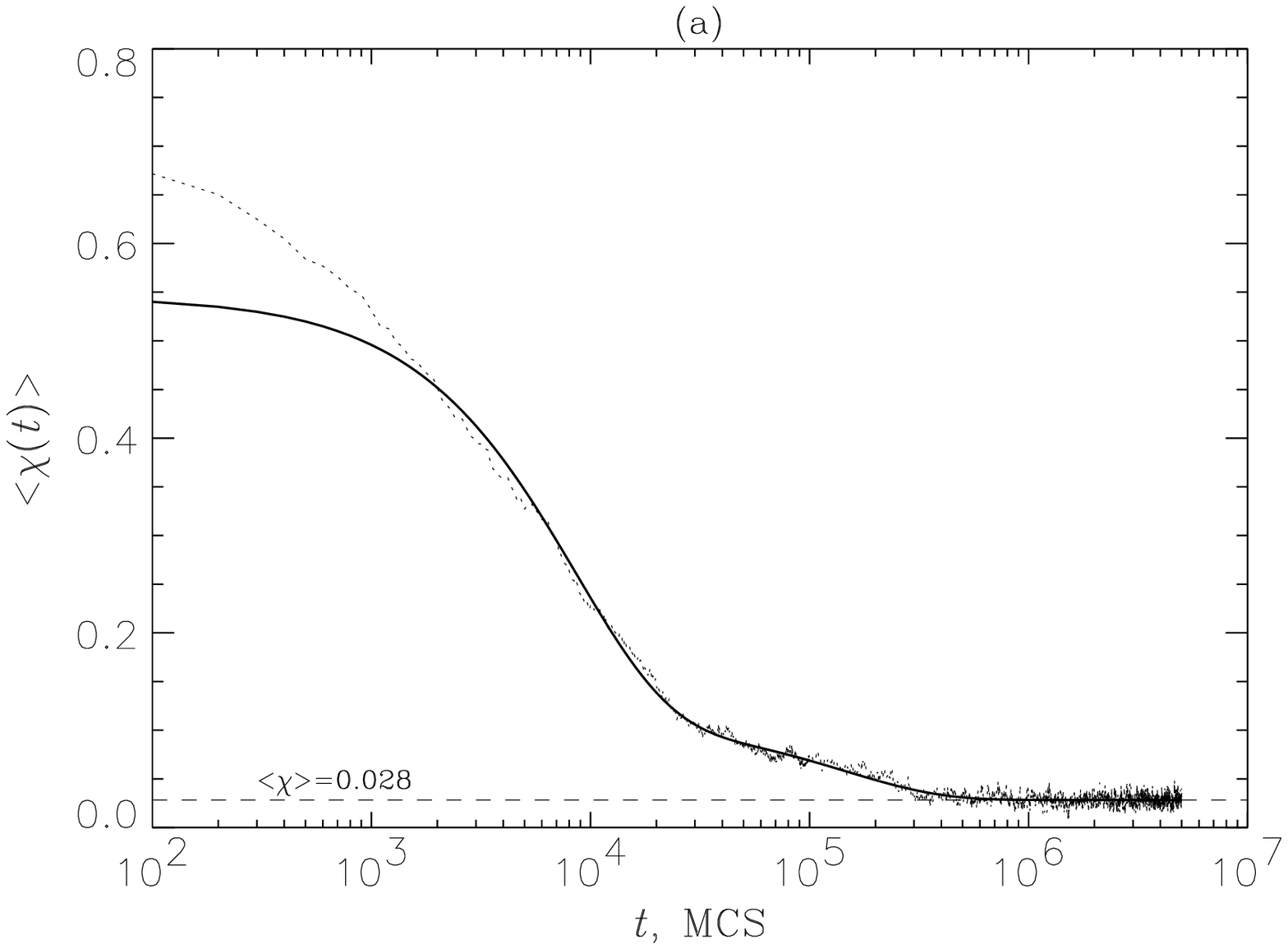,height=10cm,width=14cm}
\]
\end{minipage}

\end{center}

\begin{center}
\begin{minipage}{15cm}
\[
\psfig{figure=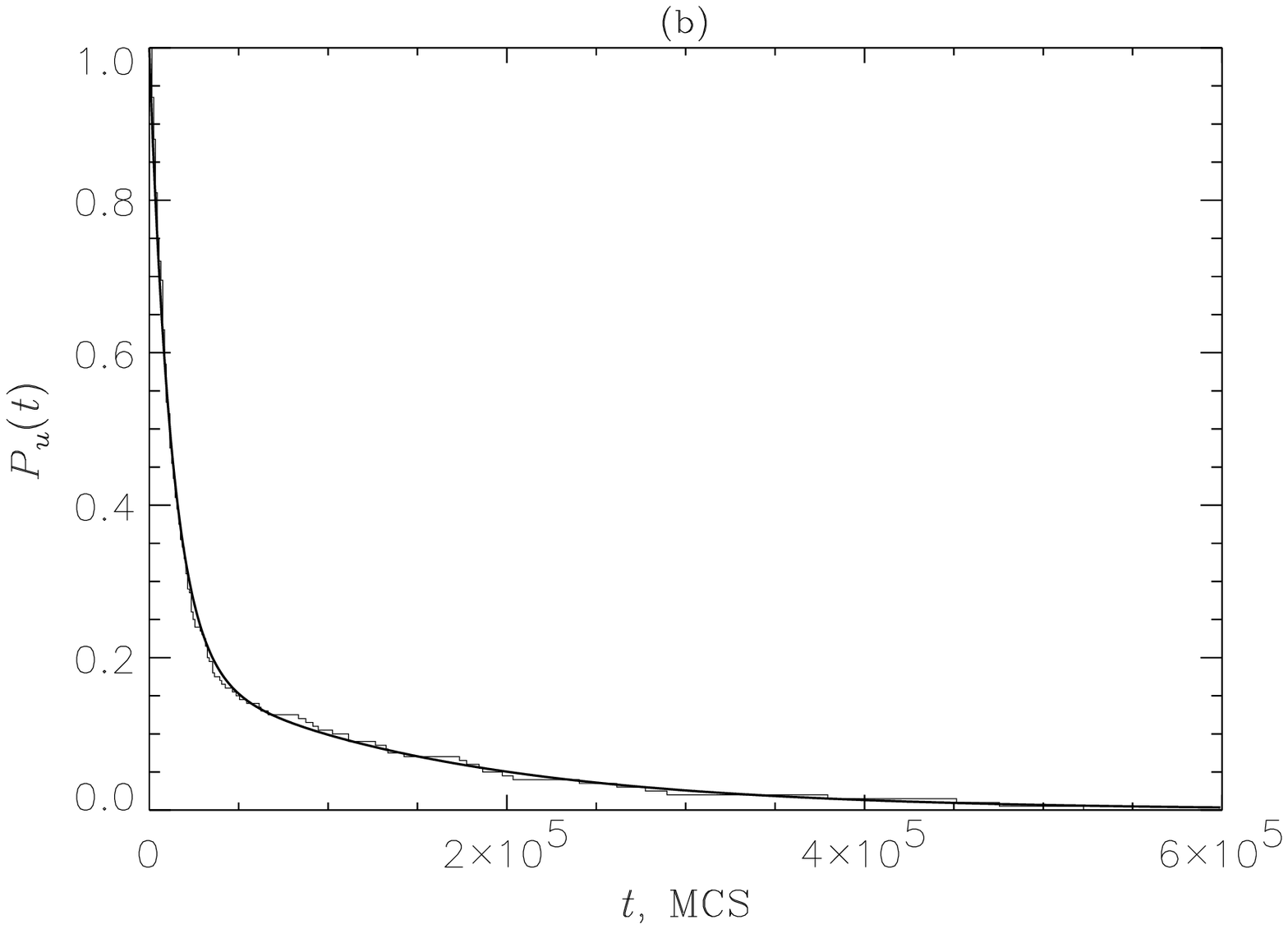,height=10cm,width=14cm}
\]
\end{minipage}

{\bf \large Fig. 24}\\
                      
\end{center}

\newpage

\begin{center}
\begin{minipage}{15cm}
\[
\psfig{figure=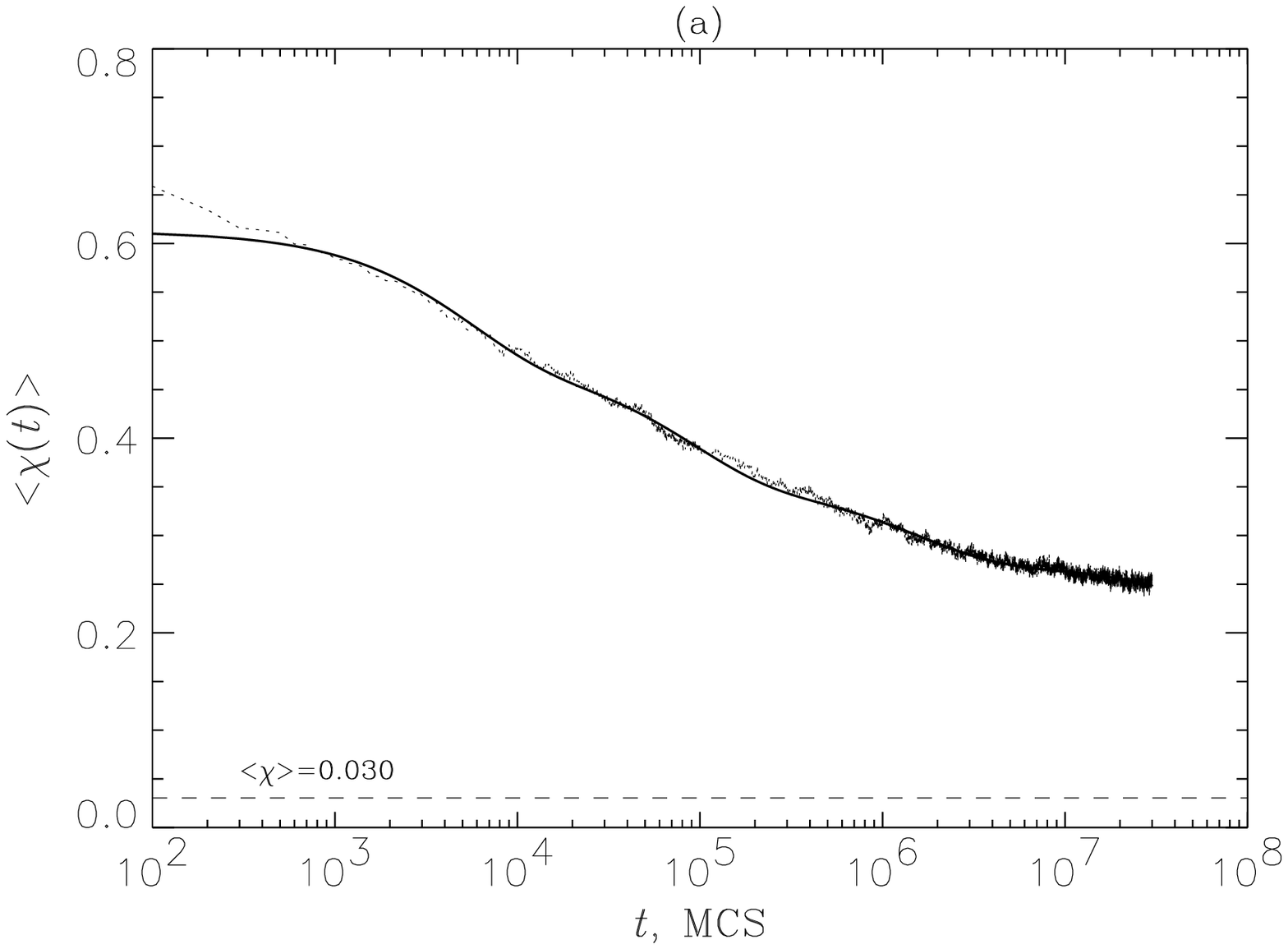,height=10cm,width=14cm}
\]
\end{minipage}

\end{center}

\begin{center}
\begin{minipage}{15cm}
\[
\psfig{figure=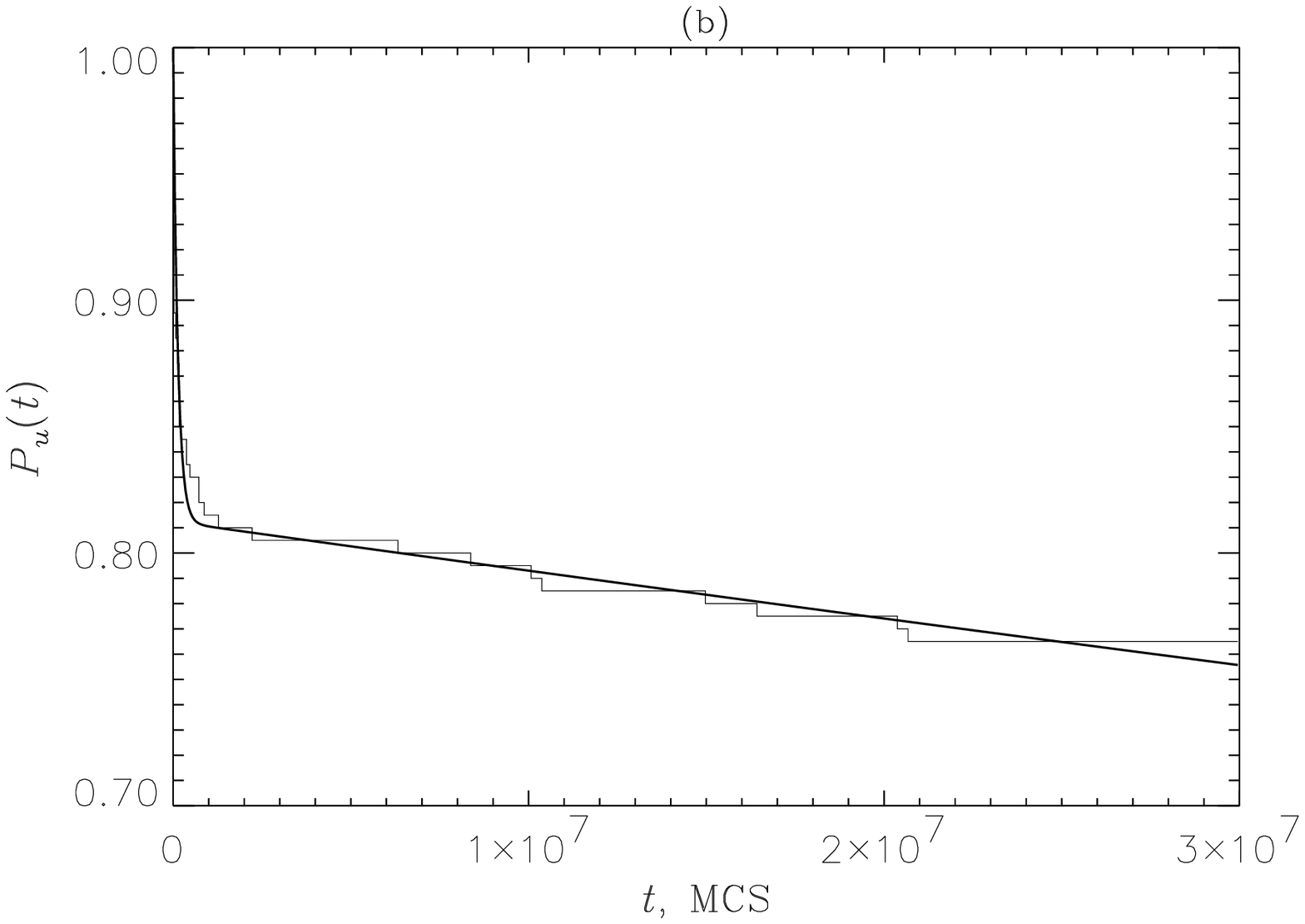,height=10cm,width=14cm}
\]
\end{minipage}

{\bf \large Fig. 25}\\

\end{center}
 
\newpage

\hspace{10cm}

\begin{center}
\begin{minipage}{18.5cm}
\[
\psfig{figure=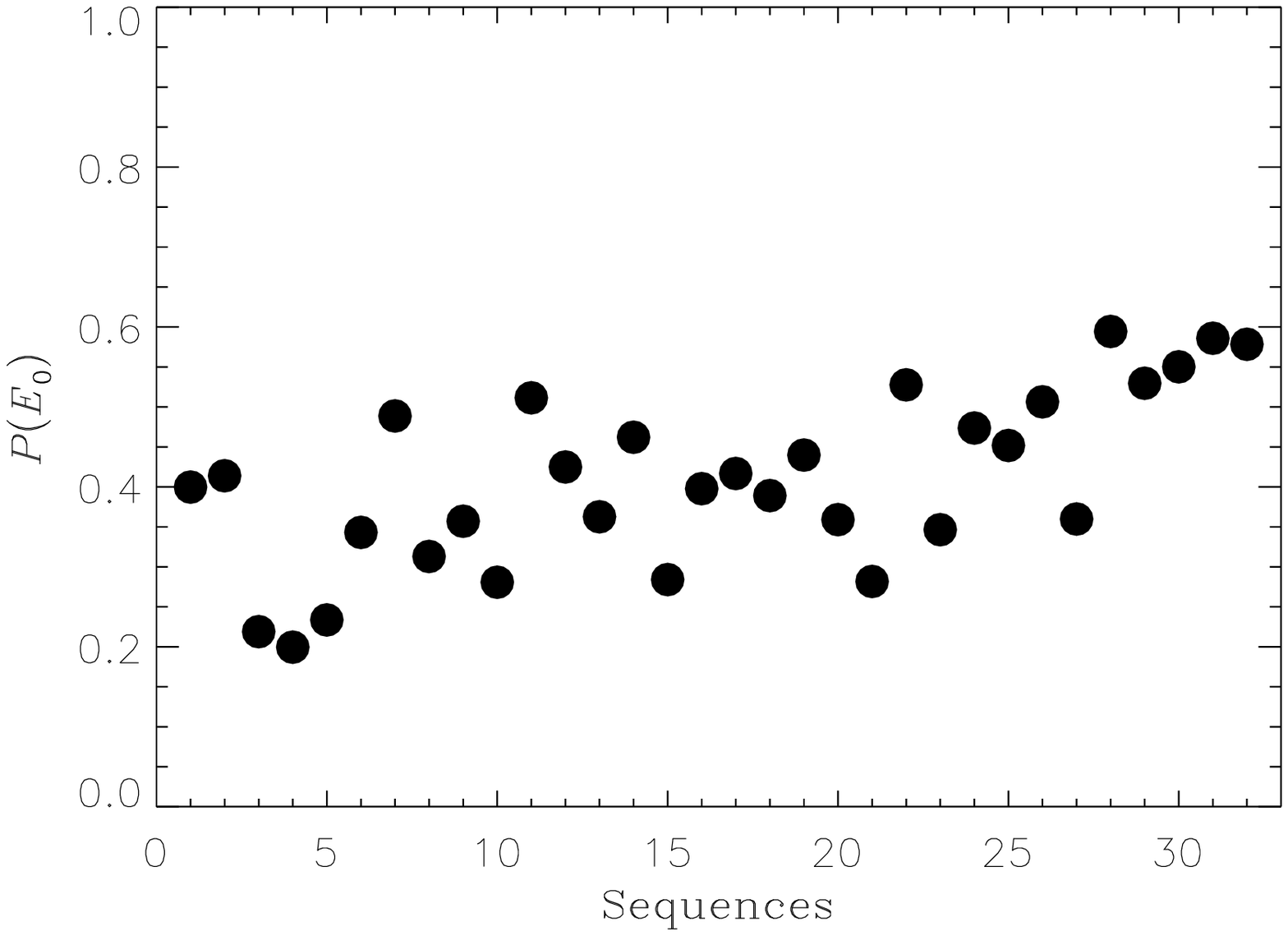,height=13.5cm,width=18.5cm}
\]
\end{minipage}

{\bf \large Fig. 26}\\
\end{center}


\begin{references}

\bibitem{Bryn95} Bryngelson, J.D., Onuchic, J.N., Socci, N.D., Wolynes,
P.G. Funnels, pathways and the energy landscape of protein folding: A
synthesis. Proteins Struct. Funct. Genet. 21:167-195, 1995. 


\bibitem{Dill} Dill, K.A., Bromberg, S., Yue, K., Fiebig, K.M., Yee, D.P.,
Thomas, P.D., Chan, H.S. Principles of protein folding
 - A perspective from simple exact models. Protein Sci. 4:561-602, 1995. 


\bibitem{Hinds94} Hinds, D.A., Levitt, M.A. Exploring conformational space
with a simple lattice model for protein structure. J. Mol. Biol.
243:668-682, 1994. 


\bibitem{Karp92} Karplus, M., Shakhnovich, E. Protein folding: 
Theoretical studies of thermodynamics and dynamics.  In: "Protein Folding".
Creighton, T.E., (ed.). New York: W.H. Freeman. 1992:127-196. 


\bibitem{Thirum94} Thirumalai, D. Theoretical perspectives on in
vitro and in vivo protein folding. In: "Statistical mechanics,
protein structure, and protein substrate interactions". Doniach, S.,   
(ed.). New York: Plenum Press. 1994:115-133. 


\bibitem{Wol} Wolynes, P.G., Onuchic, J.N., Thirumalai, D. 
Navigating the folding routes. Science 267:1619-1620, 1995.


\bibitem{Bald} Baldwin, R.L. The nature of protein folding pathways: The
classical versus the new view. J. Biomolecular NMR 5:103-109, 1995. 


\bibitem{Rad} Radford, S.E., Dobson, C.M. Insights into protein
folding using physical techniques: Studies of lysozyme and
alpha-lactalbumin. Phil. Trans. Roy. Soc. Lond. B 348:17-25, 1995.


\bibitem{Levin} Levinthal, C. In: "Mossbauer spectroscopy in
biological systems". Debrunner, P., Tsibris, J.C.M., M\"{u}nck, E.
(eds.). Urbana: University of Illinois Press. 1969:22-24. 


\bibitem{Amara} Amara, P., Straub, J.E. Folding model
proteins using kinetic and thermodynamic annealing of the classical   
density distribution. J. Phys. Chem. 99:14840-14853, 1995.


\bibitem{Bryn89} Bryngelson, J.D., Wolynes, P.G. 
Intermediates and barrier crossing in a random energy model (with
application to protein folding). J. Phys. Chem. 93:6902-6915, 1989.


\bibitem{Chan93} Chan, H.S., Dill, K.A. Energy landscapes and
the collapse dynamics of homopolymers. J. Chem. Phys. 99:2116-2127, 1993.

\bibitem{Chan94} Chan, H.S., Dill, K.A. Transition states and folding
dynamics of proteins and heteropolymers. J. Chem. Phys. 100:9238-9257,
1994. 

\bibitem{Cov} Covell, D.G., Jernigan, R.L.  Conformations of folded
proteins in restricted spans. Biochemistry 29:3287-3294, 1990. 

\bibitem{Hinds92} Hinds, D.A., Levitt, M.A. A lattice
model for protein structure prediction at low resolution. Proc. Natl.
Acad. Sci. USA 89:2536-2540, 1992.

\bibitem{Honey90} Honeycutt, J.D., Thirumalai, D. 
Metastability of the folded states of globular proteins. Proc. Natl. 
Acad. Sci. USA 87:3526-3529, 1990.

\bibitem{GuoHoney} Guo, Z., Thirumalai, D., Honeycutt,
J.D. Folding kinetics of proteins: A model study. J. Chem. Phys. 
97:525-535, 1992. 

\bibitem{Gar88} Garel, T.R., Orland, H. Mean field model for
protein folding.  Europhys. Lett. 6:307-310, 1988.

\bibitem{Leop} Leopold, P.E., Montal, M., Onuchic, J.N. 
Protein folding funnels: A kinetic approach to the sequence-structure
relationship. Proc. Natl. Acad. Sci. USA 89:8721-8725, 1992.

\bibitem{Sali94a} Sali, A., Shakhnovich, E., Karplus, M. 
How does a protein fold. Nature 369:248-251, 1994. 

\bibitem{Sali94b} Sali, A., Shakhnovich, E., Karplus, M. 
Kinetics of protein folding: A lattice model study of the requirements
for folding to the native state. J. Mol. Biol. 235:1614-1636, 1994.

\bibitem{Shakh89} Shakhnovich, E., Gutin, A.M.  Formation of unique
structure in polypeptide chains. Theoretical investigation with the aid of
a replica approach. Biophys. Chem.  34:187-199, 1989. 


\bibitem{Shakh91} Shakhnovich, E., Farztdinov, G., Gutin, A.M., 
Karplus, M. Protein folding bottlenecks: A lattice Monte
Carlo simulation. Phys. Rev. Lett. 67:1665-1668, 1991.

\bibitem{Skol90} Skolnick, J., Kolinski, A.  Simulation of the folding
of a globular protein. Science 250:1121-1125, 1990. 

\bibitem{Skol91} Skolnick, J., Kolinski, A. Dynamic Monte Carlo 
simulations of a new lattice model of globular protein folding,
structure, and dynamics.  J. Mol. Biol. 221:499-531, 1991.

\bibitem{Zwan95} Zwanzig, R. Simple model of protein folding   
kinetics. Proc. Natl. Acad. Sci. 92:9801-9804, 1995.

\bibitem{Zwan92} Zwanzig, R., Szabo, A., Bagchi, B. 
Levinthal's paradox. Proc. Natl. Acad. Sci. USA 89:20-22, 1992.

\bibitem{Cam95} Camacho, C.J., Thirumalai, D. Modeling
the role of disulfide bonds in protein folding: Entropic barriers and
pathways. Proteins Struct. Funct. Genet. 22:27-40, 1995.

\bibitem{Chan95} Chan, H.S. Kinetics of protein folding.
Nature 373:664-665, 1995.

\bibitem{Der} Derrida, B. Random-energy model: Limit of a
family of disordered models. Phys. Rev. Lett. 45:79-82, 1980.

\bibitem{Gold92} Goldstein, R.A., Luthey-Schulten, Z.A., Wolynes, P.G. 
Optimal protein-folding codes from spin-glass theory. Proc. 
Natl. Acad. Sci. USA 89:4918-4922, 1992.

\bibitem{Socci94} Socci, N.D., Onuchic, J.N. Folding kinetics
of protein-like heteropolymers. J. Chem. Phys. 101:1519-1528, 1994.

\bibitem{Cam93} Camacho, C.J., Thirumalai, D. Kinetics
and thermodynamics of folding in model proteins. Proc. Natl. Acad. Sci.
USA 90:6369-6372, 1993.

\bibitem{Thirum95} Thirumalai, D. From minimal models to real   
proteins: Time scales for protein folding kinetics. J. Physique (Paris)
I 5:1457-1467, 1995.

\bibitem{Shakh90} Shakhnovich, E., Gutin, A.M. Enumeration 
of all compact conformations of copolymers with random sequence of
links. J. Chem. Phys. 93:5967-5971, 1990.

\bibitem{White} White, S.H., Jacobs, R.E. The evolution of proteins from
random amino acid sequences. I. Evidence from the lengthwise distribution
of amino acids in modern protein sequences.  J. Mol. Evol. 36:79-95, 1993. 

\bibitem{Klim} Klimov, D.K., Thirumalai, D., 1994 (unpublished).

\bibitem{Gar94} Garel, T.R., Leibler, L., Orland, H.  Random
hydrophilic-hydrophobic copolymers. J. Physique (Paris) II 4:2139-2148,
1994. 

\bibitem{Miller} Miller, R., Danko, C.A., Fasolka, M.J., Balazs, A.C.,
Chan, H.S., Dill, K.A. Folding kinetics of proteins and copolymers. J.
Chem. Phys. 96:768-780, 1992. 

\bibitem{Binder} Binder, K.A. Theory and technical
aspects of Monte Carlo simulations. In: "Monte Carlo methods in
statistical physics".  Binder, K.A., (ed.). 
Berlin, Heidelberg, New York, Tokyo: Springer Verlag. 1986:26-30. 

\bibitem{Chan90} Chan, H.S., Dill, K.A. The effects of  
internal constraints on the configurations of chain molecules. J. Chem.
Phys. 92:3118-3135, 1990.

\bibitem{KarpSali} Karplus, M., Sali, A. Theoretical studies of protein
folding and unfolding. Curr. Opin. Struct. Biology 5:58-73, 1995. 

\bibitem{Honey92} Honeycutt, J.D., Thirumalai, D. The nature
of folded states of globular proteins. Biopolymers 32:695-709, 1992.

\bibitem{Guo} Guo, Z., Thirumalai, D. Kinetics of
protein folding: Nucleation mechanism, time scales, and
pathways. Biopolymers 36:83-103, 1995.

\bibitem{Martin} Martin, J.L. Computer enumerations. In: "Phase
transitions and critical phenomena".  Domb, C., Green, M.S. (eds.). New
York: Academic Press, 1974:102. 

\bibitem{Shakh93} Shakhnovich, E., Gutin, A.M. A new approach to the
design of stable proteins. Protein Eng. 6:793-800, 1993. 

\bibitem{Shakh94} Shakhnovich, E. Proteins with selected sequences fold
into unique native conformation. Phys. Rev. Lett. 72:3907-3910, 1994. 

\bibitem{Mount} Mountain, R.D., Thirumalai, D.  Quantitative measure of
efficiency of Monte Carlo simulations. Physica A 210:453-460, 1994. 

\bibitem{Karp95} Karplus, M., Sali, A., Shakhnovich, E.  
Kinetics of protein folding. Nature 373:665, 1995.

\bibitem{ThirumGuo} Thirumalai, D., Guo, Z.  Nucleation mechanism for
protein folding and theoretical predictions for hydrogen-exchange labeling
experiments. Biopolymers 35:137-140, 1995. 

\bibitem{Abk1} Abkevich, V.I., Gutin, A.M., Shakhnovich, E. 
Specific nucleus as the transition state for protein folding: Evidence
from the lattice model.  Biochemistry 33:10026-10036, 1994.

\bibitem{Abk2} Abkevich, V.I., Gutin, A.M., Shakhnovich, E.  Free energy
landscape of protein folding kinetics: Intermediates, traps, and multiple
pathways in theory and lattice model simulations. J. Chem. Phys. 
101:6052-6062, 1994. 

\bibitem{Otzen} Otzen, D.E., Itzhaki, L.S., Fersht, A.R.  Structure of the
transition state for the folding/unfolding of the barley chymotrypsin
inhibitor 2 and its implications for mechanisms of protein folding. Proc.
Natl. Acad. Sci. 91:10422-10425, 1994. 

\bibitem{Betan} Betancourt, M.R., Onuchic, J.N. Kinetics of protein-like
models: The energy landscape factors that determine folding. J. Chem.
Phys. 103:773-787, 1995. 

\bibitem{Socci95} Socci, N.D., Onuchic, J.N. Kinetic and thermodynamic
analysis of protein-like heteropolymers: Monte Carlo histogram technique.
J. Chem. Phys. 103:4732-4744, 1995. 

\bibitem{Bai} Bai, Y, Sosnick, T.R., Mayne, L., Englander, S.W.  Protein
folding intermediates: Native-state hydrogen exchange. Science
269:192-197, 1995. 

\bibitem{Fer} Fersht, A.R. Optimization of rates of protein   
folding: The nucleation - condensation mechanism and its implications.
Proc. Natl. Acad. Sci. 92:10869-10873, 1995.

\bibitem{Helix} Guo, Z., Thirumalai, D. Kinetics and thermodynamics of
folding of a four-helix bundle protein (unpublished). 

\end{references}
\end{document}